\documentclass[12pt]{iopart}
\usepackage{iopams}  

\expandafter\let\csname equation*\endcsname\relax
  \expandafter\let\csname endequation*\endcsname\relax

\usepackage{amssymb}
\usepackage{amsmath}

\newcommand{\moy}[1]{\left\langle #1 \right\rangle}
\newcommand{\Rtr}[0]{\boldsymbol{X}_t}
\newcommand{\RTP}[0]{\boldsymbol{X}_t}
\newcommand{\XTP}[0]{X_t}
\newcommand{\KTP}[0]{K}

\newcommand{\ex}[1]{\mathrm{e}^{#1}}

\newcommand{\epu}[0]{\ee_1}
\newcommand{\emu}[0]{\ee_{-1}}
\newcommand{\emuu}[0]{\ee_{\mu}}

\newcommand{\enu}[0]{\ee_{\nu}}

\newcommand{\ee}[0]{\boldsymbol{e}}
\newcommand{\XX}[0]{\boldsymbol{X}}
\newcommand{\ep}[0]{\epsilon}

\newcommand{\rr}[0]{\boldsymbol{r}}
\newcommand{\wt}[0]{\widetilde{w}}
\newcommand{\mt}[0]{\widetilde{m}}
\newcommand{\gt}[0]{\widetilde{g}}
\newcommand{\dd}[0]{\mathrm{d}}
\newcommand{\ii}[0]{\mathrm{i}}

\newcommand{\zz}[0]{\mathbf{0}}

\usepackage{enumerate}

\usepackage{empheq}

\usepackage{fancybox}

\usepackage{etoolbox}
\usepackage{hyperref}

\usepackage{color}

\makeatletter
\def\@mkboth#1#2{}
\newlength\appendixwidth
\preto\appendix{\addtocontents{toc}{\protect\patchl@section}}
\newcommand{\patchl@section}{%
  \settowidth{\appendixwidth}{\textbf{Appendix }}%
  \addtolength{\appendixwidth}{1.5em}%
  \patchcmd{\l@section}{1.5em}{\appendixwidth}{}{\ddt}%
}
\makeatother

\begin{document}

\title[Distribution of the position of a driven tracer in a hardcore lattice gas]{Distribution of the position of a driven tracer in a hardcore lattice gas}

\author{Pierre Illien, Olivier B\'enichou, Gleb Oshanin and Rapha\"el Voituriez}

\address{Universit\'e Pierre-et-Marie-Curie (Paris 6) - Laboratoire de Physique Th{\'e}orique de la Mati{\`e}re
Condens{\'e}e (UMR CNRS 7600), 4 place Jussieu, 75005 Paris, France}

\eads{\mailto{illien@lptmc.jussieu.fr}, \mailto{benichou@lptmc.jussieu.fr},  \mailto{oshanin@lptmc.jussieu.fr},  \mailto{voiturie@lptmc.jussieu.fr}}

\begin{abstract}

We study the position of a biased tracer particle (TP) in a bath of hardcore particles moving on a lattice of arbitrary dimension and in contact with a reservoir. Starting from the master equation satisfied by the joint probability of the TP position and the bath configuration and resorting to a mean-field-type approximation, we presented a computation of the fluctuations of the TP position  in a previous publication [Phys. Rev. E \textbf{87}, 032164 (2013)].  Counterintuitively, on a one-dimensional lattice, the diffusion coefficient of the TP was shown to be a non-monotonic function of the density of bath particles, and reaches a maximum for a nonzero value of the density. Here, we: (i) give the details of this computation and offer a physical insight into the understanding of the non-monotonicity of the diffusion coefficient; (ii) extend the mean-field-type approximation to decouple higher-order correlation functions, and obtain the evolution equation satisfied by the cumulant generating function of the position of the TP, valid in  any space dimension. In the particular case of a one-dimensional lattice, we solve this equation and obtain the probability distribution of the TP position. We  show that the position rescaled by its fluctuations is asymptotically distributed accordingly to a Gaussian distribution in the long-time limit.

\end{abstract}

\maketitle

\tableofcontents

\section{Introduction}

\subsection{Context}

Studying the dynamics of an active particle or a particle submitted to an external force travelling in a crowded environment is a frequent problem in physics and in biology. Different examples are found in biophysics, when one considers molecular motors, motile living cells or bacteria \cite{Marchetti2013,{Chou2011a}}, or in the study of biased intruders in granular systems \cite{Candelier2009} or colloidal suspensions  \cite{Habdas2004}. The determination of the dynamics of this tracer particle (TP) is consequently an important question, with different applications. In particular, new experimental tools allow the study of a medium using a microscopic probe particle submitted to an external force. This field of research, commonly named active microrheology, has been applied to  different systems in the past decades, among which biological cells \cite{Bausch1998,{Lau2003}}, complex fluids \cite{Chen2003, {Chae2005}} and colloidal suspensions \cite{Habdas2004,{Meyer2006}}.

From a theoretical point of view, the difficulty lies in the modeling of the environment of the tracer particle (TP), which is constituted of a large number of interacting bath particles. In most analytical approaches, the evolution of the position of the TP is studied with some effective description of the bath of particles \cite{Marconi2008}, which do not take into account the correlations between the position of the TP and the density profiles of the surrounding bath. In the situation where the TP and the bath particles have comparable sizes, the response of the probe to the external forcing, and in particular its fluctuations, cannot be accounted for correctly with an effective treatment.

Here, we study the diffusion of a driven tracer in a lattice gas of hardcore particles. This minimal model explicitly takes into account the bath dynamics: the driven TP performs a biased nearest-neighbor random walk, in a bath of particles performing symmetric nearest-neighbor random walks, with the restriction that each site is occupied by at most one particle. We assume that the lattice is in contact with a reservoir of particles, so that the bath particles present on the lattice may desorb back to the reservoir, and particles from the reservoir may adsorb onto vacant lattice sites. This so-called Langmuir kinetics is relevant to describe situations where a gas or a vapor is brought in contact with a solid surface, on which the gas particles may form an adsorbed layer. The transport properties of the adsorbed particles have been shown to control many different processes, such as spreading of molecular films on solid surfaces \cite{Burlatsky1996a} or dewetting \cite{Oshanin1998a,{Tosatti1994}}. The particular case where the Langmuir kinetics is coupled to a Totally Asymmetric Exclusion Process was investigated theoretically \cite{Parmeggiani2003,Parmeggiani2004}, and has been show to be relevant to describe the directional motion of molecular motors on a cytoskeletal filament, with random attachment and detachment of the motors \cite{Homard2005,Kruse2002}. \\

Studying the transport properties of a biased TP in a hardcore lattice gas is actually a complex $N$-body problem. In the situation where the density of bath particles $\rho$ is very high and where the number of particles on the lattice is conserved, the problem can be treated exactly to obtain  results at leading order in $(1-\rho)$ \cite{{Benichou2002a},Illien2013a, Benichou2013c, Illien2014}. For an arbitrary density of particles, exact results were established concerning the mean position of the TP in the one-dimensional situation \cite{Landim1998}, and concerning the validity of the Einstein relation \cite{Komorowski2005a}. The situation where the lattice is populated by an arbitrary density of particles  and where it is in contact with a reservoir of particles was  addressed  by resorting to a  mean-field-type approximation consisting of the decoupling  of relevant correlation functions, allowing the computation of  the mean position of the TP and of the bath density profiles in the long-time limit, in the case of a one-dimensional lattice  \cite{ Benichou1999} and for lattices of higher dimension \cite{{Benichou1999b},Benichou1999a,  Benichou2001, {Benichou2005}}. Numerical simulations sampling exactly the master equation of the problem revealed the accuracy of the decoupling approximation in a wide range of parameters. The situation where the bath particles are fixed and can appear and disappear with prescribed rates, known as dynamical percolation \cite{Druger1983}, was also considered  \cite{Benichou2000a}.

More recently, by extending the decoupling approximation initially proposed to study the mean position of the TP,  its fluctuations have been studied \cite{Benichou2013}. An evolution equation for the fluctuations of the TP on a lattice of arbitrary dimension was obtained. This equation was solved explicitly in the case of a one-dimensional lattice, and in the stationary limit. The analysis of the solutions revealed a striking feature of the diffusion coefficient: in a wide range of parameters, it is shown to be a nonmonotonic function of the density of particles on the lattice. Counterintuitively, the diffusion coefficient of the TP can then be enhanced by the presence of bath particles in its environment.

Recently, this nonmonotonicity was also observed in the  situation where the TP is dragged in a bath of soft particles \cite{Demery2014,Demery2015}, and then appears to be a generic feature of biased intruders in crowded environments.

\subsection{Main results of this paper and overview}

In this paper, we first give a detailed computation of the results presented in \cite{Benichou2013}: we establish the evolution equation of the fluctuations of the TP position in arbitrary dimension under the decoupling approximation, and solve it in the case of a one-dimensional lattice. We show that the non-monotonicity of the diffusion coefficient with respect to the density of particles on the lattice is correlated to a non-monotonicity of some cross-correlation functions with respect to the distance to the TP. 

The main result of this paper is the following: we go one step further and generalize the mean-field-type approximation in order to calculate the cumulant generating function of the position of the TP, and therefore its complete probability distribution. Denoting by $\boldsymbol{X}_t$ the position of the TP, by $X_t = \boldsymbol{X}_t \cdot \ee_1$ its projection along the direction of the bias, and defining the cumulant generating function $\Psi(u;t) \equiv \ln  \langle \mathrm{e}^{\mathrm{i} u X_t}\rangle$, we obtain in the long-time limit
\begin{equation}
\label{ }
\Psi(u;t) \underset{t \to \infty}{\sim} \Phi(u)t,
\end{equation}
with 
\begin{equation}
\label{ }
 \Phi(u) = \frac{p_1}{\tau}\left(\ex{\ii u \sigma}-1\right)\left[1-\frac{\moy{\ex{\ii u \XTP}\eta_{\RTP + \ee_1}}}{\moy{\ex{\ii u \XTP}}}\right]+\frac{p_{-1}}{\tau}\left(\ex{-\ii u \sigma}-1\right)\left[1-\frac{\moy{\ex{\ii u \XTP}\eta_{\RTP - \ee_{1}}}}{\moy{\ex{\ii u \XTP}}}\right],
\end{equation}
where $p_{\nu}$ is the jump probability of the TP in the direction $\nu$ of the lattice, $\tau$ is its characteristic jump time, $\sigma$ is the lattice spacing and $\eta_{\boldsymbol{r}}$ is the occupation number at site $\boldsymbol{r}$. The evolution equations for the correlation functions $\wt_{\rr}(u;t)={\moy{\ex{\ii u \XTP}\eta_{\RTP +\rr}}}/{\moy{\ex{\ii u \XTP}}}$ are obtained using an extension of the mean-field-type approximation proposed to study the mean and fluctuations of the position of the TP. This result is given by Eqs. (\ref{wtildegen}) and (\ref{wtildeboundary}), where the quantities $k_{\rr}(t)=\moy{\eta_{\RTP+\rr}}$ are the density profiles in the reference frame of the TP and are the solutions of Eqs. (\ref{systemek1}) and (\ref{systemek2}). The equations satisfied by the correlation functions $\wt_{\rr}(u;t)$ are solved in the particular case of a one-dimensional lattice, and we compute the probability distribution of the position of the TP.  We show that all the cumulants of the TP position scale as $t$. 

We also consider the  random variable obtained by rescaling the position of the TP by $\sqrt{t}$. For this random variable, all the cumulants  of order greater than $2$ vanish in the long-time limit, which shows that the position of the TP is asymptotically Gaussian. \\

The article is organized as follows. In section 2, we present the model and give the master equation governing the joint probability of the position of the TP and of the bath particles configuration. In section 3, we give the evolution equations of the first two cumulants. These equations involve the density profiles around the TP and some tracer-bath cross-correlation functions, whose evolution equations are explicitly given in a closed form by resorting to a mean-field-type approximation. In section 4, we generalize this approximation in order to obtain a closed set of equations for the cumulant generating function of the TP position. These equations are valid in arbitrary dimension, and they give in principle the whole probability distribution of the TP position. As a particular case, we also obtain the equation satisfied by the third-cumulant of the distribution. In section 5, we solve the equations obtained under the decoupling approximation in the particular case of a one-dimensional lattice, and then obtain explicit expressions for the first three cumulants, as well as an implicit determination of the cumulant generating function.  In section 6, these analytical solutions are compared with results from Monte-Carlo numerical simulations that exactly sample the master equation. We  summarize our results and give an outlook in section 7.

\section{Model and master equation}

\subsection{Model}
\label{sec:model}

\begin{figure}
	\begin{center}
		\includegraphics[width=10cm]{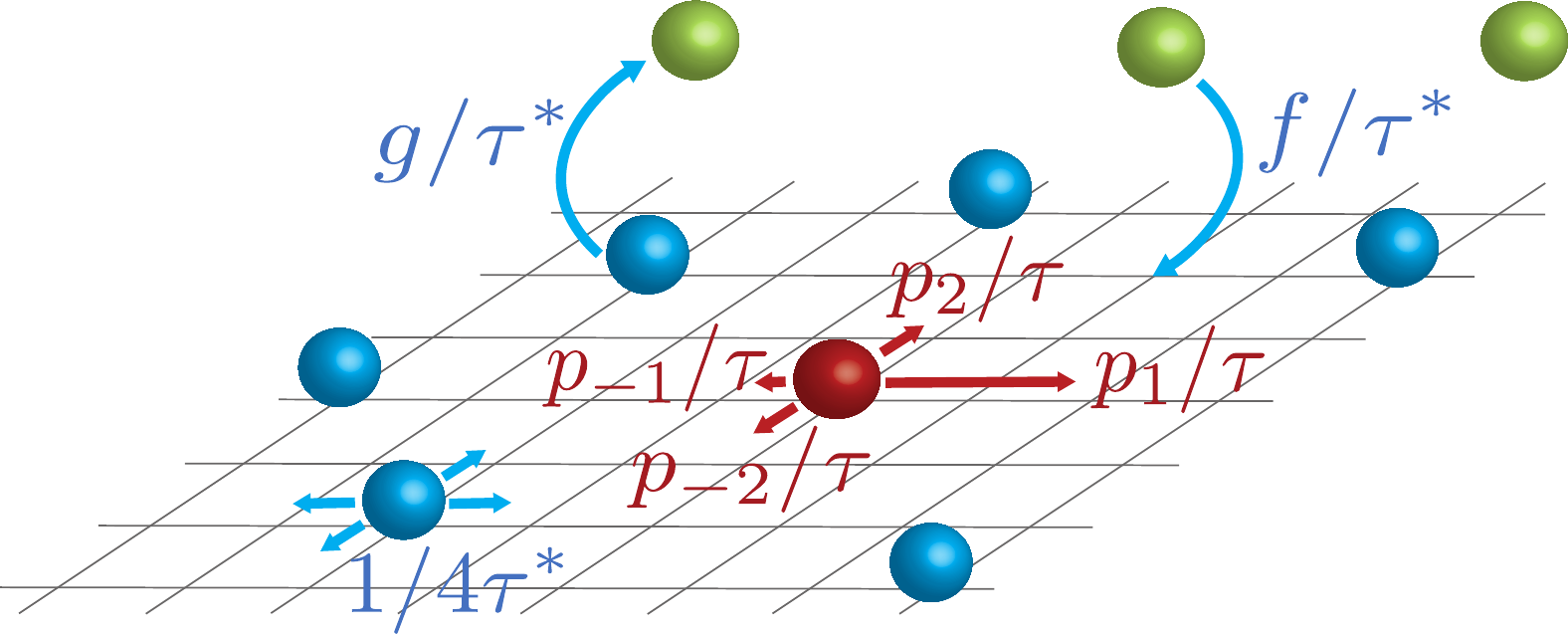}
	\end{center}
	\caption{Model and notations in the two-dimensional case.}
	\label{fig:model}
\end{figure}

We consider a $d$-dimensional  
hypercubic lattice (we denote by  $\sigma$ its spacing)
in contact with a reservoir of particles (see Fig. \ref{fig:model}).  
We adopt a continuous-time description of the system. We assume that the particles in the reservoir 
 adsorb 
onto empty lattice sites at a fixed rate $f/\tau^*$. The  particles adsorbed on the lattice 
desorb  back to the reservoir
with a rate $g/\tau^*$.
The adsorbed particles  perform symmetric nearest-neighbor random walks, and jump to any of the $2d$ neighboring sites with a rate $1/ (2d \tau^*)$. All the particles present on the lattice interact with a hardcore exclusion rule, such that each lattice site is occupied by at most one particle.

We introduce the occupation variable $\eta_{{\rr}}$, which takes two values: $1$, if the site ${\rr}$ is occupied by an adsorbed particle, and $0$, otherwise.
The mean density
of the bath particles, $ \langle\eta_{\rr}\rangle$, is equal to $\rho=f/(f+g)$ in the long-time limit. However,  the number of bath particles adsorbed on the lattice is not constant.   The case where the number of particles on the lattice is conserved can be retrieved by taking the limits $f\to0$ and $g \to 0$ with a fixed value of the density $\rho=f/(f+g)$ .

We also introduce a tracer particle (TP). The TP cannot desorb from the lattice and it is submitted to
an external force, such that it preferentially jumps in the direction of the unit vector $\ee_1$. We suppose
that the TP, whose position at time $t$ is a random variable denoted by $\XX_t$, waits a random time distributed accordingly to an exponential distribution of mean $\tau$, and attempts to jump onto one of the neighboring sites $\RTP+\ee_\mu$ ($\ee_\mu \in\{\ee_{\pm1},\cdots,\ee_{\pm d}\}$). For simplicity, we use the notation $\ee_{-\nu}\equiv -\ee_\nu$.
The probability for the TP to jump in direction $\nu$ is denoted by $p_\nu$. The equations presented below and their solutions are valid for any choice of the jump probabilities. However, it can be convenient to assume that the bias originates from an external force $\boldsymbol{F}=F  \ee_1$, so that the jump probability in direction $\nu$ writes
\begin{equation}
\label{eq:jump_prob_decoup}
p_\nu = \frac{\ex{ \frac{1}{2}\beta\boldsymbol{F}\cdot {\ee}_{ \nu} } }{ \sum_{\mu\in\{\pm 1,... \pm d\} }  \ex{ \frac{1}{2}\beta\boldsymbol{F}\cdot \ee_{ \mu}}}
\end{equation}
where $\beta=1/(k_\mathrm{B}T)$ is the inverse temperature, and will be taken equal to one. Note that this choice of $p_\nu$ fulfills the detailed balance condition. After the direction of the jump has been chosen, the TP hops to the target site if it is vacant; it remains on the same site otherwise.

\subsection{Master equation}

We begin by introducing some auxiliary definitions.
We define $\eta\equiv
\{\eta_{\rr}\}$ as the entire set of the occupation variables, which is the 
configuration of the lattice at a given time. Next, we define the joint probability
$P(\XX,\eta;t)$, which is the probability for the TP to be at site $\XX$ at time $t$ with the bath particles in  a given configuration $\eta$.
Next, we define  $\eta^{{\rr},\nu}$ as the 
configuration obtained from $\eta$ by exchanging the occupation numbers of sites   ${\rr}$ and $\rr+\ee_\nu$ (Kawasaki-type exchange \cite{Kawasaki1966} due to the hop of a bath particle) and  
${\hat \eta}^{\rr}$ as the configuration obtained from  $\eta$ by
the change $\eta_{\rr} \to 1-\eta_{\rr}$ (Glauber-type exchange \cite{Glauber1963} due to adsorption/desorption events). The time evolution of the joint
probability $P(\XX,\eta;t)$ is given by the master equation :
\begin{eqnarray}
\fl2d\tau^*\partial_t P(\XX,\eta;t)=&
\sum_{\mu=1}^d\sum_{{\rr}\neq\XX-\ee_\mu,\XX} \left[ P(\XX,\eta^{{\rr},\mu};t)-P(\XX,\eta;t)\right]\nonumber\\
&+\frac{2d\tau^*}{\tau}\sum_{\mu}p_\mu\left[\left(1-\eta_{\XX} \right)P(\XX-\ee_{\mu},\eta;t)
-\left(1-\eta_{\XX+\ee_{\mu}}\right)P(\XX,\eta;t)\right]\nonumber\\
&+2dg\sum_{{\rr}\neq \XX} \left[\left(1-\eta_{\rr}\right)P(\XX,
\hat{\eta}^{{\rr}};t)-\eta_{\rr}P(\XX,\eta;t)\right]\nonumber\\
&+2df\sum_{{\rr}\neq\XX} \left[\eta_{\rr}P(\XX,\hat{\eta}^{{\rr}};t)
-\left(1-\eta_{\rr}\right)P(\XX,\eta;t)\right].
\label{eqmaitresse}
\end{eqnarray} 
The first term of the right-hand-side of (\ref{eqmaitresse}) describes the diffusion of adsorbed particles, the second term corresponds to the diffusion of the TP, and the third and fourth terms are associated to the desorption and adsorption events of the bath particles.

If no otherwise specified, the sum over an index $\mu$ runs over the $2d$ elements $\{\pm 1,\cdots,\pm d\}$. In what follows, the brackets $\moy{ \cdot}$ denote an average over the TP position and bath particles configurations with weight $P(\XX,\eta;t)$, and $\XTP= \RTP \cdot \ee_1$ denotes the position of the TP along the direction of the external force.

\section{Equations satisfied by the first cumulants}

\subsection{Mean position}

The time evolution of the first moment 
$\langle\XTP \rangle$ 
of the TP position can be obtained by multiplying both sides of (\ref{eqmaitresse}) by
$(\XX \cdot \ee_1)$ and summing over all possible configurations $(\XX,\eta)$. An alternative way to compute $\moy{X_t}$ is to write that during an infinitesimal time interval $\Delta t$, the TP position $X_t$ evolves according to
\begin{equation}
\label{eq:bilan_pos_TP}
X_{t+\Delta t}=
\begin{cases}
X_{t}+\sigma     & \text{with probability $p_1(1-\eta_{\RTP+\ee_1})\frac{\Delta t}{\tau}$}, \\
X_{t}-\sigma     & \text{with probability $p_{-1}(1-\eta_{\RTP+\ee_{-1}})\frac{\Delta t}{\tau}$}, \\
X_{t}     & \text{with probability $1-p_1(1-\eta_{\RTP+\ee_1})\frac{\Delta t}{\tau}-p_{-1}(1-\eta_{\RTP+\ee_{-1}})\frac{\Delta t}{\tau}$},
\end{cases}
\end{equation}
and take the average of this equation. Both methods result in the following exact equation:
\begin{equation}
\label{eq:def_velocity_general}
\frac{\mathrm{d}}{\mathrm{d}t} \moy{ \XTP } =
\frac{\sigma}{\tau}\left\{p_1  \left[1-k_{\ee_1}(t)\right]-p_{-1} \left[1-k_{\ee_{-1}}(t)\right]\right\},
%\label{Xmean_first}
\end{equation}
where $k_{\rr}(t) \equiv \moy{ \eta_{\RTP+\rr} }$
is the probability of having at time $t$ an 
adsorbed particle 
at position $\rr$, 
defined in the frame of reference moving with the TP. 
In other words,  $k_{\rr}(t)$
can be thought of as being 
the density
profile as seen from the moving TP. \\

Note that the approximation  obtained by replacing all the local average densities $k_{\rr}(t)$ by the global density $\rho$ in Eq. (\ref{eq:def_velocity_general})  (that we will refer to as the \emph{trivial mean-field} approximation) yields
\begin{equation}
\frac{\mathrm{d}}{\mathrm{d}t} \moy{ \XTP } =
\frac{\sigma}{\tau}(p_1-p_{-1})(1-\rho).
\end{equation}

A more accurate determination of the evolution of the  mean position of the TP is given by Eq. (\ref{eq:def_velocity_general}) and  relies on the calculation of the quantities $k_{\ee_{\pm 1}}(t)$, which are the mean density of bath particles at the sites in the vicinity of the TP. This actually requires the computation of the density profile $k_{\rr}(t)$ for arbitrary $\rr$. The evolution equations for $k_{\rr}(t)$ may be obtained by multiplying the master equation (\ref{eqmaitresse}) by $\eta_{\XX+\rr}$ and summing over all the configurations of $(\XX,\eta)$. We get the following equation :
\begin{eqnarray}
 2d\tau^*\partial_t k_{\rr}(t)&=&\sum_\mu\left(\nabla_\mu-\delta_{\rr,\emuu}\nabla_{-\mu}\right)k_{\rr}(t)-2d(f+g)k_{\rr}(t)+2df \nonumber\\ 
&&+\frac{2d\tau^*}{\tau}\sum_\nu p_\nu \moy{(1-\eta_{\RTP+\ee_\nu})\nabla_\nu\eta_{\RTP+\rr}},
\label{eqknoapprox}
\end{eqnarray}
where we define the operator $\nabla_\mu$ acting on any space-dependent function $f$:
\begin{equation}
\label{eq:def_nabla}
\nabla_\mu f(\rr)=f(\rr+\ee_\mu)-f(\rr).
\end{equation}
Eq. (\ref{eqknoapprox}) is not closed with respect to the density profiles $k_{\rr}$, but involves the correlation functions $\moy{\eta_{\RTP+\ee_\mu}\eta_{\RTP+\rr}}$. The evolution equations for these correlation functions can be obtained from the master equation (\ref{eqmaitresse}), and one can show that they actually involve higher-order correlation functions. Consequently, we face the problem of solving an infinite hierarchy of coupled equations for the correlation functions. We then resort to an  approximation, obtained by writing the occupation variables as $\eta_{\boldsymbol{R}} = \moy{\eta_{\boldsymbol{R}}} +\delta \eta_{\boldsymbol{R}}$, and by discarding the terms of order $(\delta \eta_{\boldsymbol{R}})^2$. We obtain
\begin{eqnarray}
\moy{\eta_{\RTP+{ \rr}}\eta_{\RTP+\ee_{ \mu}}} &=& \moy{(\moy{\eta_{\RTP+\rr}} + \delta \eta_{\RTP+\rr})(\moy{\eta_{\RTP+\ee_\mu}} + \delta \eta_{\RTP+\ee_\mu})}\\
& \simeq & \moy{\eta_{\RTP+{ \rr}}}\moy{\eta_{\RTP+\ee_{ \mu}}} \\
&\simeq & k_{\rr}(t) k_{\ee_\mu}(t),
\label{decouplage_k}
\end{eqnarray}
which is valid for $\rr \neq \ee_\mu$. In the particular case where $\rr = \ee_\mu$, recalling that $\eta_{\RTP+\ee_\mu} \in \{0,1\}$, we get
\begin{eqnarray}
\moy{({\eta_{\RTP+\ee_{ \mu}}})^2} & = & \moy{{\eta_{\RTP+\ee_{ \mu}}}} \\
 & = & k_{\ee_\mu}(t).
\end{eqnarray}

 The approximation (\ref{decouplage_k}) then relies on the \emph{decoupling} of the correlation functions $\moy{\eta_{\RTP+{ \rr}}\eta_{\RTP+\ee_{ \mu}}} $. It can be seen as a mean-field type approximation, and will be generalized later to study  the other cumulants of the TP position. It will be referred to as the \emph{decoupling} approximation in what follows. This approximation will be shown to be very accurate later on. \\

Using this approximation in Eq. (\ref{eqknoapprox}), we obtain
\begin{equation}
%\boxed{
2d\tau^*\partial_tk_{\rr}(t)=\widetilde{L}k_{\rr}(t)+2df, 
%}
\label{systemek1}
\end{equation}
if $\rr\neq\ee_\nu$. For the  sites ${ \rr} = \ee_\nu$ 
with $\nu=\{\pm1,\pm2,\ldots,\pm d\}$ we find
\begin{equation}
%\boxed{
2d\tau^*\partial_tk_{\ee_\nu}(t)=[\widetilde{L}+A_\nu(t)]k_{\ee_\nu}(t)+2df,  
%}
\label{systemek2}
\end{equation}
where 
$\widetilde{L}$ is the operator  
\begin{equation}
\widetilde{L}\equiv\sum_\mu A_\mu(t)\nabla_\mu-2d(f+g),
\end{equation}
and the coefficients $A_{\mu}(t)$ are defined by
\begin{equation}
A_\mu(t)  \equiv1+\frac{2d\tau^*}{\tau}p_\mu[1-k_{\ee_\mu}(t)].
\label{eq:def_A_general}
\end{equation}
The occupation number of the origin is taken equal to zero by convention. Note that Eq. (\ref{systemek2}) represents, from the mathematical point of view, 
the boundary conditions for the general evolution equation (\ref{systemek1}), imposed on the sites in the
immediate vicinity of the TP. Eqs. (\ref{systemek1}) and (\ref{systemek2}) together with Eq. (\ref{eq:def_A_general}) thus constitute
a closed system of equations which suffices to compute the density profiles $k_{\rr}(t)$.  These equations were first obtained by B\'enichou et al., and solved in the case of a one-dimensional lattice \cite{Benichou1999} and of higher-dimensional lattices \cite{Benichou2001,Benichou1999a}.

\subsection{Fluctuations of the TP position}

The time evolution of the second moment $\langle {\XTP}^2\rangle$ is obtained by multiplying the master equation by $(\XX\cdot \ee_1)^2$, and averaging over the TP position and the bath configuration; or, alternatively, averaging the balance equation (\ref{eq:bilan_pos_TP}). The details of this calculation are given in \ref{app:master_eq_sec_moment}. We get 
\begin{align}
\frac{\mathrm{d}}{\mathrm{d}t} \langle {\XTP}^2\rangle\equiv&\frac{\mathrm{d}}{\mathrm{d}t}  \moy{(\RTP\cdot \ee_1)^2} \nonumber\\
=&\frac{2 \sigma}{\tau}\left\{ p_1  \left[\langle \XTP \rangle-g_{\ee_1}(t)\right]-p_{-1} \left[\langle \XTP \rangle-g_{\ee_{-1}}(t)\right]\right \}\nonumber\\
&+\frac{\sigma^2}{\tau}
\left\{ p_1  \left[ 1 -k_{\ee_1}(t)\right]+p_{-1} \left[1-k_{\ee_{-1}}(t)\right]\right\},
\label{X2mean}
\end{align}
where $g_{ \rr}(t)\equiv \moy{\XTP\,\eta_{\RTP+{ \rr}}}$.
Knowing that 
\begin{equation}
\frac{\mathrm{d}}{\mathrm{d}t} \langle {\XTP}\rangle^2=2\langle \XTP \rangle \left(\frac{\mathrm{d}}{\mathrm{d}t} \langle \XTP \rangle \right) ,
\end{equation}
and using Eq. (\ref{eq:def_velocity_general}), we can deduce an expression for the second cumulant of the TP position in the first direction: 
\begin{align}
\frac{\mathrm{d}}{\mathrm{d}t} \left(\moy{{\XTP}^2}-\moy{{\XTP}}^2\right) =&-\frac{2 \sigma}{\tau}\left[p_1  \widetilde{g}_{\ee_1}(t)-p_{-1} \widetilde{g}_{\ee_{-1}}(t)\right] \nonumber \\
&+\frac{\sigma^2}{\tau}
\left\{ p_1  \left[ 1 -k_{\ee_1}(t) \right]+p_{-1} \left[ 1-k_{\ee_{-1}}(t) \right]\right\},
\label{dispersion}
\end{align}
where
\begin{equation}
\widetilde{g}_{\rr}(t)\equiv\moy{\delta\XTP\delta\eta_{\RTP+{\rr}}},
\label{defgtilde}
\end{equation}
with $\delta\XTP\equiv \XTP-\moy{\XTP}$ and $\delta\eta_{\RTP+{\rr}} =\eta_{\RTP+{\rr}}-\moy{\eta_{\RTP+{\rr}}}$. It then suffices to determine $\gt_{\ee_{\pm1}}(t)$ to compute the fluctuations of the TP position.

 In the trivial mean-field approximation, one has $k_{\rr}=\rho$ and $\gt_{\rr}=0$ for any $\rr$. Eq. (\ref{dispersion}) then reduces to
\begin{equation}
\frac{\mathrm{d}}{\mathrm{d}t} \left(\moy{{\XTP}^2}-\moy{{\XTP}}^2\right) =\frac{\sigma^2}{\tau}
(p_1+p_{-1})(1-\rho).
\label{eq:fluct_trivial}
\end{equation}
Note that Eq. (\ref{eq:fluct_trivial}) corresponds to the standard result for the fluctuations of the position of a biased random walker, where the time $t$ is renormalized by the fraction of unoccupied sites $1-\rho$. In the particular case of a one-dimensional lattice where $p_1+p_{-1}=1$, the fluctuations of the TP position do not depend on the bias. This is in contrast with the case where the TP has a  discrete time evolution \cite{Benichou2013g}.

The evolution equations for $\gt_{\rr}(t)$ are obtained by multiplying the master equation (\ref{eqmaitresse}) by $\delta X \eta_{\XX+\rr}$ and summing over all the configurations of $(\XX,\eta)$. The details of the calculation are given in  \ref{sec:master_eq_fluc_gt_gt}. We get the following equation 
\begin{align}
2d\tau^*\partial_t\gt_{\rr}(t)=&\sum_\mu\left(\nabla_\mu-\delta_{{\rr},{\boldsymbol e_\mu}}\nabla_{-\mu}\right)\gt_{\rr}(t)-2d(f+g)\gt_{\rr}(t)\nonumber\\
&+\frac{2d\tau^*}{\tau}\sum_{\mu} p_\mu \moy{\delta\XTP(1-\eta_{\RTP+\ee_\mu})\nabla_\mu\eta_{\RTP+\rr}}\nonumber\\
&+\frac{2d\tau^*}{\tau}\sigma\left[p_1 \moy{ (1-\eta_{\RTP+\ee_1})\eta_{\RTP+\rr+\ee_1}} - p_{-1}\moy{ (1-\eta_{\RTP+\ee_{-1}})\eta_{\RTP+\rr+\ee_{-1}}}\right]\nonumber\\
&-\frac{2d\tau^*}{\tau}\sigma\left[p_1(1-k_{\ee_1})-p_{-1}(1-k_{\ee_{-1}})\right]k_{\rr}(t).
\label{eqgtnoapprox}
\end{align}
We then notice that this evolution equation involves higher-order correlation functions, of the form $\moy{\delta\XTP\eta_{\RTP+{\rr}}\eta_{\RTP+\ee_\mu}}$. As previously, their computation would lead to an infinite hierarchy of coupled equations. We then write an extension of the decoupling approximation (\ref{decouplage_k}), obtained by writing $\eta_{\boldsymbol{R}} = \moy{\eta_{\boldsymbol{R}}} +\delta \eta_{\boldsymbol{R}}$ and discarding the terms of order $(\delta \eta_{\boldsymbol{R}})^2$. We  find
\begin{eqnarray}
\moy{\delta\XTP\eta_{\RTP+{\rr}}\eta_{\RTP+\ee_{\mu}}}  &\simeq&  \moy{\eta_{\RTP+{\rr}}} \moy{\delta\XTP
	\eta_{\RTP+\ee_{\mu}}} +\moy{\delta \XTP
	\eta_{\RTP+\rr}} \moy{\eta_{\RTP+\ee_\mu}}, \\
&=& k_{\rr}(t)\gt_{\ee_\mu}(t)+k_{\ee_\mu}(t)\gt_{\rr}(t),
\label{decouplage2}
\end{eqnarray}
which is valid for $\rr \neq \ee_\mu$. For $\rr = \ee_\mu$, using the relation $({\eta_{\RTP+\ee_{\mu}}})^2=\eta_{\RTP+\ee_{\mu}}$, we obtain
\begin{equation}
\moy{\delta\XTP({\eta_{\RTP+{\ee_\mu}}})^2}  = \gt_{\ee_\mu}.
\end{equation}

If $\rr \notin \{ \ee_{\pm1},\dots,\ee_{\pm d}\}$, expandind the term of second line of Eq. (\ref{eqgtnoapprox}), we obtain
\begin{eqnarray}
\fl \frac{2d\tau^*}{\tau}\sum_{\mu} p_\mu \moy{\delta\XTP(1-\eta_{\RTP+\ee_\mu})\nabla_\mu\eta_{\RTP+\rr}}   \nonumber \\
\fl =\frac{2d\tau^*}{\tau}\sum_{\mu} p_\mu \left[  \moy{\delta X_t \eta_{\RTP+\rr+\ee_\mu}} - \moy{\delta X_t \eta_{\RTP+\rr}}-\moy{\delta X_t \eta_{\RTP+\ee_\mu}\eta_{\RTP+\rr+\ee_\mu}}+\moy{\delta X_t \eta_{\RTP+\ee_\mu} \eta_{\RTP+\rr}}  \right],\nonumber\\
\end{eqnarray}
and using the definitions of $\gt_{\rr}$ (Eq. (\ref{defgtilde})) as well as the decoupling approximation (Eq. (\ref{decouplage2})), we obtain
\begin{eqnarray}
\label{gt_rewrite_0}
\frac{2d\tau^*}{\tau}\sum_{\mu} p_\mu \moy{\delta\XTP(1-\eta_{\RTP+\ee_\mu})\nabla_\mu\eta_{\RTP+\rr}}  \\
= \frac{2d\tau^*}{\tau}\sum_{\mu} p_\mu \left[   \nabla_\mu \gt_{\rr}(t) - \gt_{\ee_\mu}(t) \nabla_\mu k_{\rr}(t) - k_{\ee_\mu}  (t)\nabla_{\mu} \gt_{\rr} (t)\right].
\end{eqnarray}
 Then, we gather the first three terms of the r.h.s. of Eq. (\ref{eqgtnoapprox}) and obtain
 \begin{eqnarray}
\fl \sum_\mu\nabla_\mu\gt_{\rr}(t)-2d(f+g)\gt_{\rr}(t)+\frac{2d\tau^*}{\tau}\sum_{\mu} p_\mu \moy{\delta\XTP(1-\eta_{\RTP+\ee_\mu})\nabla_\mu\eta_{\RTP+\rr}} \nonumber\\
\fl = \sum_{\mu }\left\{1+\frac{2d \tau^*}{\tau} \left[1-k_{\ee_\mu} (t)\right] \right\} \nabla_{\mu} \gt_{\rr}(t) -2d(f+g)\gt_{\rr}(t) - \frac{2d \tau^*}{\tau}  \sum_{\mu} p_\mu \gt_{\ee_\mu}(t) \nabla_{\mu} k_{\rr}(t) \nonumber\\
\fl = \widetilde{L} \gt_{\rr}(t) - \frac{2d \tau^*}{\tau}  \sum_{\mu} p_\mu \gt_{\ee_\mu}(t) \nabla_{\mu} k_{\rr}(t).
 \label{gt_rewrite_1}
\end{eqnarray}
 The last two terms of Eq. (\ref{eqgtnoapprox}) are recast using the decoupling approximation from Eq. (\ref{decouplage_k}). One obtains
 \begin{eqnarray}
\fl \frac{2d\tau^*}{\tau}\sigma\left[p_1 \moy{ (1-\eta_{\RTP+\ee_1})\eta_{\RTP+\rr+\ee_1}} - p_{-1}\moy{ (1-\eta_{\RTP+\ee_{-1}})\eta_{\RTP+\rr+\ee_{-1}}}\right] \nonumber\\
 -\frac{2d\tau^*}{\tau}\sigma\left\{p_1[1-k_{\ee_1}(t)]-p_{-1}[1-k_{\ee_{-1}}(t)]\right\}k_{\rr} (t)\\
\fl  =\frac{2d\tau^*}{\tau}\sigma\left[p_1(k_{\rr+\ee_1}-k_{\ee_1}k_{\rr+\ee_1})-p_{-1}(k_{\rr+\ee_{-1}}-k_{\ee_{-1}}k_{\rr+\ee_{-1}})\right]  \nonumber\\
 -\frac{2d\tau^*}{\tau}\sigma\left\{p_1[1-k_{\ee_1}(t)]-p_{-1}[1-k_{\ee_{-1}}(t)]\right\}k_{\rr}(t) \\
 \fl = \frac{2d\tau^*}{\tau}\sigma\left\{p_1[1-k_{\ee_1}(t)]\nabla_1 k_{\rr}(t)-p_{-1}[1-k_{\ee_{-1}}(t)]\nabla_{-1} k_{\rr}(t)\right\}.
 \label{gt_rewrite_2}
\end{eqnarray}
 Using Eqs. (\ref{eqgtnoapprox}), (\ref{gt_rewrite_1})  and (\ref{gt_rewrite_2}), we finally obtain the following equation for the evolution of the correlation functions $\gt_{\rr}$:
\begin{align}
2d\tau^*\partial_t\gt_{\rr}(t)=&\widetilde{L}\gt_{\rr}(t)+\frac{2d\tau^*}{\tau}\sigma\left\{
p_1[1-k_{\ee_1}(t)]\nabla_1 k_{\rr}(t)-p_{-1}[1-k_{\ee_{-1}}(t)]\nabla_{-1} k_{\rr}(t)\right\} \nonumber \\
&-\frac{2d\tau^*}{\tau}\sum_\mu p_\mu\gt_{\ee_\mu}(t)\nabla_\mu k_{\rr}(t),
\label{systemegtilde1}
\end{align}
which holds for all $\rr$, except for  $\rr=\{{\bf 0},\ee_{\pm 1},\ldots,\ee_{\pm d}\}$.

On the other hand, for the special sites ${\rr} = {\ee_{\nu}}$ 
with $\nu=\{\pm1,\ldots, \pm d\}$, the term in the second line of Eq. (\ref{eqgtnoapprox}) can be rewritten as
\begin{eqnarray}
\frac{2d\tau^*}{\tau}\sum_{\mu} p_\mu \moy{\delta\XTP(1-\eta_{\RTP+\ee_\mu})\nabla_\mu\eta_{\RTP+\ee_\nu}} \\ 
= \frac{2d\tau^*}{\tau}\sum_{\mu} p_\mu \moy{\delta\XTP(1-\eta_{\RTP+\ee_\mu})(\eta_{\RTP+\ee_\nu+\ee_\mu}-\eta_{\RTP+\ee_\nu})}\\
 = \frac{2d\tau^*}{\tau}\sum_{\mu\neq \pm \nu} p_\mu \moy{\delta\XTP(1-\eta_{\RTP+\ee_\mu})(\eta_{\RTP+\ee_\nu+\ee_\mu}-\eta_{\RTP+\ee_\nu})}\\
 +\frac{2d\tau^*}{\tau} p_\nu \moy{\delta\XTP(1-\eta_{\RTP+\ee_\nu})(\eta_{\RTP+2\ee_\nu}-\eta_{\RTP+\ee_\nu})}\\
  +\frac{2d\tau^*}{\tau} p_{-\nu} \moy{\delta\XTP(1-\eta_{\RTP+\ee_{-\nu}})(\eta_{\RTP}-\eta_{\RTP+\ee_\nu})}.
\end{eqnarray}
The first term can be written in terms of $k_{\rr}$ and $\gt_{\rr}$ using Eq. (\ref{gt_rewrite_0}). The last two terms are rewritten using the decoupling approximation (Eq. (\ref{decouplage2})), the property $(\eta_{\rr})^2 = \eta_{\rr}$, and the conventions $k_{\zz}=\gt_{\zz} = 0$. We obtain
\begin{align}
 &\frac{2d\tau^*}{\tau}\sum_{\mu} p_\mu \moy{\delta\XTP(1-\eta_{\RTP+\ee_\mu})\nabla_\mu\eta_{\RTP+\ee_\nu}}\nonumber \\
=&\frac{2d\tau^*}{\tau}\sum_{\mu \neq \pm \nu} p_\mu \left[   \nabla_\mu \gt_{\rr} (t)- \gt_{\ee_\mu}(t) \nabla_\mu k_{\rr}(t) - k_{\ee_\mu} (t) \nabla_{\mu} \gt_{\rr}(t) \right] \nonumber\\
&+ \frac{2d\tau^*}{\tau} p_\nu \left\{  \gt_{2 \ee_\nu}(t) - [k_{\ee_\nu}(t) \gt_{2 \ee_\nu}(t)+k_{2 \ee_\nu}(t) \gt_{ \ee_\nu}(t)] \right\}\nonumber\\
&+\frac{2d\tau^*}{\tau} p_{-\nu} \left\{  -\gt_{ \ee_\nu}(t) + [k_{\ee_{-\nu}} (t)\gt_{ \ee_\nu}(t)+k_{ \ee_\nu}(t) \gt_{ \ee_{-\nu}}(t) ] \right\}
\label{details_gt_BC}
\end{align}
Using Eq. (\ref{details_gt_BC}), and after straightforward computations, the first three terms of the r.h.s. of Eq. (\ref{eqgtnoapprox}) yield
 \begin{eqnarray}
\fl \sum_\mu \left( \nabla_\mu -\delta_{\ee_\nu,\ee_\mu} \nabla_{-\mu}\right)\gt_{\rr}(t)-2d(f+g)\gt_{\ee_\nu}(t)+\frac{2d\tau^*}{\tau}\sum_{\mu} p_\mu \moy{\delta\XTP(1-\eta_{\RTP+\ee_\mu})\nabla_\mu\eta_{\RTP+\ee_\nu}} \nonumber\\
\fl = [ \widetilde{L}  +A_\nu(t)  ] \gt_{\ee_\nu}(t) -\frac{2d\tau^*}{\tau} p_\nu\gt_{\ee_\nu}(t) k_{\ee_\nu}(t) -\frac{2d\tau^*}{\tau}\sum_{\mu}p_\mu\gt_{\ee_\mu}(t)\nabla_\mu k_{\ee_\nu}(t).
\end{eqnarray}
Finally, using again Eq. (\ref{gt_rewrite_2}) to rewrite the last two terms of Eq. (\ref{eqgtnoapprox}), we obtain the equation verified by $\gt_{\ee_\nu}(t)$ for $\nu \in \{ \pm 1 ,\dots, \pm d\}$:
\begin{align}
2d\tau^*\partial_t\gt_{\ee_\nu}(t)  = & [\widetilde{L}+A_\nu(t)]\gt_{\ee_\nu}(t)+\frac{2d\tau^*}{\tau}\sigma\left\{
p_1[1-k_{\ee_1}(t)]\nabla_1 k_{\ee_\nu}(t)-p_{-1}[1-k_{\ee_{-1}}(t)]\nabla_{-1} k_{\ee_\nu}(t)\right\} \nonumber \\
&-\frac{2d\tau^*}{\tau} p_\nu\gt_{\ee_\nu}(t) k_{\ee_\nu}(t) -\frac{2d\tau^*}{\tau}\sum_{\mu}p_\mu\gt_{\ee_\mu}(t)\nabla_\mu k_{\ee_\nu}(t).
\label{systemegtilde2}
\end{align}
Eqs. (\ref{systemegtilde1}) and (\ref{systemegtilde2}) then form a closed system of equations for the quantities $\gt_{\rr}(t)$, provided that the quantities $k_{\ee_\nu}(t)$ are known. Note that Eqs. (\ref{systemegtilde1}) and (\ref{systemegtilde2}) are linear in the correlation functions $\gt_{\rr}(t)$, which can be written explicitly in terms of the density profiles $k_{\rr}(t)$. 

 The quantities $\gt_{\ee_{\pm 1}}(t)$ can be deduced from these equations, and one can compute the evolution of the fluctuations of the TP position using Eq. (\ref{dispersion}).

\subsection{Stationary values}

We turn to the limit $t\to\infty$. We assume that the quantities $k_{\rr}(t)$ and $\gt_{\rr}(t)$ have stationary values, so that
\begin{eqnarray}
\lim_{t \to \infty} \partial_t k_{\rr}(t) & = &0, \\
\lim_{t \to \infty} \partial_t \gt_{\rr}(t)& = &0.
\end{eqnarray}
The existence of these stationary values will be shown afterwards. We will use the simplified notations:
\begin{eqnarray}
k_{\rr} & = & \lim_{t\to\infty} k_{\rr}(t), \\
\gt_{\rr} & = & \lim_{t\to\infty} \gt_{\rr}(t), \\
A_{\mu} & = & \lim_{t\to\infty}A_\mu(t).  
\end{eqnarray}
We also define the observables:
\begin{eqnarray}
V&\equiv&\lim_{t\to\infty}\frac{\mathrm{d}}{\mathrm{d}t} \moy{\XTP},\label{eq:def_V_1}\\
K&\equiv&\lim_{t\to\infty} \frac{1}{2}\frac{\mathrm{d}}{\mathrm{d}t} \left(\moy{{\XTP}^2}-\moy{{\XTP}}^2\right) \label{eq:def_K_1}
\label{definitions}
\end{eqnarray}
so that $V$ and $K$ represent respectively the  velocity and the diffusion coefficient of the TP in the stationary state. Using Eq. (\ref{eq:def_velocity_general}), the velocity $V$  can be written in terms of the functions $k_{\rr}$:   
\begin{equation}
V=\frac{\sigma}{\tau}\left[p_1  \left(1-k_{\ee_1}\right)-p_{-1} \left(1-k_{\ee_{-1}}\right)\right].
\label{eq:V_stat}
\end{equation}
Similarly, using Eq. (\ref{dispersion}), the diffusion coefficient can be written
\begin{equation}
\label{KTP_def}
K=\frac{\sigma^2}{2d\tau}\left[p_1\left(1-k_{\ee_1}\right)+p_{-1}\left(1-k_{\ee_{-1}}\right)\right]-\frac{\sigma}{d\tau}\left(p_1\widetilde{g}_{\ee_{1}}-p_{-1}\widetilde{g}_{\ee_{-1}}\right).
\end{equation}

The stationary values of the density profiles $k_{\rr}$ (in particular $k_{\ee_{\pm1}}$ and therefore the velocity $V$) are obtained by solving Eqs. (\ref{systemek1}) and (\ref{systemek2}) with $\partial_t k_{\boldsymbol{r}}(t)=0$. Similarly, solving  Eqs. (\ref{systemegtilde1}) and (\ref{systemegtilde2}) with $\partial_t \gt_{\boldsymbol{r}}(t)=0$, one obtains the stationary values of  $\gt_{\ee_{\pm1}}$ and the diffusion coefficient $K$.

Note that these stationary equations are valid in  \emph{any dimension}, and allow to compute the velocity and diffusion coefficient of the TP under the approximations  presented above (Eqs. (\ref{decouplage_k}) and (\ref{decouplage2})). Their solutions will be presented in the case of a one-dimensional system in section \ref{sec:solutions_1D}.

\section{Cumulant generating function}

\subsection{Governing equations}
\label{sec:governing_eq_cgf}

In the previous sections, using a decoupling approximation, we were able to determine the stationary equations satisfied by the quantities $k_{\rr}=\moy{\eta_{\rr}}$ and $\gt_{\rr}=\moy{\delta X_t \delta \eta_{\XTP+\rr}}$, which are involved in the expression of the stationary velocity $V$ (Eq. (\ref{eq:V_stat})) and of the stationary diffusion coefficient $K$ (Eq. (\ref{KTP_def})) of the TP.
Here, we aim at calculating the higher-order cumulants of $\XTP$, defined by
\begin{equation}
\kappa_n(t) \equiv \frac{1}{\ii^n}\left.\frac{\partial ^ n \Psi(u;t)}{ \partial u^n}\right|_{u=0},
\end{equation}
where the quantity
\begin{equation}
\label{eq:def_CGF}
 \Psi(u;t) \equiv \ln \moy{\ex{\ii u \XTP}}
\end{equation}
 is known as the second characteristic function (or cumulant generating function) of $\XTP$. Using the balance equation (\ref{eq:bilan_pos_TP}), we get the relation
\begin{eqnarray}
\fl \moy{\ex{\ii u X_{t+\Delta t}}} = \moy{\ex{\ii u (X_{t}+\sigma)} \frac{\Delta t}{\tau} p_1 (1-\eta_{\RTP + \ee_1})}+\moy{\ex{\ii u (X_{t}-\sigma)} \frac{\Delta t}{\tau} p_{-1} (1-\eta_{\RTP + \ee_{-1}})}\nonumber\\
+\moy{\ex{\ii u X_{t}} \left[1-\frac{\Delta t}{\tau}( p_1 (1-\eta_{\RTP + \ee_1})+p_{-1} (1-\eta_{\RTP + \ee_{-1}})\right]}.
\label{eq:bilan_exp_moy}
\end{eqnarray}
Note that this equation involves two different averages: an average over the direction of the step taken by the TP, and an average on the realizations. Eq. (\ref{eq:bilan_exp_moy}) leads to
\begin{equation}
\frac{\mathrm{d}}{\mathrm{d}t}\moy{\ex{\ii u \XTP}} = \frac{p_1}{\tau}\left(\ex{\ii u \sigma}-1\right)\moy{\ex{\ii u X_{t}}(1-\eta_{\RTP + \ee_1})}+\frac{p_{-1}}{\tau}\left(\ex{-\ii u \sigma}-1\right)\moy{\ex{\ii u X_{t}}(1-\eta_{\RTP + \ee_{-1}})},
\end{equation}
and, using the definition of $\Psi(u;t)$ from Eq. (\ref{eq:def_CGF}):
\begin{align}
\frac{\mathrm{d}\Psi}{\mathrm{d}t}& = \frac{1}{\moy{\ex{\ii u \XTP}}} \frac{\mathrm{d}}{\mathrm{d}t}\moy{\ex{\ii u \XTP}} \\
&=\frac{p_1}{\tau}\left(\ex{\ii u \sigma}-1\right)\left[1-\frac{\moy{\ex{\ii u \XTP}\eta_{\RTP + \ee_1}}}{\moy{\ex{\ii u \XTP}}}\right]+\frac{p_{-1}}{\tau}\left(\ex{-\ii u \sigma}-1\right)\left[1-\frac{\moy{\ex{\ii u \XTP}\eta_{\RTP + \ee_{-1}}}}{\moy{\ex{\ii u \XTP}}}\right].
\end{align}
We define the following correlation functions
\begin{equation}
\label{eq:def_wt}
w_{\rr}(u;t) \equiv \moy{\ex{\ii u X_{t}}\eta_{\RTP + \rr}}\mathrm{~~~and~~~}\wt_{\rr} (u;t)\equiv \frac{\moy{\ex{\ii u X_{t}}\eta_{\RTP + \rr}}}{\moy{\ex{\ii u X_{t}}}}.
\end{equation}
Finally, we obtain the following evolution equation for the cumulant generating function of the TP position:
\begin{equation}
\frac{\mathrm{d}\Psi}{\mathrm{d}t} = \frac{p_1}{\tau}\left(\ex{\ii u \sigma}-1\right)\left[1-\wt_{\ee_1}(u;t)\right]+\frac{p_{-1}}{\tau}\left(\ex{-\ii u \sigma}-1\right)\left[1-\wt_{\ee_{-1}}(u;t)\right].
\label{dpsidt}
\end{equation}
Assuming that the quantities $\wt_{\ee_1}(u;t)$ and $\wt_{\ee_{-1}}(u;t)$ reach stationary values when $t\to\infty$ (their existence will be shown a posteriori), the second characteristic function has the following asymptotic behavior :
\begin{equation}
\label{eq:psi_phi}
\Psi(u;t)\underset{t\to\infty}{\sim}\Phi(u)t,
\end{equation}
with
\begin{equation}
\label{eq:phi_stat}
\Phi(u)=\frac{p_1}{\tau}\left(\ex{\ii u \sigma}-1\right)\left[1-\widetilde{w}_{\ee_1}(u)\right]+\frac{p_{-1}}{\tau}\left(\ex{-\ii u \sigma}-1\right)\left[1-\widetilde{w}_{\ee_{-1}}(u)\right].
\end{equation}
 The relation (\ref{eq:psi_phi}) indicates that all the cumulants of $X_t$ are linear in time in the long-time limit. In particular, this implies that the $n$-th moment of the rescaled variable $Z_t = (X_t-\moy{X_t})/\sqrt{\moy{{X_t}^2}-\moy{X_t}^2}$ scales as $t^{1-n/2}$ in the long-time limit. All the moments of $Z_t$ of order greater than $2$ vanish when $t\to\infty$, and $Z_t$ is distributed accordingly to a \emph{Gaussian distribution} at large times.\\

This calculation allows us to compute the full distribution of $X_t$ in the long-time limit. Assuming that the $u$-dependance of $\wt_{\ee_{\pm1}}(u)$ is known, we can derive from the previous equations  the probability density function (p.d.f.) $P_t(x)\equiv \mathrm{Prob}[X_t=x]$ as follows. The quantity $\moy{\ex{\ii u \XTP}}=\ex{\Psi(u;t)}$ is defined by
\begin{equation}
\label{ }
\moy{\ex{\ii u \XTP}} = \sum_{x=-\infty}^\infty P_t(x) \ex{\ii u x}.
\end{equation}
$\ex{\Psi(u;t)}$ is then the Fourier transform of the p.d.f. $P_t(x)$, which can be obtained by the inverse Fourier transform:
\begin{equation}
\label{ }
P_t(x) = \int_{-\pi}^\pi \frac{\dd u}{2 \pi} \ex{-\ii u x} \ex{\Psi(u;t)},
\end{equation}
and, using the expression of the long-time limit of $\Psi(t)$  (Eqs. (\ref{eq:psi_phi}) and (\ref{eq:phi_stat})),
\begin{equation}
\label{pdf_computation}
P_t(x) \underset{t\to \infty}{\sim} \int_{-\pi}^\pi \frac{\dd u}{2 \pi} \exp\left\{ \frac{p_1 t}{\tau}(\ex{\ii u \sigma}-1)[1-\wt_{\ee_1}(u)]+\frac{p_{-1} t}{\tau}(\ex{-\ii u \sigma}-1)[1-\wt_{\ee_{-1}}(u)] -\ii u x   \right\}.
\end{equation}
Consequently, it suffices to determine the $u$-dependence of $\wt_{\ee_{\pm 1}}(u;t)$ to obtain the p.d.f. of the TP position. In what follows, we establish the evolution equations for the quantities $\wt_{\rr}(u;t)$ starting again from the master equation (\ref{eqmaitresse}).

\subsection{Evolution equations of the quantities $\wt_{\rr}(u;t)$}
\label{sec:evolution_w}

The evolution equation of the correlation functions $w_{\rr}(u;t)$ (defined by Eq. (\ref{eq:def_wt})) can be obtained by multiplying both sides of the master equation (\ref{eqmaitresse}) by the quantity $\eta_{\XX+\rr}\ex{\ii u X}$ and averaging with respect to the bath configuration $\eta$ and the TP position $\RTP$. Extending the method used to derive the evolution equations of the correlation functions $\gt_{\rr}(t)$ starting from the master equation (\ref{eqmaitresse}), it is found that $w_{\rr}(u;t)$ obeys the following exact equation (see \ref{sec:master_eq_fluc_gt_gt}): 
\begin{eqnarray}
\fl 2d\tau^*\partial_tw_{\rr}(u;t)=\left(\sum_\mu\nabla_\mu-\delta_{{\rr},{\emuu}}\nabla_{-\mu}\right)w_{\rr}(u;t)-2d(f+g)w_{\rr}(u;t)+2df\moy{\ex{\ii u X_{t}}}\nonumber\\
+\frac{2d\tau^*}{\tau}\sum_{\mu}p_\mu \moy{ \ex{\ii u X_{t}}(1-\eta_{\RTP+\emuu})\nabla_\mu\eta_{\RTP+{\rr}}} \nonumber\\
+\frac{2d\tau^*}{\tau}\sum_{\epsilon=\pm1}p_\epsilon\left(\ex{\ii u \epsilon\sigma}-1\right)\moy{\ex{\ii u X_{t}}(1-\eta_{\RTP+\ee_\epsilon})\eta_{\RTP+{\rr}+\ee_\epsilon}}.
\label{wgeneral}
\end{eqnarray}
We then make the following decoupling hypothesis:
\begin{align}
\moy{\left(\ex{\ii u X_{t}} -\moy{\ex{\ii u X_{t}}}\right) \eta_{\Rtr+\rr}\eta_{\Rtr+\emuu}} \simeq& \moy{\left(\ex{\ii u X_{t}} -\moy{\ex{\ii u X_{t}}}\right) \eta_{\Rtr+\rr}}\moy{\eta_{\Rtr+\emuu}} \nonumber \\
&+\moy{\left(\ex{\ii u X_{t}} -\moy{\ex{\ii u X_{t}}}\right) \eta_{\Rtr+\emuu}}\moy{\eta_{\Rtr+\rr}},
\end{align}
which is valid for $\rr\neq\ee_\mu$. 
This is equivalent to
\begin{equation}
\moy{\ex{\ii u X_{t}} \eta_{\Rtr+\rr}\eta_{\Rtr+\ee_\mu}}= w_{\rr}(u;t)k_{\ee_\mu}(t)+k_{\rr}(t)w_{\ee_\mu}(u;t)-\moy{\ex{\ii u X_{t}}}k_{\rr}(t)
k_{\ee_\mu}(t).
\label{decouplage3}
\end{equation}
For $\rr=\ee_\mu$, one gets
\begin{align}
\moy{\ex{\ii u X_{t}} (\eta_{\Rtr+\ee_\mu})^2}&=\moy{\ex{\ii u X_{t}} \eta_{\Rtr+\ee_\mu}} \\
&=w_{\ee_\mu}(u;t).
%\label{decouplage3}
\end{align}
This decoupling approximation is an extension of the approximations (\ref{decouplage_k}) and (\ref{decouplage2}): it is obtained by writing the occupation variables as $\eta_{\boldsymbol{R}} =  \moy{\eta_{\boldsymbol{R}} } + \delta \eta_{\boldsymbol{R}} $, and discarding the terms of order $(\delta \eta_{\boldsymbol{R}} )^2$. Note that expanding Eq. (\ref{decouplage3}) at order 0 and 1 in $u$, we retrieve the decoupling approximations made for the correlation functions $\moy{\eta_{\RTP+{ \rr}}\eta_{\RTP+\ee_{ \mu}}}$ (Eq. (\ref{decouplage_k})) and $\moy{\delta X_t\eta_{\RTP+{ \rr}}\eta_{\RTP+\ee_{ \mu}}}$  (Eq. (\ref{decouplage2})). Using this approximation in Eq. (\ref{wgeneral}), and following a procedure similar to the one used to derive Eqs. (\ref{systemegtilde1}) and (\ref{systemegtilde2}) from Eq. (\ref{eqgtnoapprox}) we obtain for $\rr \neq \ee_\nu$,
\begin{eqnarray}
\fl 2d\tau^*\partial_tw_{\rr}(u;t)
=\widetilde{L}w_{\rr}(u;t)+2df\moy{\ex{\ii u X_{t}}}+\frac{2d\tau^*}{\tau}\sum_\mu p_\mu\left[ k_{\emuu}(t)\moy{\ex{\ii u X_{t}}}-w_{\emuu}(u;t)\right]\nabla_\mu k_{\rr}(t)\nonumber \\
\fl +\frac{2d\tau^*}{\tau} \sum_{\ep=\pm1} p_{\ep}\left(\ex{\ii u \ep \sigma}-1\right)\left\{ w_{\rr+\ee_\ep}(u;t)[1-k_{\ee_\ep}(t)]+k_{\rr+\ee_\ep}(t)\left[\moy{\ex{\ii u X_{t}}}k_{\ee_\ep}(t)-w_{\ee_\ep}(u;t)\right]\right\}.\nonumber\\
\label{eq:evolution_w_bulk}
\end{eqnarray}
For $\rr = \ee_\nu$, the evolution equation becomes :
\begin{eqnarray}
\fl 2d\tau^*\partial_tw_{\enu}(u;t)
=[\widetilde{L}+A_\nu(t)]w_{\enu}(u;t)+2df\moy{\ex{\ii u X_{t}}}\nonumber\\
\fl +\frac{2d\tau^*}{\tau}\sum_{\mu} p_\mu\left[ k_{\emuu}(t)\moy{\ex{\ii u X_{t}}}-w_{\emuu}(u;t)\right]\nabla_\mu k_{\enu}(t)   +\frac{2d\tau^*}{\tau}p_\nu\left[ k_{\enu}(t) \moy{\ex{\ii u X_{t}}}-w_{\enu}(u;t) \right]k_{\enu}(t)  \nonumber \\
\fl +\frac{2d\tau^*}{\tau} \sum_{\ep = \pm 1} p_\ep \left(\ex{\ii u \ep \sigma}-1\right)\left\{ w_{\ee_{\nu}+\ee_\ep}(u;t)[1-k_{\ee_\ep}(t)]+k_{\ee_{\nu}+\ee_{\ep}}(t) \left[\moy{\ex{\ii u X_{t}}}k_{\ee_\ep}(t)-w_{\ee_\ep}(u;t)\right]\right\}.\nonumber\\
\label{eq:evolution_w_BC}
\end{eqnarray}
These equations are conveniently written in terms of the variable $\widetilde{w}_{\rr}(u;t)$ by noticing that :
\begin{align}
 \frac{\partial \wt_{\rr}}{\partial t} =& \frac{1}{\moy{\ex{\ii u X_{t}}}}\frac{\partial {w}_{\rr}(u;t)}{\partial t}-\frac{{w}_{\rr}(u;t)}{\moy{\ex{\ii u X_{t}}}}\frac{\partial \moy{\ex{\ii u X_{t}}}}{\partial t} \nonumber\\
=&\frac{1}{\moy{\ex{\ii u X_{t}}}}\frac{\partial {w}_{\rr}}{\partial t}-\wt_{\rr}(u;t)\frac{\partial \Psi}{\partial t}  \nonumber \\
=&\frac{1}{\moy{\ex{\ii u X_{t}}}}\frac{\partial {w}_{\rr}}{\partial t}-\wt_{\rr}(u;t) \left\{ \frac{p_1}{\tau}\left(\ex{\ii u \sigma}-1\right)\left[1-\wt_{\ee_1}(u;t)\right]+\frac{p_{-1}}{\tau}\left(\ex{-\ii u \sigma}-1\right)\left[1-\wt_{\ee_1}(u;t)\right]\right\}. \nonumber\\
\end{align}
Finally, we divide the evolution equations (\ref{eq:evolution_w_bulk}) and (\ref{eq:evolution_w_BC}) by $\moy{\ex{\ii u X_{t}}}$ and obtain the evolution equations for $\wt_{\rr}(u;t)$, for $\rr\neq \ee_\nu$ :
\begin{eqnarray}
\fl 2d\tau^*\partial_t\widetilde{w}_{\rr}(u;t)
=\widetilde{L}\widetilde{w}_{\rr}(u;t)+2df+\frac{2d\tau^*}{\tau}\sum_\mu p_\mu\left[ k_{\emuu}(t)-\widetilde{w}_{\emuu}(u;t)\right]\nabla_\mu k_{\rr}(t) \nonumber \\
\fl +\frac{2d\tau^*}{\tau} \sum_{\ep = \pm 1}p_\ep\left(\ex{\ii u \ep \sigma}-1\right)\left\{\nabla_\ep \wt_{\rr}(u;t)-k_{\ee_\ep}(t)[\widetilde{w}_{\rr+\ee_\ep}(u;t)-k_{\rr+\ee_\ep}(t)]-\wt_{\ee_\ep}(u;t)[k_{\rr+\ee_\ep}(t)-\wt_{\rr}(u;t)]\right\}, \nonumber \\
\label{wtildegen}
\end{eqnarray}
and for $\rr = \ee_\nu$ :
\begin{eqnarray}
\fl 2d\tau^*\partial_t\widetilde{w}_{\enu}(u;t)
=[\widetilde{L}+A_\nu(t)]\widetilde{w}_{\enu}(u;t)+2df\nonumber \\
\fl +\frac{2d\tau^*}{\tau}\sum_{\mu} p_\mu\left[ k_{\emuu}(t)-\widetilde{w}_{\emuu}(u;t)\right]\nabla_\mu k_{\enu}(t)+\frac{2d\tau^*}{\tau}p_\nu\left[ k_{\enu}(t)-\widetilde{w}_{\enu}(u;t)\right]k_{\enu}(t)\nonumber\\
\fl +\frac{2d\tau^*}{\tau} \sum_{\ep = \pm 1}p_\ep\left(\ex{\ii u \ep \sigma}-1\right)\left\{\nabla_\ep \wt_{\ee_\nu}(u;t)-k_{\ee_\ep}(t)[\widetilde{w}_{\ee_\nu+\ee_\ep}(u;t)-k_{\ee_\nu+\ee_\ep}(t)]-\wt_{\ee_\ep}(u;t)[k_{\ee_\nu+\ee_\ep}(t)-\wt_{\ee_\nu}(u;t)]\right\}
 \nonumber \\
\label{wtildeboundary}
\end{eqnarray}

These equations can in principle be solved in the stationary limit ${t\to\infty} $, by setting $\partial_t \wt_{\rr}(u;t)=0$ in Eqs. (\ref{wtildegen}) and (\ref{wtildeboundary}) and obtaining the values of $\wt_{\rr}(u)$ satisfying these stationary equations. In particular, this allows us to obtain the $u$-dependence of the functions $\wt_{\ee_{\pm 1}}(u)$ and to deduce $P_t(x)$ from Eq. (\ref{pdf_computation}). These equations are valid in any dimension, and their solution gives the cumulant generating function and the probability distribution function of the TP position.

This resolution will be made explicit in the case of a one-dimensional lattice in section \ref{sec:cgf1d}. We also notice that the functions $\wt_{\rr}(u;t)$ can be expanded in powers of $u$  to compute higher-order cumulants. In particular, we give in the next section the evolution equation satisfied by the third cumulant of the position of the TP.

\subsection{Third-order cumulant}

In this section, we study the third cumulant of the distribution of $X_t$, which characterizes its skewness. We use Eqs. (\ref{wtildegen}) and (\ref{wtildeboundary}), describing the evolution of $\wt_{\rr}(u;t)$, to calculate the third order cumulant. We define the coefficient $\gamma$ by the relation :
\begin{equation}
\label{eq:def_gamma}
\gamma\equiv\lim_{t\to\infty}\frac{1}{6} \frac{\dd}{\dd t}{\moy{\left(\XTP-\moy{\XTP}\right)^3}}.
\end{equation}
Recalling the definition of $\Psi(t)$ from Eq. (\ref{eq:def_CGF}), we get the following expansion in powers of $u$:
\begin{equation}
\label{ }
\Psi(t) \underset{u\to0}{=} \ii u \moy{X_t} + \frac{(\ii u)^2}{2} \moy{\left(\XTP-\moy{\XTP}\right)^2} + \frac{(\ii u)^3}{6}\moy{\left(\XTP-\moy{\XTP}\right)^3}+\dots
\end{equation}
In the long-time limit, and using the definitions of $V$ (Eq. (\ref{eq:def_V_1})), $K$ (Eq. (\ref{KTP_def})) and $\gamma$ (Eq. (\ref{eq:def_gamma})), one gets
\begin{equation}
\label{eq:exp_dpsidt}
\lim_{t\to\infty} \frac{\dd \Psi}{\dd t} = \ii u V+ (\ii u)^2  K + (\ii u )^3 \gamma +\dots
\end{equation}
We define the correlation function $\widetilde{m}_{\rr}(t)$ by the relation
\begin{equation}
\widetilde{m}_{\rr}(t) \equiv\moy{(\XTP-\langle \XTP\rangle)^2\eta_{\RTP+{\rr}}}-k_{\rr}(t)\left[ \moy{{X_t}^2} - \moy{X_t}^2\right],
\label{defmtilde}
\end{equation}
so that the expansion of $\widetilde{w}_{\rr}(t)$ in powers of $u$ writes
\begin{equation}
\wt_{\rr}(t)=k_{\rr}(t)+iu\widetilde{g}_{\rr}(t)+\frac{(iu)^2}{2}\widetilde{m}_{\rr}(t)+\mathcal{O}\left(u^3\right),
\label{eq:dev_wt}
\end{equation}
where we used the definitions of $k_{\rr}(t)=\moy{\eta_{\RTP+\rr}}$ and $\gt_{\rr}(t)=\moy{\delta\XTP\delta\eta_{\RTP+{\rr}}}$. Expanding both sides of Eq. (\ref{dpsidt}) up to order 3 in $u$ in the limit $t\rightarrow \infty$ and using Eq. (\ref{eq:exp_dpsidt}), one gets the following expression for $\gamma$ :
\begin{equation}
\gamma=\frac{p_1\sigma}{\tau}\left[\frac{1}{6}\sigma^2(1-k_{\epu})-\frac{1}{2}\sigma\widetilde{g}_{\epu}-\frac{1}{2}\widetilde{m}_{\epu}\right]-\frac{p_{-1}\sigma}{\tau}\left[\frac{1}{6}\sigma^2(1-k_{\emu})+\frac{1}{2}\sigma\widetilde{g}_{\emu}-\frac{1}{2}\widetilde{m}_{\emu}\right].
\label{alpha_def}
\end{equation}

According to Eq. (\ref{eq:dev_wt}), the general evolution equations for $\widetilde{w}_{\rr}(u;t)$ (Eqs. (\ref{wtildegen}) and (\ref{wtildeboundary})) expanded at order 2 in $u$ then give the evolution equations of $\widetilde{m}_{\rr}(t)$. We get:

\begin{itemize}

	\item for $\rr\neq \mathbf{e_{\boldsymbol\nu}}$~:
	\begin{eqnarray}
	\label{eq:mt_bulk}
	\fl 2d\tau^*\partial_t\widetilde{m}_{\rr}(t)
	=\widetilde{L}\widetilde{m}_{\rr}(t)-\frac{2d\tau^*}{\tau}\sum_\mu p_\mu\widetilde{m}_{\emuu}(t)\nabla_\mu k_{\rr}(t)\nonumber \\
	\fl +\frac{2d\tau^*}{\tau}\sum_{\ep= \pm1 } p_\ep \sigma\left\{2\ep \left[(1-k_{\ee_\ep}(t))\nabla_\ep\widetilde{g}_{\rr}(t)-\widetilde{g}_{\ee_\ep}(t)\nabla_\ep k_{\rr}(t)\right]+\sigma(1-k_{\ee_\ep}(t))\nabla_\ep k_{\rr}(t)\right\}. \nonumber\\
	\end{eqnarray}
	
	\item for $\rr = \enu$ :
	\begin{eqnarray}
	\label{eq:mt_BC}
	\fl2d\tau^*\partial_t\widetilde{m}_{\enu}(t)
	=[\widetilde{L}+A_\nu(t)]\widetilde{m}_{\ee_\nu}(t)-\frac{2d\tau^*}{\tau}\sum_{\mu} p_\mu\widetilde{m}_{\emuu}(t) \nabla_\mu k_{\ee_\nu}(t) -\frac{2d\tau^*}{\tau}p_\nu\widetilde{m}_{\enu}(t)k_{\enu}(t)\nonumber \\
	\fl +\frac{2d\tau^*}{\tau} \sum_{\ep = \pm 1}p_\ep \sigma\left\{2\ep \left[(1-k_{\ee_\ep}(t))\nabla_\ep\widetilde{g}_{\enu}(t)-\widetilde{g}_{\ee_\ep}(t)\nabla_\ep k_{\enu}(t)\right]+\sigma(1-k_{\ee_{\ep}}(t))\nabla_\ep k_{\enu}(t)\right\}. \nonumber\\
	\end{eqnarray}

\end{itemize}

In the stationary limit, one computes from Eqs. (\ref{eq:mt_bulk}) and (\ref{eq:mt_BC})  the quantities $\widetilde{m}_{\ee_{\pm1}}$ and then the coefficient $\gamma$ from Eq. (\ref{alpha_def}). This will be made explicit in the case of a one-dimensional lattice in section \ref{sec:sol_mt_1D}.

\section{First cumulants and distribution of the TP position in one dimension}
\label{sec:solutions_1D}

In this section, we focus on the one-dimensional version of the general model presented in section \ref{sec:model}. This situation is related to a number of lattice models of interacting particles which have been widely studied in the mathematical and physical literature.  In particular, in the situation where the number of particles on the lattice is conserved and where the TP is not biased, the model corresponds to the well-known single file problem, for which several results have been derived exactly \cite{Harris1965,Levitt1973,Arratia1983,{Lizana2008},{Barkai2010},Krapivsky2014,{Sadhua},{Sabhapandit2015}}.

In what follows, we consider the equations derived in the previous section in the particular case of a one-dimensional lattice. We first recall the solutions of the equations  satisfied by the density profiles $k_{\rr}$ that were obtained in previous studies, and give a detailed resolution of the equations verified by the correlation functions $\gt_{\rr}$ and $\wt_{\rr}$.

\subsection{Solution of the equation on $k_{\rr}$ in one dimension}
\label{velocity_1D}

The solutions of the equations verified by $k_{\rr}$  have already been presented \cite{Benichou1999}, and we recall them here for completeness. In one dimension, we adopt the simplified notation $k_{\rr}=k_{n\ee_1}\equiv k_n$.  The stationary limit of the general equation verified by the density profiles (Eq. (\ref{systemek1})) is then a second order recurrence relation on the quantities $k_n$. Its solution has the following form:
\begin{equation}
\label{dprofiles}
k_{n} =
\begin{cases}
\rho + K_{+} {r_1}^n    & \text{for $n>0$}, \\
\rho + K_{-} {r_2}^n    & \text{for $n<0$},
\end{cases}
\end{equation}
where 
\begin{equation}
\label{r1}
r_{\substack{1\\2}}= \frac{A_{1} + A_{-1} + 2 (f + g) \mp \sqrt{\Big(A_{1} + A_{-1} + 2 (f + g)\Big)^2 - 4 A_{1} A_{-1}}}{2 A_{1}},
%\label{def_r12}
\end{equation} 
while the amplitudes $K_{\pm}$ are given respectively by
\begin{equation}
\label{Kp}
K_{+} = \rho \frac{A_{1} - A_{-1}}{A_{-1} - A_{1} r_1} ,
\end{equation}
and
\begin{equation}
\label{Km} 
K_{-} = \rho \frac{A_{1} - A_{-1}}{A_{-1}/r_2 - A_{1} } .
\end{equation}
We notice that $r_2>r_1$, which indicates that the density profile behind the TP decreases slower than in front of it. We also note that $K_+>0$  and $K_-<0$ which indicates that there is a jammed region ahead of the TP, and, on the contrary, a depleted region behind it.

One obtains a closed sed of  two  non-linear equations determining implicitly the  
parameters $A_{1}$ and $A_{-1}$, from which one can  compute
the TP stationary velocity, related to $A_{\pm 1}$ through 
\begin{equation}
\label{VTPsimple}
V = \frac{\sigma}{2 \tau^*} (A_{1} - A_{-1}).
\end{equation}
Substituting Eq. (\ref{dprofiles}) into the definition of $A_\mu$ (Eq. (\ref{eq:def_A_general})), we find 
\begin{eqnarray}
A_{1} &=& 1 + \frac{2p_1 \tau^*}{\tau} \left[1 - \rho - \rho \frac{A_{1} - A_{-1}}{A_{-1} /r_1 - A_{1}}
\right], \label{A1_implicit}\\
A_{-1} &=& 1 + \frac{2p_{-1} \tau^*}{\tau} \left[1 - \rho - \rho \frac{A_{1} - A_{-1}}{A_{-1}  - A_{1} r_2}
\right].\label{Am1_implicit}
\end{eqnarray}
For a given set of parameters ($f$, $g$, $\sigma$, $\tau$, $\tau^*$ and $p_1$), the numerical resolution of this system leads to the values of $A_1$ and $A_{-1}$ and then, using Eq. (\ref{VTPsimple}), to the value of the stationary
velocity of the TP. This approximated value of the stationary velocity will be compared to numerical simulations in section \ref{sec:1d_num_velocity}.

\subsection{Solution of the equation on $\gt_{\rr}$ in one dimension}
\label{diffcoeff_1D}

We now go one step further and determine the diffusion coefficient $K$. This in turn requires the knowledge of the  functions $\widetilde{g}_{\rr}$. For simplicity, we adopt the notations $\gt_{\rr}=\gt_{n\ee_1}=\gt_n$. Using the expressions of $k_n$ from Eq. (\ref{dprofiles}), the general equations satisfied by $\widetilde{g}_n$ (Eqs. (\ref{systemegtilde1}) and  (\ref{systemegtilde2}))  become:
\begin{itemize}
	\item for $n>1$:
\begin{eqnarray}
\label{gt_1d_bulk1}
&&A_1(\widetilde{g}_{n+1}-\widetilde{g}_{n})+A_{-1}(\widetilde{g}_{n-1}-\widetilde{g}_{n})-2(f+g)\widetilde{g}_{n}\nonumber\\
&+&\frac{2\tau^*}{\tau}\sigma\left\{p_1K_+r_1^n\left(1-\rho-K_+r_1-\frac{\widetilde{g}_{1}}{\sigma}\right)(r_1-1)\right.\nonumber\\
&-&\left.p_{-1}K_+r_1^n\left(1-\rho-K_+r_1+\frac{\widetilde{g}_{-1}}{\sigma}\right)(r_1^{-1}-1)\right\} =0,
\end{eqnarray}
	\item for $n<-1$:
\begin{eqnarray}
\label{gt_1d_bulkm1}
&&A_1(\widetilde{g}_{n+1}-\widetilde{g}_{n})+A_{-1}(\widetilde{g}_{n-1}-\widetilde{g}_{n})-2(f+g)\widetilde{g}_{n}\nonumber\\
&+&\frac{2\tau^*}{\tau}\sigma\left\{p_1K_-r_2^n\left(1-\rho-K_+r_1-\frac{\widetilde{g}_{1}}{\sigma}\right)(r_2-1)\right.\nonumber\\
&-&\left.p_{-1}K_-r_2^n\left(1-\rho-K_+r_1+\frac{\widetilde{g}_{-1}}{\sigma}\right)(r_2^{-1}-1)\right\} =0,
\end{eqnarray}
	\item  $\widetilde{g}_{1}$ and $\widetilde{g}_{-1}$ may be computed using the  boundary conditions given by Eq. (\ref{systemegtilde2}):
\begin{eqnarray}
\label{CL1}
\fl A_1\widetilde{g}_{2}-\widetilde{g}_{1}\left(A_{-1}+2(f+g)+\frac{2\tau^*}{\tau}p_1(\rho+K_+r_1^2)\right)
+\frac{2\tau^*}{\tau}p_{-1}(\rho+K_+r_1)\widetilde{g}_{-1}\nonumber\\
\fl =-\frac{2\tau^*}{\tau}\sigma p_1(1-\rho-K_+r_1)(\rho+K_+r_1^2)\nonumber\\
\fl+\frac{2\tau^*}{\tau}\sigma (p_1(1-\rho-K_+r_1)-p_{-1}(1-\rho-K_-r_2^{-1}))(\rho+K_+r_1),
\end{eqnarray}
\begin{eqnarray}
\label{CL2}
\fl A_{-1}\widetilde{g}_{-2}-\widetilde{g}_{-1}\left(A_{1}+2(f+g)+\frac{2\tau^*}{\tau}p_{-1}(\rho+K_-r_2^{-2})\right)
+\frac{2\tau^*}{\tau}p_{1}(\rho+K_-r_2^{-1})\widetilde{g}_{1}\nonumber\\
\fl =\frac{2\tau^*}{\tau}\sigma p_{-1}(1-\rho-K_-r_2^{-1})(\rho+K_-r_2^{-2})\nonumber\\
\fl +\frac{2\tau^*}{\tau}\sigma (p_1(1-\rho-K_+r_1)-p_{-1}(1-\rho-K_-r_2^{-1}))(\rho+K_-r_2^{-1}).
\end{eqnarray}
\end{itemize}
The general solution of Eqs. (\ref{gt_1d_bulk1}) and (\ref{gt_1d_bulkm1}) can be written:
\begin{equation}
\label{npositive}
\widetilde{g}_n=\alpha r_1^n-\frac{W}{A_1r_1-A_{-1}r_1^{-1}}nr_1^n\;\;{\rm for}\;n>0,
\end{equation}
and 
\begin{equation}
\label{nnegative}
\widetilde{g}_n=\beta r_2^n-\frac{W'}{A_1r_2-A_{-1}r_2^{-1}}nr_2^n\;\;{\rm for}\;n<0,
\end{equation}
where $\alpha$ and $\beta$ are constants to be determined, and where
\begin{equation}
W\equiv K_+\frac{2\tau^*}{\tau}\sigma\left\{p_1\left(1-\rho-K_+r_1-\frac{\widetilde{g}_{1}}{\sigma}\right)(r_1-1)-
p_{-1}\left(1-\rho-\frac{K_-}{r_1}+\frac{\widetilde{g}_{-1}}{\sigma}\right)(r_1^{-1}-1)\right\},
\label{def_W}
\end{equation}
\begin{equation}
W'\equiv K_-\frac{2\tau^*}{\tau}\sigma\left\{p_1\left(1-\rho-K_+r_1-\frac{\widetilde{g}_{1}}{\sigma}\right)(r_2-1)-
p_{-1}\left(1-\rho-\frac{K_-}{r_1}+\frac{\widetilde{g}_{-1}}{\sigma}\right)(r_2^{-1}-1)\right\}.
\label{def_Wp}
\end{equation}
Substituting Eq. (\ref{npositive}) into Eq. (\ref{CL1}),  Eq. (\ref{nnegative}) into Eq. (\ref{CL2}), and writing 
Eq. (\ref{npositive}) for $n=1$ and Eq. (\ref{nnegative}) for $n=-1$, we obtain a linear system of four equations satisfied by  $\alpha$, $\beta$, $\widetilde{g}_{1}$ and $\widetilde{g}_{-1}$, which is straightforward to solve. The explicit expressions of $\gt_1$ and $\gt_{-1}$ are given in \ref{sec:J1}. Note that they rely on the determination of the quantities  $K_{\pm}$ and $r_1,r_2$, which are  determined numerically for a given set of parameters with the method detailed in section \ref{velocity_1D}. Finally, for a given set of parameters, one can deduce the values of $\gt_{\pm1}$ and  the value of the diffusion coefficient using Eq. (\ref{KTP_def}). Note that this calculation also gives access to the spatial dependence of the cross-correlations functions $\gt_{n}$ through Eqs. (\ref{npositive}) and (\ref{nnegative}).\\

In section \ref{sec:1d_num_K}, we investigate the dependence of $K$ on the different parameters of the problem, and we confront the analytical prediction from the decoupling approximation to results from numerical simulations. We give a first insight into the understanding of the counter-intuitive non-monotonic dependence of $K$ over the density $\rho$ that was described in \cite{Benichou2013}.

\subsection{Solution of the equation on $\widetilde{m}_{\rr}$ in one dimension}
\label{sec:sol_mt_1D}

We finally solve the equations satisfied by the correlation functions $\widetilde{m}_{\ee_{\pm 1}}$, from which we will compute the coefficient $\gamma$, related to the third cumulant of the distribution, and defined by Eq. (\ref{defmtilde}). Starting from the general equations verified by $\widetilde{m}_{\rr}$ and  valid in any dimension (Eqs. (\ref{eq:mt_bulk}) and (\ref{eq:mt_BC})), we study the one-dimensional case. For simplicity, we write $\widetilde{m}_{\rr}=\widetilde{m}_{n\epu}=\widetilde{m}_n$. The quantities $\widetilde{m}_n$ are the solutions of the equations presented below:

\begin{itemize}
	
	\item for $\rr\neq \ee_\nu$, using Eq. (\ref{eq:mt_bulk}), one gets :
	\begin{equation}
	A_1 (\widetilde{m}_{n+1}-\widetilde{m}_n)+A_{-1}(\widetilde{m}_{n-1}-\widetilde{m}_n)-2(f+g)\widetilde{m}_n=S(n)
	\label{mn}
	\end{equation}
	where $S(n)$ can be expressed explicitly in terms of the functions $k_n$ and $\widetilde{g}_n$, determined respectively in sections \ref{velocity_1D} and \ref{diffcoeff_1D}. For $n>0$, we  write $S(n)$ under the following form :
	\begin{equation}
	S(n)=(C_1^+ +nC_2^+)r_1^n.
	\end{equation}
	The explicit expressions of $C_1^+$ and $C_2^+$ are given in \ref{app:cumul3}. Consequently, the solution of Eq. (\ref{mn}) reads:
	\begin{equation}
	\mt_n=\Gamma r_1^n+(a_+ n^2+b_+ n)r_1^n,
	\label{solmpos_main}
	\end{equation}
	with
	\begin{eqnarray}
	a_+&=&\frac{1}{2}\frac{C_2^+}{A_1r_1-A_{-1}r_1^{-1}},\\
	b_+&=&\frac{1}{A_1r_1-A_{-1}r_1^{-1}}[C_1^+-a_+(A_1r_1+A_{-1}r_1^{-1}].
	\end{eqnarray}
	
	For $n<0$, a similar resolution leads to:
	\begin{equation}
	\mt_n=\delta r_2^n+(a_- n^2+b_- n)r_2^n,
	\label{solmneg_main}
	\end{equation}
	with
	\begin{eqnarray}
	a_-&=&\frac{1}{2}\frac{C_2^-}{A_1r_2-A_{-1}r_2^{-1}},\\
	b_-&=&\frac{1}{A_1r_2-A_{-1}r_2^{-1}}[C_1^- -a_-(A_1r_2+A_{-1}r_2^{-1}],
	\end{eqnarray}
	and where the explicit expressions of $C_1^-$ and $C_2^-$ are given in  \ref{app:cumul3}.

	\item for $\rr=\epu$, we obtain from Eq. (\ref{eq:mt_BC}):
	\begin{equation}
	0=A_1\mt_2-\mt_1\left(A_{-1}+2(f+g)+\frac{2\tau^*}{\tau}p_1k_2\right)+\frac{2\tau^*}{\tau}p_{-1}\mt_{-1}k_1+\varphi_1,
	\label{BC1}
	\end{equation}
	with
	\begin{eqnarray}
	\fl \varphi_1=\frac{2d\tau^*}{\tau}p_1\left\{2\sigma\left[(\widetilde{g}_2-\widetilde{g}_1)(1-k_1)-\widetilde{g}_1(k_2-k_1)\right] + \sigma^2(k_2-k_1)(1-k_1)\right\}\nonumber \\
	+\frac{2d\tau^*}{\tau}p_{-1}\left\{-2\sigma\left[-\widetilde{g}_1(1-k_{-1})+\widetilde{g}_{-1}k_1\right] - \sigma^2 k_1(1-k_{-1})\right\}.
	\end{eqnarray}
	
	\item for $\rr=\emu$ we obtain from Eq. (\ref{eq:mt_BC}):
	\begin{equation}
	0=A_{-1}\mt_{-2}-\mt_{-1}\left(A_{1}+2(f+g)+\frac{2\tau^*}{\tau}p_{-1}k_{-2}\right)+\frac{2\tau^*}{\tau}p_{1}\mt_{1}k_{-1}+\varphi_{-1},
	\label{BC-1}
	\end{equation}
	with
	\begin{eqnarray}
	\fl \varphi_{-1}=\frac{2d\tau^*}{\tau}p_{-1}\left\{-2\sigma\left[-\widetilde{g}_{-1}(1-k_1)+\widetilde{g}_1k_{-1}\right] - \sigma^2 k_{-1}(1-k_1)\right\}\nonumber \\
	\fl+\frac{2d\tau^*}{\tau}p_{-1}\left\{2\sigma\left[(\widetilde{g}_{-2}-\widetilde{g}_{-1})(1-k_{-1})-\widetilde{g}_{-1}(k_{-2}-k_{-1})\right] + \sigma^2(k_{-2}-k_{-1})(1-k_{-1})\right\}. \nonumber \\
	\end{eqnarray}

\end{itemize}
Writing  Eq. (\ref{solmpos_main}) for $n=1$,  Eq. (\ref{solmneg_main}) for $n=-1$, and considering  the boundary conditions given by Eqs. (\ref{BC1}) and (\ref{BC-1}), one obtains a linear system of four equations with unknowns $\widetilde{m}_1$, $\widetilde{m}_{-1}$, $\Gamma$ and $\delta$ :
\begin{equation}
\begin{pmatrix} 
r_1     & 0 & M_{13} & M_{14} \\ 
0 & r_2^{-1} & M_{23} & M_{24} \\
A_1 r_1^2 & 0 & M_{33} & M_{34} \\
0 & A_{-1}r_2^{-2} & M_{43} & M_{44} 
\end{pmatrix}
\begin{pmatrix} 
\Gamma \\
\delta \\
\mt_1 \\
\mt_{-1} 
\end{pmatrix}
=
\begin{pmatrix} 
Y_1 \\
Y_2 \\
Y_3 \\
Y_4 
\end{pmatrix}
\label{systemem}
\end{equation}
where the expressions of the quantities $M_{ij}$ and $Y_j$ are given in \ref{app:cumul3}. Finally:
\begin{align}
\mt_1=&\frac{1}{\det M} \left[(M_{44}M_{22}-M_{42}M_{24})(M_{11}Y_3+M_{31}Y_1) \right.\nonumber\\
& \left.+(M_{34}M_{11}-M_{31}M_{14})(M_{42}Y_2+M_{22}Y_4)\right],\label{eq:exp_m1}\\
\mt_{-1}=&\frac{1}{\det M} \left[(M_{43}M_{22}-M_{42}M_{23})(M_{11}Y_3+M_{31}Y_1) \right.\nonumber\\
& \left.+(M_{33}M_{11}-M_{13}M_{31})(M_{42}Y_2+M_{22}Y_4)\right].\label{eq:exp_mm1}
\end{align}

The procedure to compute the coefficient $\gamma$ for a given set of parameters is the following. With the method presented in section \ref{velocity_1D}, one can compute numerically the quantities $K_{\pm}$, $r_1$, $r_2$ and $k_{\pm 1}$ for a given set of parameters. Using the analytical expressions of $\gt_{\pm 1}$, $\alpha$ and $\beta$ in terms of these quantities given in \ref{sec:J1}, one computes $\widetilde{m}_{\pm1}$  with Eqs. (\ref{eq:exp_m1}) and (\ref{eq:exp_mm1}). The coefficient $\gamma$ is deduced from its definition (\ref{alpha_def}).\\

\subsection{Solution of the equations on $\wt_{\rr}$ in one dimension}

We now turn to the resolution of the equations satisfied by the correlation functions $\wt_{\rr}$ in the specific case of a 1D lattice and in the stationary limit.

Starting from Eqs. (\ref{wtildegen}) and (\ref{wtildeboundary}) and assuming that there exists non-trivial stationary solutions, one gets the following equations satisfied by $\wt_n \equiv \wt_{n \ee_1}$:
\begin{eqnarray}
\fl B_1\wt_{n+1}-B_2\wt_n+B_3\wt_{n-1} = -(B_4 K_+ r_1^n+B_5) &\mathrm{~~for~~}& n > 1 \label{implicit1}\\
\fl B_1\wt_{n+1}-B_2\wt_n+B_3\wt_{n-1}=-(C_4 K_- r_2^n+B_5)  &\mathrm{~~for~~}&  n<-1 \label{implicit2}\\
\fl D_1 \wt_2 - D_2\wt_1+D_3 \wt_{-1}+D_4+D_5\wt_1^2+D_6\wt_1\wt_{-1}=0  &\mathrm{~~for~~}&  n=1 \label{implicit3}\\
\fl E_1 \wt_1 - E_2\wt_{-1} +E_3 \wt_{-2}+E_4+D_6\wt_{-1}^2 +D_5\wt_1\wt_{-1}=0  &\mathrm{~~for~~}& n=-1 \label{implicit4}
\end{eqnarray}
where the expressions of the different coefficients $B_i$, $C_i$, $D_i$, $E_i$ are given in \ref{detailedimplicit}. Eqs. (\ref{implicit1}) and (\ref{implicit2}) are associated to the characteristic equation
\begin{equation}
B_1 X^2-B_2 X+B_3=0,
\end{equation}
which has the solutions
\begin{equation}
q_{1,2}=\frac{B_2\pm\sqrt{B_2^2-4B_1B_3}}{2B_1}.
\end{equation}
An expansion in powers of $u$ shows that
\begin{equation}
q_1=r_1+\mathcal{O}(u)\mathrm{~~~and~~~}q_2=r_2+\mathcal{O}(u).
\end{equation}
At order zero in $u$,  Eqs. (\ref{implicit1}) and (\ref{implicit2}) are equivalent to the stationary limit of Eqs. (\ref{systemek1}) and (\ref{systemek2}), so that their solutions must coincide at this order. Consequently, the general solution of Eq. (\ref{implicit1}) is of the form
\begin{equation}
\widetilde{w}_n = \alpha_+ q_1^n
\end{equation}
The particular solution of Eq. (\ref{implicit1}) is easily calculated, and for $n>0$, we find
\begin{equation}
\widetilde{w}_n = \alpha_+q_1^n-\frac{B_4K_+}{B_1r_1-B_2+B_3r_1^{-1}}r_1^n-\frac{B_5}{B_1-B_2+B_3}.
\label{wtpos}
\end{equation}
With similar arguments, we find for $n<0$
\begin{equation}
\widetilde{w}_n = \alpha_-q_2^n-\frac{C_4K_-}{B_1r_2-B_2+B_3r_2^{-1}}r_2^n-\frac{B_5}{B_1-B_2+B_3}.
\label{wtneg}
\end{equation}
Finally, writing Eq. (\ref{wtpos}) (resp. Eq. (\ref{wtneg})) for $n=1$ (resp. $n=-1$) and using Eqs. (\ref{implicit3}) and (\ref{implicit4}), we find the following nonlinear system of four equations whose unknowns are $\widetilde{w}_1$, $\widetilde{w}_{-1}$, $\alpha_+$ and $\alpha_-$ :
\begin{equation}
\label{eq:system_w_alpha}
\left \{
\begin{array}{l}
\widetilde{w}_1 = \alpha_+q_1-\dfrac{B_4K_+}{B_1r_1-B_2+B_3r_1^{-1}}r_1-\dfrac{B_5}{B_1-B_2+B_3}\\
\widetilde{w}_{-1} = \alpha_-q_2^{-1}-\dfrac{C_4K_-}{B_1r_2-B_2+B_3r_2^{-1}}r_2^{-1}-\dfrac{B_5}{B_1-B_2+B_3}\\
D_1 \left[\alpha_+q_1^2-\dfrac{B_4K_+}{B_1r_1-B_2+B_3r_1^{-1}}r_1^2-\dfrac{B_5}{B_1-B_2+B_3} \right]\\
\indent - D_2\wt_1+D_3 \wt_{-1}+D_4+D_5\wt_1^2+D_6\wt_1\wt_{-1}=0\\
E_1 \wt_1 - E_2\wt_{-1} +E_3 \left[ \alpha_-q_2^{-2}-\dfrac{C_4K_-}{B_1r_2-B_2+B_3r_2^{-1}}r_2^{-2}-\dfrac{B_5}{B_1-B_2+B_3}\right]\\
\indent +E_4+D_6\wt_{-1}^2 +D_5\wt_1\wt_{-1}=0.
\end{array}
\right.
\end{equation}
The numerical resolution of this system of equations for specific values of $u$ allows us to calculate $\widetilde{w}_1$ and $\widetilde{w}_{-1}$ as functions of $u$, and to deduce the stationary cumulant generating function $\Psi$ as a function of $u$. Using Eq. (\ref{pdf_computation}), one can calculate the probability distribution $P_t(x)$, valid in the asymptotic regime $t\to\infty$. \\

In the next section, we analyze the solutions obtained for the velocity $V$, the diffusion coefficient $K$ and the coefficient $\gamma$ to study their dependence on the different parameters of the problem. These results, which were obtained using our decoupling approximation, are confronted to numerical simulations which exactly sample the master equation (\ref{eqmaitresse}).

\section{One-dimensional lattice: results and discussion}
\label{sec:simus_1d}

\subsection{Algorithm and numerical methods}
\label{sec:algorithm_1d}

In order to verify the accuracy of the approximation involved in the computation of the cumulants of the TP position, we perform numerical simulations. We use a kinetic Monte-Carlo (or Gillespie) algorithm \cite{Young1966,Gillespie1976} in order to get an exact sampling of the master equation (\ref{eqmaitresse}) describing the dynamics of the system. The details of the numerical methods are given in \ref{app:numericalmeth}.

\subsection{Velocity}
\label{sec:1d_num_velocity}

\begin{figure}
	\begin{center}
		\includegraphics[width=12cm]{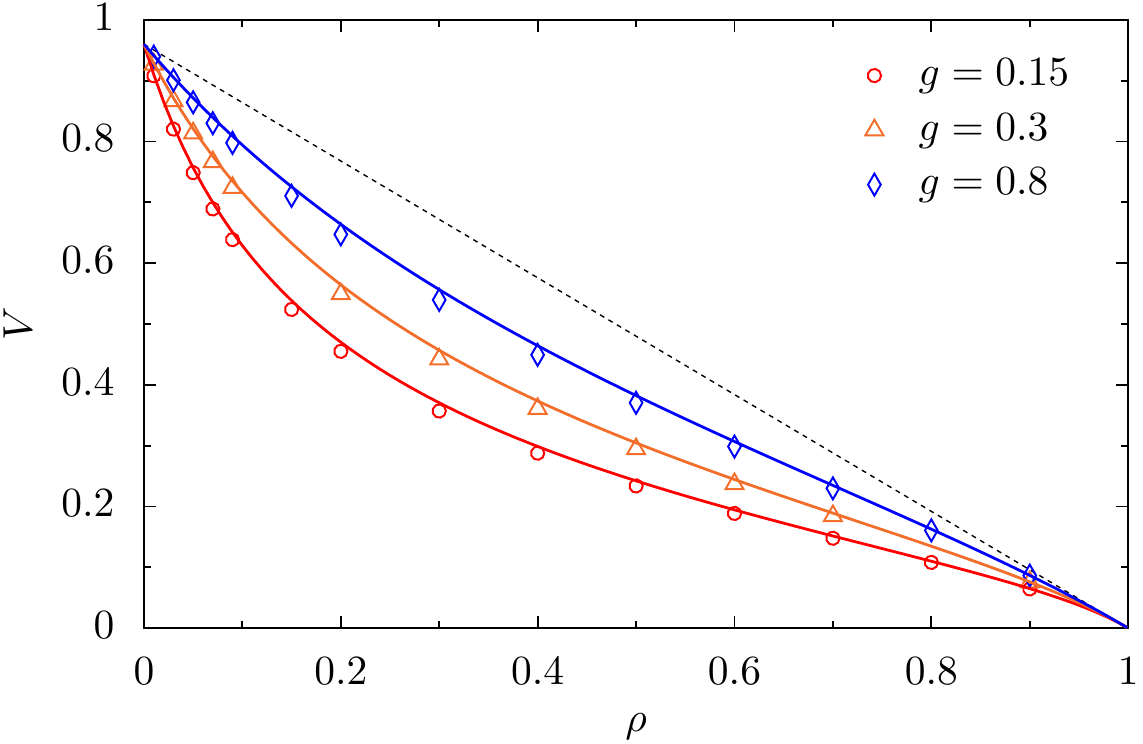}
		\includegraphics[width=12cm]{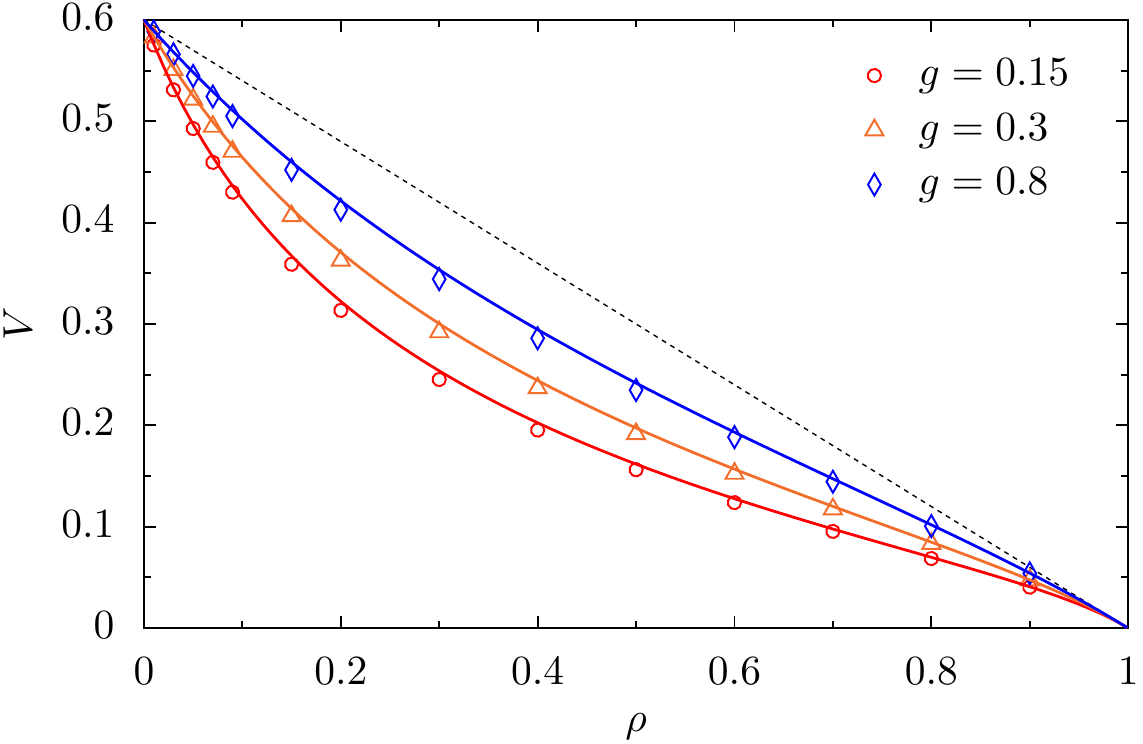}
		\caption{Stationary velocity $V$ of the TP as a function of the density for different values of the desorption rate $g$ obtained from numerical simulations (symbols) and from the decoupling approximation (lines). The bias is $p_1-p_{-1}=0.96$ (top) and $p_1-p_{-1}=0.6$ (bottom), the waiting times are $\tau=\tau^*=1$. The dashed line is the trivial mean-field solution $V=\frac{\sigma}{\tau}(p_1-p_{-1})(1-\rho)$.}
		\label{fig:simu_V}
	\end{center}
\end{figure}

For completeness, we present results for the velocity of the TP, which had already been presented in \cite{Benichou1999}. We study here the terminal velocity reached by the TP as a function of the density $\rho$, for different values of the bias. As $\rho$ is in fact fixed by the values of $f$ and $g$, we decide to vary $f$ for different values of $g$ in order to explore the whole range of parameters. Results are presented in Fig. \ref{fig:simu_V}. As expected, the velocity of the TP is a decreasing function of the bath density. We also confront the result from our decoupling approximation to the trivial mean-field solution (dashed line).

The discrepancy between the results from numerical simulations, which correspond to an exact sampling of the master equation (\ref{eqmaitresse}), and the solution obtained using the decoupling approximation (\ref{decouplage_k}) is very small, and the agreement is particularly good close to $\rho =0$ and $\rho=1$. The decoupling approximation is then very accurate for the estimation of the velocity of the TP.

\subsection{Diffusion coefficient}
\label{sec:1d_num_K}

\subsubsection{Results}

\begin{figure}
	\begin{center}
		\includegraphics[width=12cm]{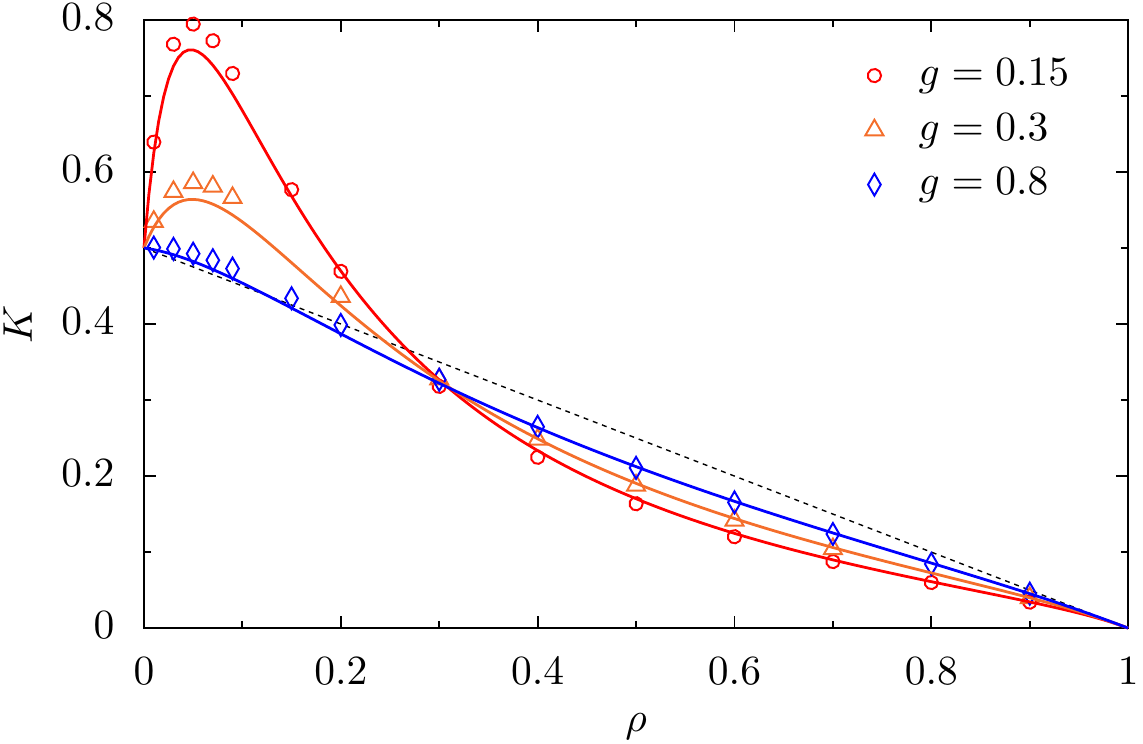}
		\includegraphics[width=12cm]{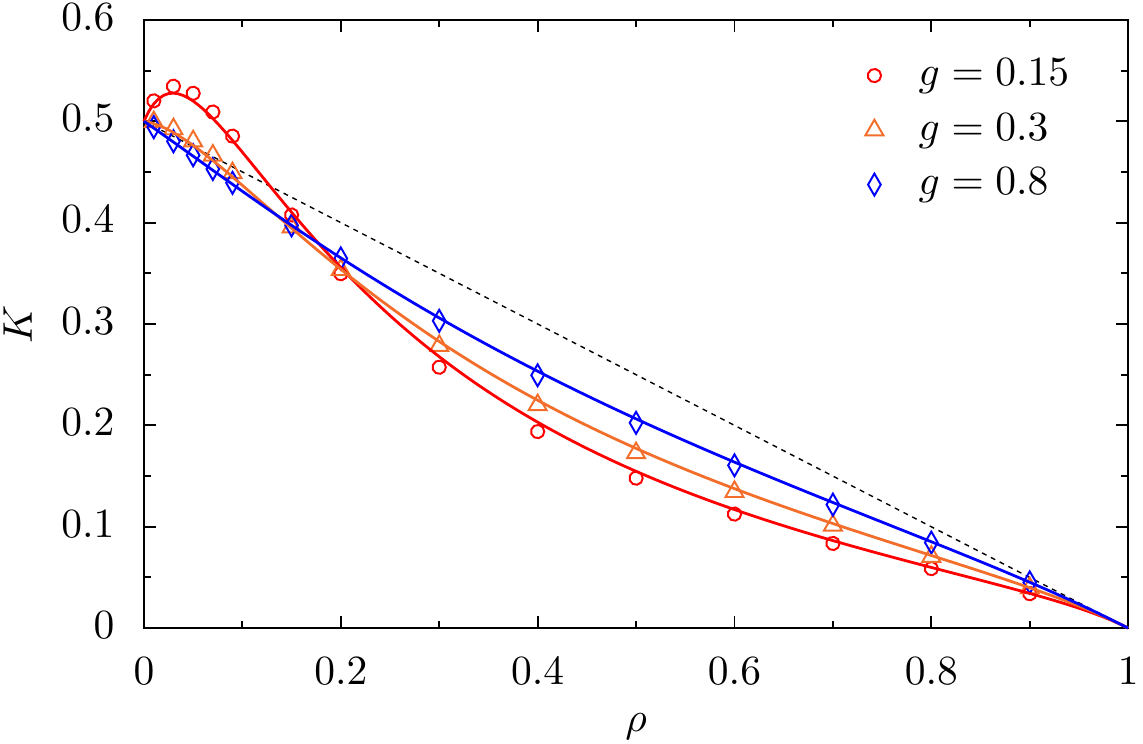}
		\caption{Stationary diffusion coefficient $K$ of the TP as a function of the density for different values of the desorption rate $g$ obtained from numerical simulations (symbols) and from the decoupling approximation (lines). The bias is $p_1-p_{-1}=0.96$ (top) and $p_1-p_{-1}=0.6$ (bottom), the waiting times are $\tau=\tau^*=1$. The dashed line is the trivial mean-field solution $K=\frac{\sigma^2}{2\tau}(1-\rho)$.}
		\label{fig:simu_K}
	\end{center}
\end{figure}

In a similar way, we study the diffusion coefficient as a function of the density $\rho$, for different values of $p_1$ and $g$. We confront the analytical predictions to results from numerical simulations in Fig. \ref{fig:simu_K}. The dependence of the diffusion coefficient on the density of particles was first investigated in \cite{Benichou2013}. In the simulations results presented in that publication, the numerical errors were underestimated, and we present here refined results.

The analytical predictions as well as the numerical simulations reveal the existence of a striking effect: the diffusion coefficient can be a non-monotonic function of the density of bath particles. Conterintuitively, increasing the density of particles surrounding the TP may increase the TP diffusion coefficient in some range of parameters. This result is surprising, since one naturally expects that the diffusion coefficient will be maximal when there are no hardcore bath particles (i.e., when $\rho=0$). Consequently, this means that the diffusion of the TP may be enhanced by the presence of bath particles on the lattice.  Moreover, we observe that for given values of $p_1$ and $g$, this extremum of the function $\KTP(\rho)$ may appear if the bias $p_1-p_{-1}$ is large enough. This effect could be investigated in experimental situations (e.g. in microrheology) and could have interesting applications.

The discrepancy between the results from numerical simulations, which correspond to an exact sampling of the master equation (\ref{eqmaitresse}), and the solution obtained using the decoupling approximations (Eqs. (\ref{decouplage_k}) and (\ref{decouplage2}))  is small, except in the domain $0.05 \lesssim\rho\lesssim0.15$. However, the approximate solution still gives a good qualitative description of the evolution of $K$. The agreement is particularly good close to $\rho =0$ and $\rho=1$. \\

In what follows, we find the criterion on the parameters $g$ and $p_1$ allowing the emergence of a maximum for $K(\rho)$. In addition, we show that the non-monotonicity of the diffusion coefficient is actually correlated to the non-monotonicity of the cross-correlation functions $\gt_n$ in the domain $n<0$ (i.e. behind the TP).

\subsubsection{Criterion for the existence of a maximum value of $\KTP$}
\label{sec:max_K_origin}

\begin{figure}
	\begin{center}
		\includegraphics[width=8cm]{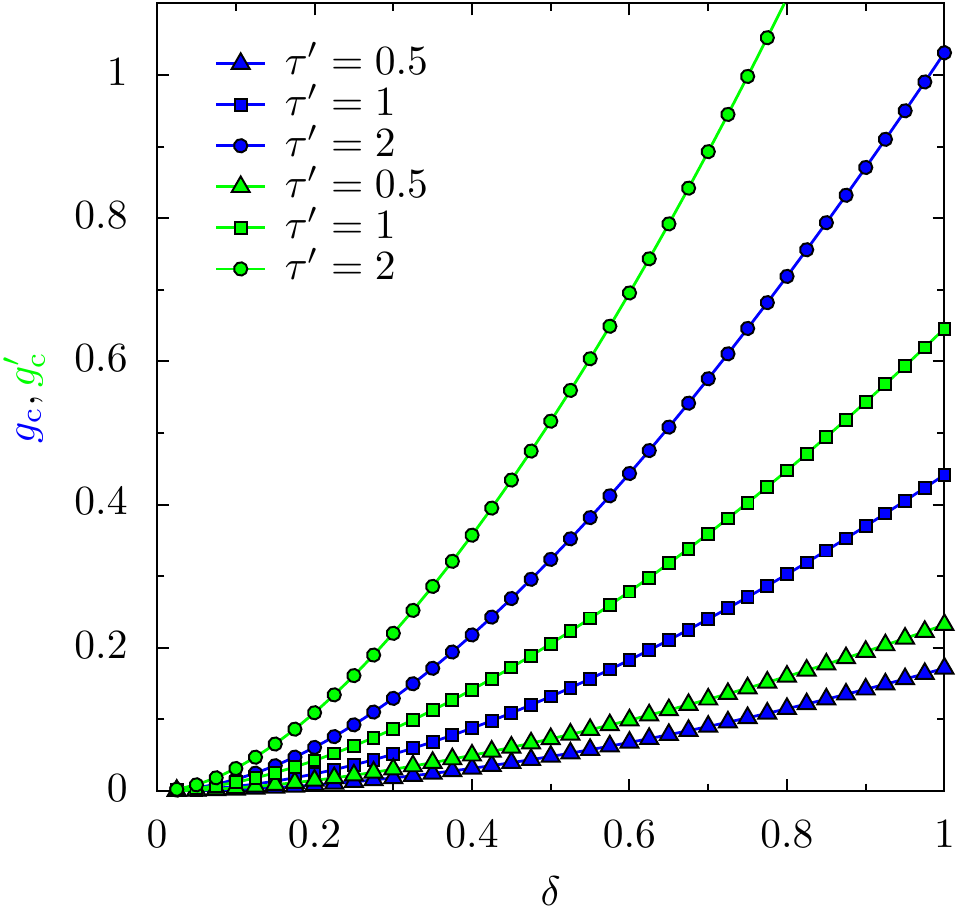}
		\caption{Critical value of the desorption rate $g_\mathrm{c}$ as a function of $\delta$, for different values of the ratio $\tau'=\tau^*/\tau$, with $\sigma=1$ and $\tau=1$ obtained by the study of the behavior of $K$ at $\rho\to0$ (curves in blue). The region below the curves correspond to the range of parameters where $\KTP(\rho)$ is nonmonotonic. The critical value of the desorption rate $g'_\mathrm{c}$ for which the cross-correlations $\gt$ have the property $\gt_{-1}>\gt_{-2}$ is represented in green. The region below the curves correspond to the range of parameters where $\gt_{-1}>\gt_{-2}$.}
		\label{fig_gcrit}
	\end{center}
\end{figure}

In this section, we determine an explicit criterion for the existence of this maximum. This is equivalent to determine the parameters for which $\KTP(\rho)$ has a positive derivative at the origin. We will then solve the equation:
\begin{equation}
\label{ }
\left.\frac{\dd \KTP}{\dd \rho}\right|_{\rho=0} = 0.
\end{equation}
If $g$ is fixed, as the density $\rho$ is related to $f$ and $g$ through $\rho=f/(f+g)$, this is equivalent to considering $\KTP$ as a function of $f$ and solving
\begin{equation}
\label{ }
\left.\frac{\dd \KTP}{\dd f}\right|_{f=0} = 0.
\end{equation}
In what follows,  we obtain the leading order term of $\KTP$ in an expansion in powers of $f$, the other parameters being constant. For simplicity, we introduce the quantities $\tau'\equiv\tau^*/\tau$ and $\delta\equiv p_1-p_{-1}$. Assuming that the quantities $A_{\pm 1}$ have the following expansions
\begin{equation}
\label{ }
A_{\pm1} \underset{f\to0}{=} A^{(0)}_{\pm1} + A^{(1)}_{\pm1} f +\mathcal{O}(f^2),
\end{equation}
and using Eqs. (\ref{A1_implicit}) and (\ref{Am1_implicit}), we obtain
\begin{equation}
\label{A_small_f}
A_{\pm1} \underset{f\to0}{=} 1+(1\pm\delta)\tau'  -\frac{\tau'}{g^2}\frac{1\pm\delta}{1+\tau'(1\pm\delta)}\left(\delta^2{\tau'}^2+g(1+\tau')\pm\delta\sqrt{\delta^2{\tau'}^2+g^2+2g(1+\tau')}\right)f+\mathcal{O}(f^2).
\end{equation}
We can deduce from the expansions of $A_{\pm1}$ the expansions of $K_{\pm}$, $r_1$ and $r_2$ using Eqs. (\ref{r1}), (\ref{Kp}) and (\ref{Km}). We then obtain the expansions of $k_{\pm1}$ and $\gt_{\pm1}$ in powers of $f$ using the results from sections \ref{velocity_1D} and \ref{diffcoeff_1D}, and finally an expansion of $\KTP$ with Eq. (\ref{KTP_def}). These general expressions are too lengthy to be reproduced here, but we give them in the case $\tau'=1$, which is the case we considered in our simulations:
\begin{equation}
\label{ }
\KTP\underset{f\to0}{=}\frac{\sigma^2}{2\tau}\left[1+\frac{N(g,\delta)}{D(g,\delta)}f+\mathcal{O}(f^2)\right]
\end{equation}
where
\begin{eqnarray}
\fl N(g,\delta)  =  2\left[  (\delta^2-4)g^3+(3\delta^4-16)g^2+\delta^2(3\delta^4-5\delta^2+4)g+\delta^4(\delta^4-3\delta^2+4)  \right] \nonumber\\
\fl- \sqrt{\delta^2+g^2+4g}\left[  2(\delta^4-4)(\delta^2-3)g^2+\delta^2(\delta^4-\delta^2+4)g+4\delta^4(\delta^2-2)  \right],\\
\fl D(g,\delta)  =  2(\delta-2)^2(\delta+2)^2(\delta^2+g^2+4g)g^3.
\end{eqnarray}
For any value of $\delta$ and $g$, $D(g,\delta)>0$. For a given value of the bias $\delta$, the critical value of $g$ (denoted by $g_\mathrm{c}$) allowing $\KTP$ to reach a maximal value is then the solution of the equation:
\begin{equation}
\label{eq_gcrit}
N(g_\mathrm{c}, \delta)=0,
\end{equation}
which can be determined numerically. We present the numerical solutions of Eq. (\ref{eq_gcrit}) in Fig. \ref{fig_gcrit} (curves in blue). On this figure, we also give the solutions obtained for other values of $\tau'$. We conclude that, for a fixed value of the bias, $K$ is nonmonotonic if the desorption rate $g$ is large enough (or, for a fixed value of the desorption rate, if the bias is large enough).

\subsubsection{Influence of the cross-correlations functions on the non-monotonicity}
\label{sec:crit_1}

\begin{figure}
	\begin{center}
		\includegraphics[width=11cm]{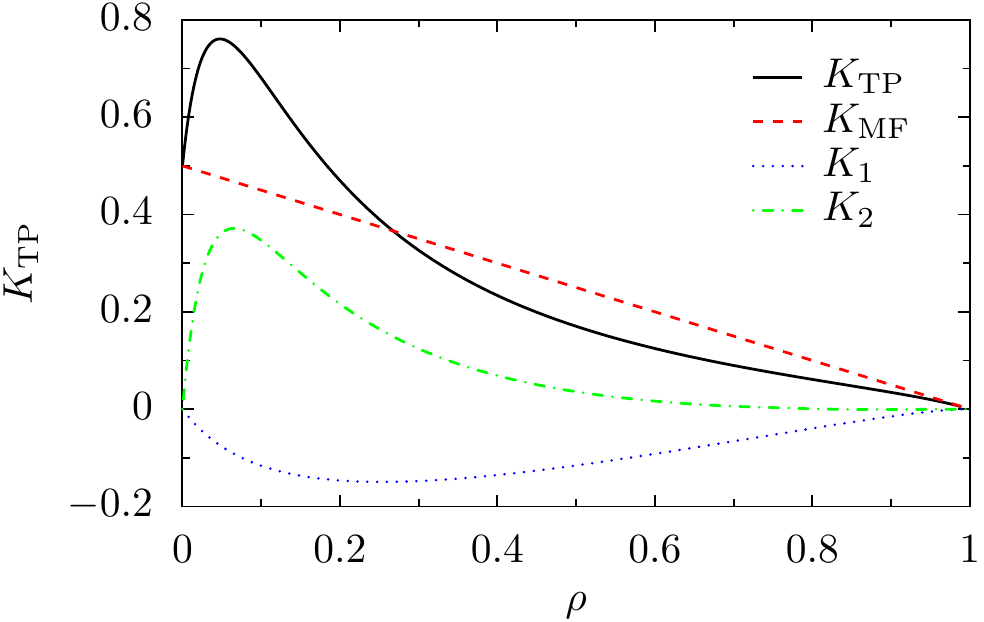}
		\caption{Contributions in the expressions of $\KTP$ as a function of the density, for the parameters $p_1=0.98$ and $g=0.15$.}
		\label{KTP_contribs}
	\end{center}
\end{figure}

We now aim to give a first insight into a better understanding of the non-monotonicity of the diffusion coefficient. We first notice that the diffusion coefficient $\KTP$ (Eq. (\ref{KTP_def})) may be separated into three contributions:
\begin{equation}
\label{ }
\KTP=K_{\rm MF} +K_1+K_2,
\end{equation}
with
\begin{eqnarray}
K_{\rm MF} & = & \frac{\sigma^2}{\tau}(1-\rho), \\
K_1 & = & -\frac{\sigma^2}{\tau}[p_1(k_1-\rho)+p_{-1}(k_{-1}-\rho)],\\
K_2 &=& -\frac{2\sigma}{\tau}(p_1 \gt_1-p_{-1}\gt_{-1}).
\end{eqnarray}
The first term is the trivial mean-field approximation of the problem, obtained by taking the average local densities $k_{\rr}$ equal to $\rho$ and the cross-correlation functions $\gt_{\rr} = \moy{\delta X_t \delta \eta_{\XTP+\rr}}$ equal to zero. $K_1$ may be seen as a contribution from the inhomogeneous density profiles, and $K_2$ a contribution from the cross-correlations $\gt_{\pm1 }$. For a given set of parameters which gives rise to a non-monotonic behavior ($p_1=0.98$ and $g=0.15$), we plot in Fig. \ref{KTP_contribs} $\KTP$ as well as the three contributions. The origin of the non-monotonicity of $\KTP$ with respect to the density $\rho$ can then be attributed to $K_2$.

\begin{figure}
	\begin{center}
		\includegraphics[width=13cm]{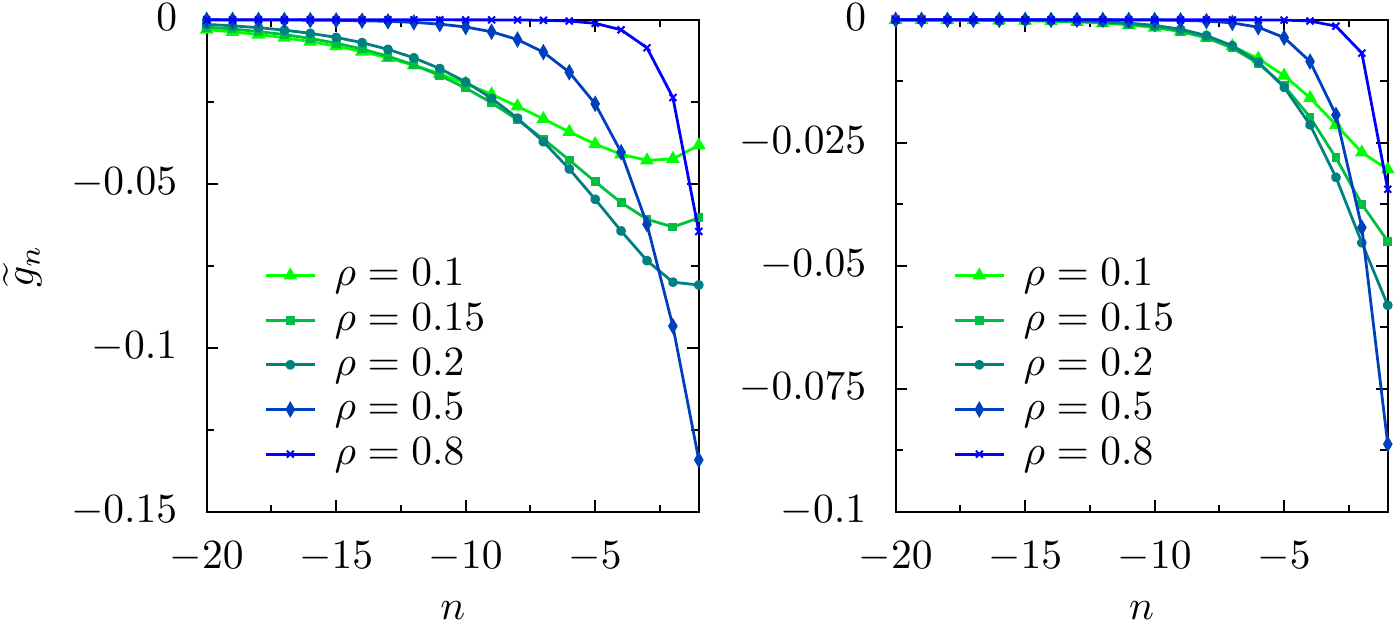}
		\caption{Cross-correlations functions $\gt$ as a function of the distance to the tracer $n$ for $g=0.2$ (left) and $g=0.6$ (right), and for different values of the density $\rho$. In both cases,  $p_1=0.98$.}
		\label{fig:gtilde_vary_rho}
	\end{center}
\end{figure}

\begin{figure}
	\begin{center}
		\includegraphics[width=7cm]{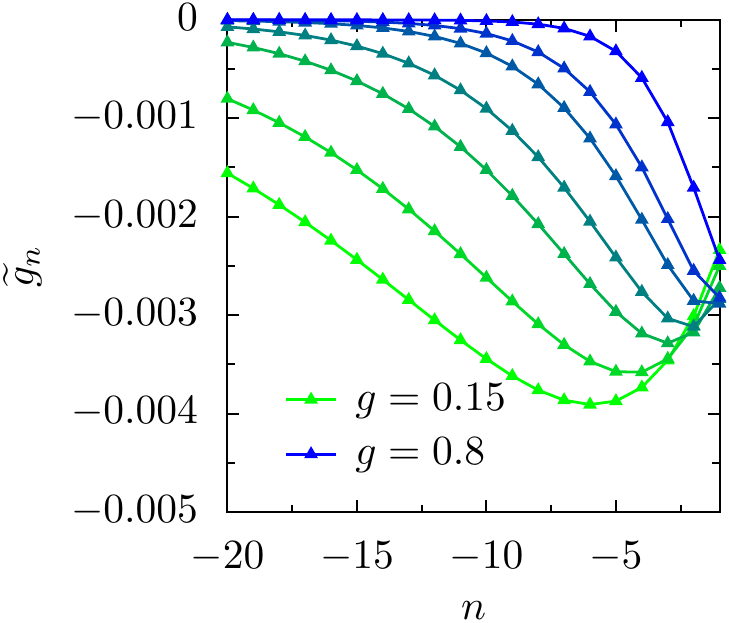}
		\caption{Cross-correlations functions $\gt$ as a function of the distance to the tracer $n$ for $\rho=0.01$, $p_1=0.98$, and for different values of the desorption parameter $g$.}
		\label{fig:gtilde_vary_g}
	\end{center}
\end{figure}

The cross-correlation functions $\gt_n$ have nontrivial behaviors with respect to the distance $n$ to the TP. In particular, they appear to have a non-monotonous behavior in the domain $n<0$, i.e. behind the TP, for some values of the parameters. On Fig. \ref{fig:gtilde_vary_rho}, we plot the functions $\gt_n$ for different values of the density $\rho$, for two sets of parameters: one for which the diffusion $\KTP$ is known to be a non-monotonic function of the density ($p_1=0.98$ and $g=0.2$), and one for which it is monotonic ($p_1=0.98$ and $g=0.6$). On Fig. \ref{fig:gtilde_vary_g}, we plot the functions $\gt_n$ for different values of the desorption parameter $g$, for $\rho=0.01$, for $p_1=0.98$. The non-monotonicity of $\gt_n$ with the distance to the TP then seems to be correlated with that of $\KTP$ with the density, as it occurs for $g$ small enough.

In order to get a more quantitative comparison of these two effects (the non-monotonicity of $\KTP$ with respect to the density $\rho$ and that of $\gt_n$ with respect to the distance to the TP $n$ in the domain $n<0$),  we first determine, for a fixed value of the bias $p_1-p_{-1}$, and at leading order in $f$, the critical value of $g'_{\rm c}$ giving a non-monotonic behavior for $\gt_n$, i.e. for which $\gt_{-1}>\gt_{-2}$.

Using the small $f$ expansion of $A_{\pm 1}$ (Eq. (\ref{A_small_f})), we easily deduce from the definitions of $\gt_n$ in the domain $n<0$ (Eq. (\ref{nnegative})) an expression for $\gt_{-1}-\gt_{-2}$ at leading order in $f$. The general expression is too lengthy to be given here. We present it in the particular case where $\tau'=1$:
\begin{equation}
\label{ }
\gt_{-1}-\gt_{-2} \underset{f\to 0}{=} 2\sigma f \frac{N'(g,\delta)}{D'(g,\delta)}+\mathcal{O}(f^2)
\end{equation}
with
\begin{align}
 N'(g,\delta) =& (\delta-g) \left[(1+\delta)g^2-2(\delta^2-2\delta-2)g+\delta^2(\delta-1)   \right]    \nonumber \\
&-\sqrt{\delta^2+g^2+4g}\left[(1+\delta)g^2-(3\delta+2)(\delta-1)g+\delta^2(\delta-1)   \right],  \\
D'(g, \delta)  =&  2g\sqrt{\delta^2+g^2+4g}\Big\{     \left[2g^2+(2\delta+4)g+\delta(\delta+2) \right] \sqrt{\delta^2+g^2+4g} 
 \nonumber\\
&  +2g^3+2(\delta+4)g^2+2(\delta+2)(\delta+1)g+\delta^2(\delta+2)\Big\}.
\end{align}
Then, for a fixed value of the bias $\delta$, the critical value of $g$ canceling $\gt_{-1}-\gt_{-2} $ is the solution of the equation
\begin{equation}
\label{ }
N'(g'_{\rm c}, \delta)=0.
\end{equation}
The numerical solutions for $g'_{\rm c}$ as a function of $\delta$, for different values of the ratio $\tau'$ are represented in Fig. \ref{fig_gcrit} and confronted to the values $g_{\rm c}$ obtained by the criterion on $\KTP$. The two functions $g_{\rm c}$ and $g'_{\rm c}$ are comparable as long as $\tau'$ is not too large. We also show that their expansions for $\delta \to 0$ are identical at leading order, and that they go to zero as $(\sqrt{17}+1)\delta^2/8$.

This study indicates that the emergence of a maximal value for $K(\rho)$ for $\rho>0$ is correlated to the observation of a minimal value in the cross-correlation functions $\gt_{n}$ in the domain $n<0$. \\

In \ref{sec:crit_2}, for given values of the jump probability $p_1$ and of the desorption parameter $g$, we study the range of density $[0,\widetilde{\rho}]$  for which the diffusion coefficient is greater than $1/2$, which is its value when there is no bath particle. We also determine the range of density $[0,\widetilde{\rho}']$ for which $\gt_{-1}-\gt_{-2}<0$. The range of parameters for which $K>1/2$ is then shown to be correlated to the range of parameters for which $\gt_{-1}-\gt_{-2}<0$.\\

These two  studies (from sections \ref{sec:crit_1} and \ref{sec:crit_2})  show that the non-monotonicity of the diffusion coefficient with respect to the density (emergence of a maximum value of $K$ for $\rho>0$) and the non-monotonicity of the cross-correlation function $\gt_n$ behind the TP are correlated. A more detailed study of the cross-correlation functions $\gt_n$, whose behavior affects the fluctuations of the TP position, could allow us to have a more physical understanding of the phenomenon highlighted in this section.

\subsection{Third cumulant}

The third cumulant of the distribution of $X_t$ gives information about its asymmetry. We introduced earlier the coefficient $\gamma$, defined by
\begin{equation}
\label{eq:def_gamma_recall}
\gamma =  \lim_{t\to\infty} \frac{1}{6}\frac{\mathrm{d}}{\mathrm{d}t} \moy{({\XTP}-\moy{\XTP})^3} .
\end{equation}
With this definition, if $\gamma>0$ (resp. $\gamma<0$), the distribution of $X_t$ is expected to be skewed to the right (resp. to the left). Let us recall the situation of a biased random walker on a lattice in the absence of exclusion interactions. If the particle is more likely to jump to the right ($p_1>p_{-1}$), its  third cumulant will be positive, indicating a distribution skewed to the right. Here, we study the influence of the presence of bath particles and of the different parameters of our model on the sign of the third cumulant of $X_t$.

\begin{figure}
	\begin{center}
		\includegraphics[width=12cm]{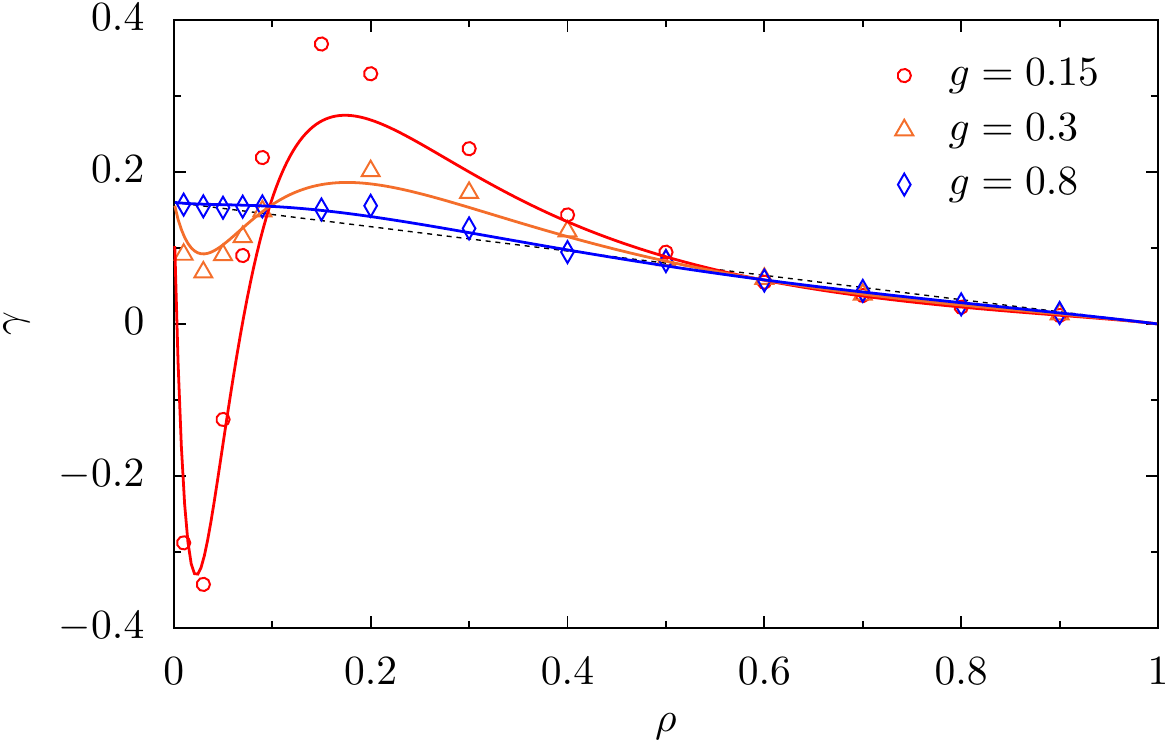}
		\includegraphics[width=12cm]{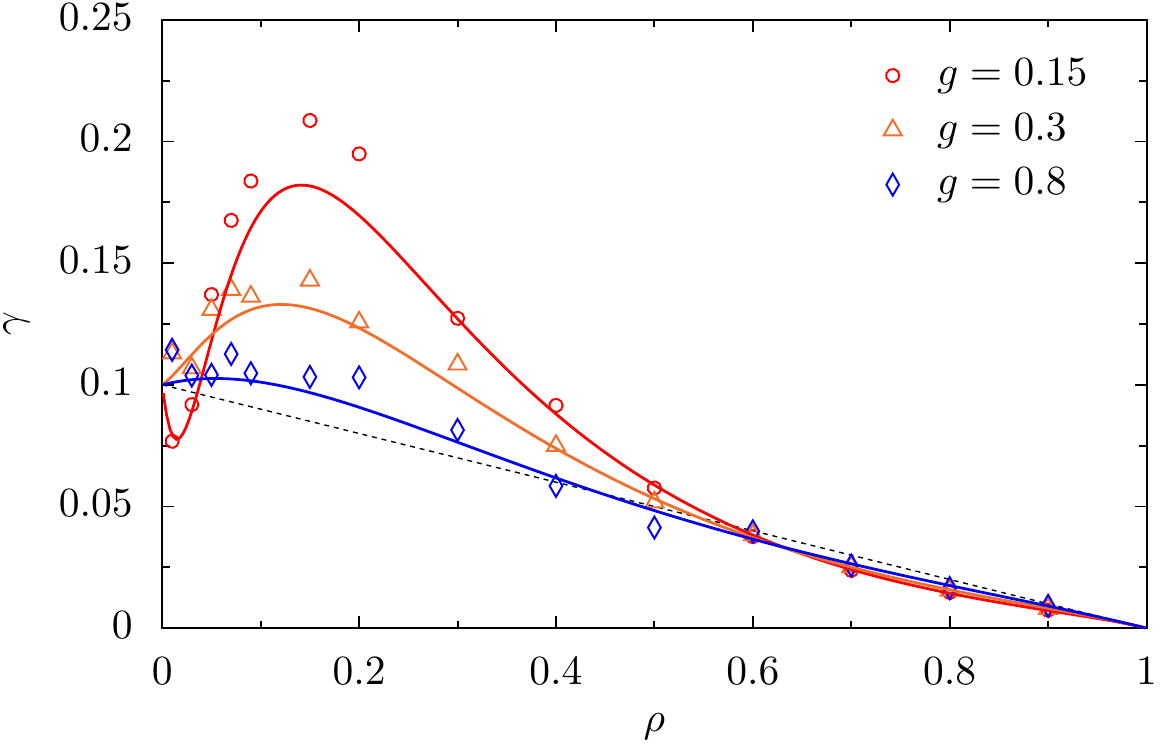}
		\caption{Coefficient $\gamma$  defined in Eq. (\ref{eq:def_gamma_recall}) as a function of the density for different values of the desorption rate $g$ obtained from the decoupling approximation. The bias is $p_1-p_{-1}=0.96$ (top) and $p_1-p_{-1}=0.6$ (bottom). The waiting times are $\tau=\tau^*=1$. The dashed lines are the trivial mean-field solutions: $\frac{\sigma^3}{6\tau}(p_1-p_{-1})(1-\rho)$.}
		\label{fig:cumulant3_approx}
	\end{center}
\end{figure}

We use the solutions for $\widetilde{m}_{\pm1}$ (Eqs. (\ref{eq:exp_m1}) and (\ref{eq:exp_mm1})) obtained from the decoupling approximation in order to compute the coefficient $\gamma$  from its definition (Eq. (\ref{alpha_def})).  For two values of the bias ($p_1=0.98$ and $0.8$), we study the coefficient $\gamma$ as a function of the density $\rho$, for different values of the desorption rate $g$. The solutions of the equations obtained from the decoupling approximation are compared to results from numerical simulations. The curves are presented in Fig. \ref{fig:cumulant3_approx}. For high values of the desorption rate, $\gamma(\rho)$ is monotonic and decreases when the density of bath particles increases. However, for small values of the desorption rate, the function becomes nonmonotonic, and one observes the emergence of a minimum and a maximum value, different from the trivial extrema at $\rho=0$ and $\rho=1$. Finally, if $g$ is small enough, there exists an interval of density for which $\gamma$ becomes negative, which means that the distribution of $X_t$ may actually be negatively skewed, in opposition with the situation where there is no bath particles on the lattice.

Confronting these predictions with results from numerical simulations, we show that the decoupling approximation  offers a good prediction of the behavior of the third cumulant of the TP position in a wide range of parameters.

\subsection{Cumulant generating function and propagator}
\label{sec:cgf1d}

\begin{figure}
	\begin{center}
		\includegraphics[width=12cm]{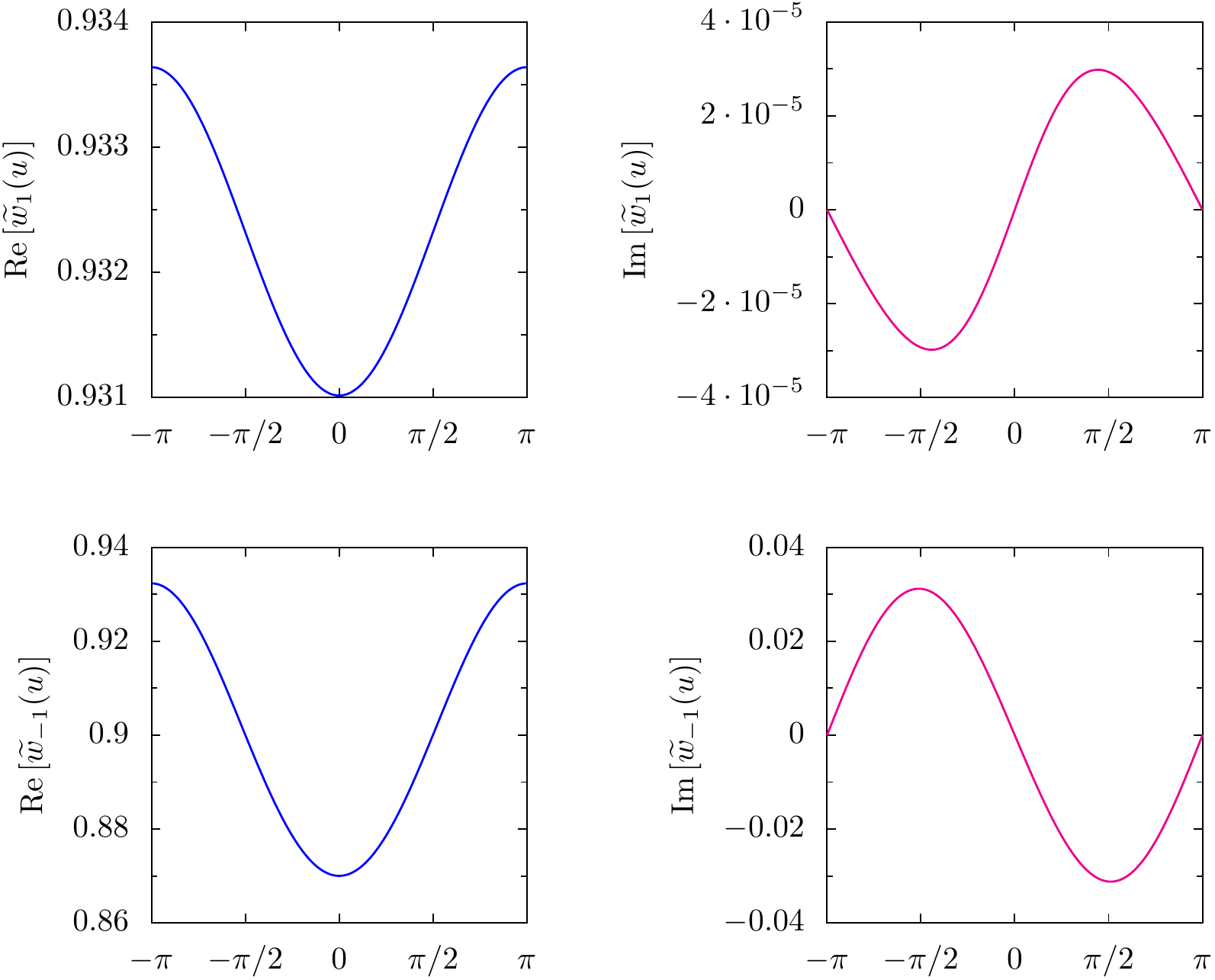}
		\caption{Real and imaginary parts of the generalized correlation functions $\wt_{\pm1}(u)$, obtained from the resolution of the system (\ref{eq:system_w_alpha}), for the parameters $\rho=0.9$, $g=0.15$, $p_1=0.98$, $\tau=\tau^*=\sigma=1$.}
		\label{solutions_wtilde}
	\end{center}
\end{figure}

For a given set of parameters, we solve the system of equations  which determines $\wt_{\pm 1}$ implicitly (Eq. (\ref{eq:system_w_alpha})), and we obtain the numerical values of $\wt_{\pm1}(u)$ for $u$ varying in $[-\pi, \pi]$. We represent the real and imaginary parts of $\wt_{\pm1}(u)$ as functions of $u$ in Fig. \ref{solutions_wtilde}. The p.d.f. $P_t(x)$ in the long-time limit is obtained from Eq. (\ref{pdf_computation}).

\begin{figure}
	\begin{center}
		\includegraphics[width=13cm]{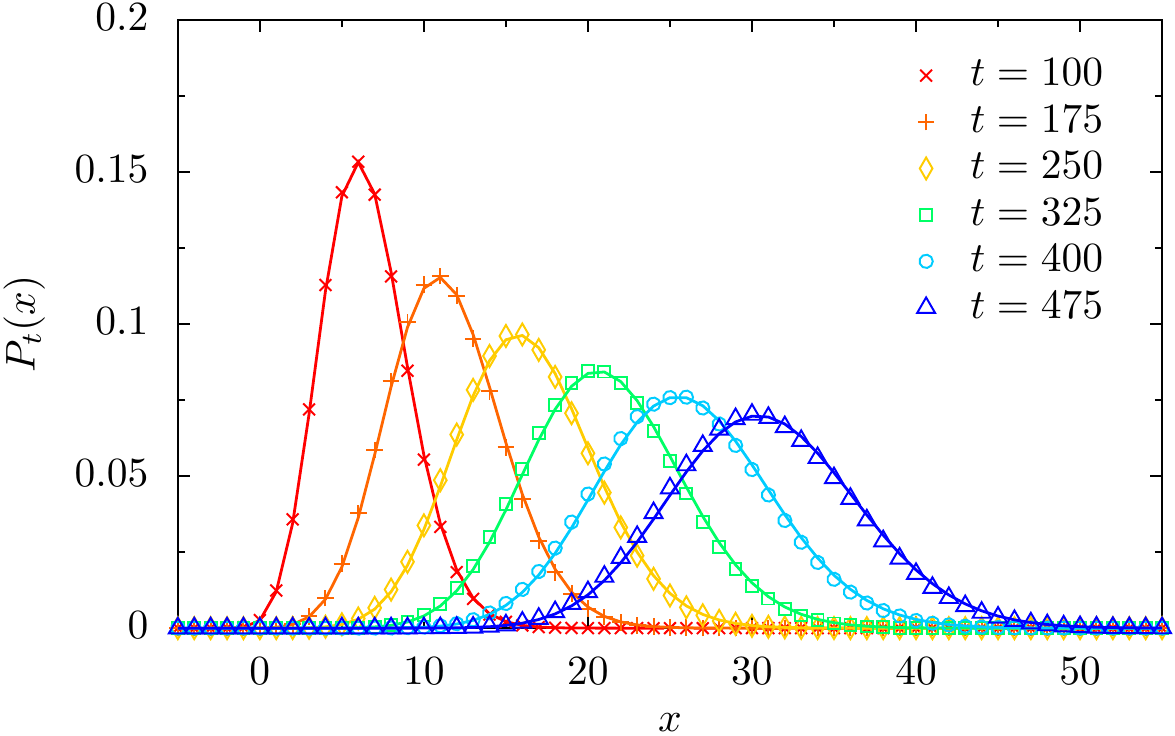}
		\caption{Probability distribution function of the TP position $P_t(x)=\mathrm{Prob}[X_t=x]$, for different times. The results from numerical simulations (symbols) are compared to the expression (\ref{pdf_computation}), the functions $\wt_{\pm1}(u)$ being numerically computed as solutions of the system (\ref{eq:system_w_alpha}). The parameters are $g=0.15$, $p_1=0.98$, $\rho=0.9$.}
		\label{distri_example}
	\end{center}
\end{figure}

We compare the results from this calculation with data obtained from Monte Carlo simulations for a given set of parameters (see Fig. \ref{distri_example}).
We observe a good agreement between the analytical prediction obtained from the decoupling approximation and the results from numerical simulations. 
We see that the prediction from the decoupling approximation tends to be shifted to the right for large times: this is expected from the analysis of the velocity of the TP, which was shown to be overestimated by the approximation (Fig. \ref{fig:simu_V}). 

\begin{figure}
	\begin{center}
		\includegraphics[width=13cm]{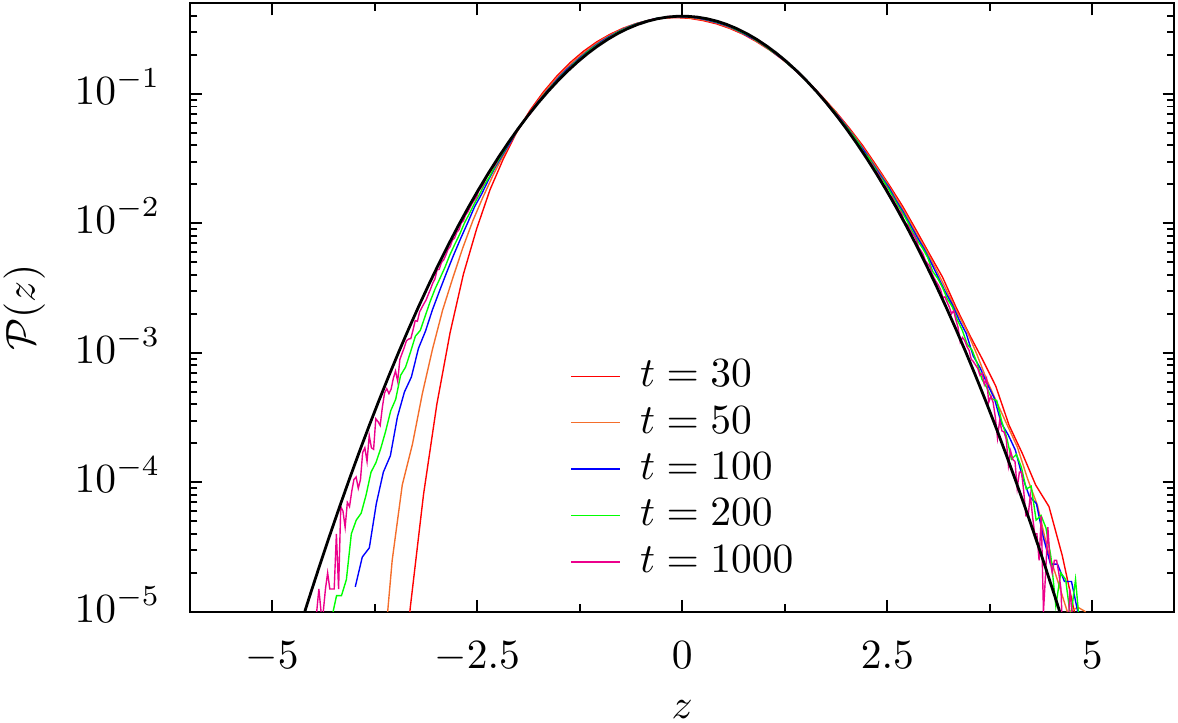}
		\caption{Probability distribution of the rescaled variable  $Z_t=(X_t-\moy{X_t})/\sqrt{\mathrm{Var}(X_t)}$ at different times, obtained with the following parameters: $\rho=0.1$, $g=0.15$, $p_1=0.98$, $\tau=\tau^*=\sigma=1$. The black line is the normal distribution $\exp(-x^2/2)/\sqrt{2\pi}$.}
		\label{distri_gauss}
	\end{center}
\end{figure}

As emphasized in section \ref{sec:governing_eq_cgf}, the rescaled variable $Z_t=(X_t-\moy{X_t})/\sqrt{\mathrm{Var}(X_t)}$ is expected to be distributed accordingly to a Gaussian distribution in the long-time limit. We plot on Fig. \ref{distri_gauss} the distribution $\mathcal{P}$ defined by
\begin{equation}
\label{ }
\mathcal{P}_t(z)=\mathrm{Prob}\left[Z_t= z   \right],
\end{equation}
 and compare it with the normal distribution $\ex{-x^2/2}/\sqrt{2\pi}$. At sufficiently long times, the distribution of the random variable $Z_t$ converges to the normal distribution of mean zero and unit variance. The rescaled position of the TP is then asymptotically Gaussian in the long-time limit.

\section{Conclusion}

We studied the diffusion of a biased tracer particle (TP) in a hardcore lattice gas in contact with a reservoir of particles. From the general master equation of the problem, we gave a detailed derivation of the equation satisfied by the fluctuations of the position of the TP and presented in a previous publication. This equation involve the density profiles around the TP and cross-correlation functions, whose evolution equations are obtained in a closed form by resorting to mean-field type approximation.

Going one step further, we extended this approximation to higher-order correlation functions in order to obtain the evolution equation verified by the cumulant generating function of the TP position. This equation then yields the entire probability distribution of the TP on a lattice of arbitrary dimension. We also obtained the equation satisfied by the third-cumulant of the distribution, which gives information about its asymmetry.

We then solved these equations in the particular case of a one-dimensional lattice. We recall the result from \cite{Benichou2013}: the diffusion coefficient of the TP is a nonmonotonic function of the density of bath particles. Counterintuitively, it reaches a maximum value for a nonzero value of the density. Thus, the presence of bath particles on the lattice may actually enhance its diffusion coefficient. Here, we showed that this effect is related to an anomaly in the behavior of  bath-tracer cross-correlation functions, that are nonmonotonic functions of the distance to the TP. Another surprising observation arises by analyzing the third cumulant of the TP position, which was shown to be nonmonotonic and to take negative values in a wide range of parameters. These analytical predictions were confronted with exact numerical samplings of the master equation, which indicate that the approximation we used is accurate in a wide range of parameters.

We finally solved the equation satisfied by the cumulant generating function, deduced the probability distribution and showed that the position of the TP rescaled by its fluctuations is Gaussian-distributed in the long-time limit.\\

The equations presented in this paper are very general, and allow to compute the cumulants of the TP position and therefore its distribution under a mean-field-type approximation which was shown to be  accurate in a broad range of parameters. These equations are valid for lattices of arbitrary dimension. We leave to future work the study of their solutions on higher-dimensional lattices. It could be interesting to see if the results obtained on the one-dimensional lattice -- enhanced diffusion coefficient, nonmonotonic and negative third cumulant, convergence of the rescaled distribution to a Gaussian distribution -- can be extended to lattices of higher dimension, or observed in experimental systems.

\section*{Acknowledgments}

OB acknowledges financial support from the European Research Council Starting Grant FPTOpt-277998.

\appendix

\section{Evolution equations of $\moy{{X_t}^2}$ }
\label{app:master_eq_sec_moment}

In this appendix, we give an explicit derivation of  Eq. (\ref{X2mean}), which governs the evolution of the second moment of $X_t$. We multiply the master equation (\ref{eqmaitresse}) by $(\XX\cdot \ee_1)^2$ and average over all the bath configurations $\eta$ and all the positions of the TP $\XX$. We consider separately each term of the master equation:
\begin{itemize}
  \item the left-hand-side term of (\ref{eqmaitresse}) gives the contribution:
  \begin{eqnarray}
C_\mathrm{L} & = & \sum_{\XX,\eta} (\XX \cdot \ee_1)^2 2d\tau^*\partial_t P(\XX,\eta;t) \\
 & = & 2d\tau^*\partial_t \left(  \sum_{\XX,\eta}  (\XX \cdot \ee_1)^2 P(\XX,\eta;t)  \right) \\
 &=& 2d\tau^* \frac{\dd \moy{{X_t}^2}}{\dd t}
\end{eqnarray}
  \item the first term of the right-hand-side yields
  \begin{eqnarray}
\label{ }
\fl C_1 = \sum_{\XX,\eta} (\XX \cdot \ee_1)^2 \sum_{\mu=1}^d\sum_{{\rr}\neq\XX-\ee_\mu,\XX} \left[ P(\XX,\eta^{{\rr},\mu};t)-P(\XX,\eta;t)\right] \\
\fl = \sum_{\XX} (\XX \cdot \ee_1)^2 \sum_{\mu=1}^d\sum_{{\rr}\neq\XX-\ee_\mu,\XX} \sum_{\eta} \left[ P(\XX,\eta^{{\rr},\mu};t)-P(\XX,\eta;t)\right].
\end{eqnarray}
Recalling that $\eta^{{\rr},\mu}$ a
configuration obtained from $\eta$ by exchanging the occupation
variables of two neighboring 
sites ${\rr}$ and $\rr+\ee_\mu$, we obtain  
\begin{equation}
\label{ }
 \sum_{\eta} P(\XX,\eta^{{\rr},\mu};t)=  \sum_{\eta} P(\XX,\eta;t),
\end{equation}
and we conclude that $C_1=0$.
  \item the second term of the right-hand-side of Eq. (\ref{eqmaitresse}) gives the contribution:
  \begin{align}
C_2 =& \frac{2d\tau^*}{\tau}\sum_{\mu}p_\mu\left[  \sum_{\XX,\eta} (\XX\cdot \ee_1)^2 \left(1-\eta_{\XX} \right)P(\XX-\ee_{\boldsymbol \mu},\eta;t) \right.\nonumber\\
&\left.-\sum_{\XX,\eta} (\XX\cdot \ee_1)^2\left(1-\eta_{\XX+\ee_{\mu}}\right)P(\XX,\eta;t)\right].
\label{eq:C2_inter}
\end{align}
We consider for instance the term corresponding to $\mu=1$, and consider the first sum over $\XX$ and $\eta$, in which we make the change of variable $\XX\leftarrow \XX+\ee_1$:
\begin{eqnarray}
\sum_{\XX,\eta} (\XX\cdot \ee_1)^2 \left(1-\eta_{\XX} \right)P(\XX-\ee_{1},\eta;t)  \\
=  \sum_{\XX,\eta} (\XX\cdot \ee_1 + \sigma)^2 \left(1-\eta_{\XX+\ee_1} \right)P(\XX,\eta;t) \\
  =   \sum_{\XX,\eta} [(\XX\cdot \ee_1)^2 + \sigma^2 + 2 \sigma (\XX\cdot \ee_1) ]^2 \left(1-\eta_{\XX+\ee_1} \right)P(\XX,\eta;t).
\end{eqnarray}
Finally, we get
\begin{eqnarray}
\fl   \sum_{\XX,\eta} (\XX\cdot \ee_1)^2 \left(1-\eta_{\XX} \right)P(\XX-\ee_{\mu},\eta;t)
-\sum_{\XX,\eta} (\XX\cdot \ee_1)^2\left(1-\eta_{\XX+\ee_{\mu}}\right)P(\XX,\eta;t) \\
\fl =  \sigma^2 \sum_{\XX,\eta} (1-\eta_{\XX+\ee_1}) P(\XX,\eta;t)+2\sigma \sum_{\XX,\eta} (\XX\cdot \ee_1) (1-\eta_{\XX+\ee_1})P(\XX,\eta;t).
\end{eqnarray}
Following the same procedure for the term $\mu=-1$ and noticing that the terms obtained for $\mu=\pm 2 ,\dots,\pm d$ in (\ref{eq:C2_inter}) cancel, we finally get
\begin{align}
C_2  =& \frac{2d \tau^*}{\tau} \left\{  p_1[\sigma^2(1-k_{\ee_1}(t))+2\sigma (\moy{X_t}-g_{\ee_1}(t))]  \right. \nonumber\\
&+\left. p_{-1}[\sigma^2(1-k_{\ee_{-1}}(t))-2\sigma (\moy{X_t}-g_{\ee_{-1}}(t))]  \right\},
\end{align}
where we define
\begin{equation}
\label{ }
g_{\rr} (t)= \moy{X_t \eta_{\XX_t+\rr}}.
\end{equation}
\item the third term of the right-hand-side of Eq. (\ref{eqmaitresse}) yields
\begin{align}
C_3=&2dg   \sum_{\XX,\eta} (\XX\cdot \ee_1)^2 \sum_{{\rr}\neq \XX} \left[\left(1-\eta_{\rr}\right)P(\XX,
\hat{\eta}^{{\rr}};t)-\eta_{\rr}P(\XX,\eta;t)\right] \\ 
=& 2dg   \sum_{\XX} (\XX\cdot \ee_1)^2 \sum_{{\rr}\neq \XX} \sum_{\eta}\left[\left(1-\eta_{\rr}\right)P(\XX,
\hat{\eta}^{{\rr}};t)-\eta_{\rr}P(\XX,\eta;t)\right].
\end{align}
Recalling that $\hat{\eta}^{\rr}$ is the configuration obtained from $\eta$ with the change $\eta_{\rr}\leftarrow 1-\eta_{\rr}$, we have the following equality
\begin{equation}
\label{ }
\sum_{\eta}\left(1-\eta_{\rr}\right)P(\XX,
\hat{\eta}^{{\rr}};t)=\sum_{\eta}\eta_{\rr}P(\XX,\eta;t),
\end{equation}
which yields $C_3=0$.
\item for the same reason, the fourth term will  have a zero contribution after multiplying by $(\XX\cdot \ee_1)^2$ and averaging over $\XX$ and $\eta$.
\end{itemize}
Finally, bringing together the different contributions originating from the different terms of Eq. (\ref{eqmaitresse}), we obtain
\begin{align}
\label{ }
2d \tau^*\frac{\mathrm{d}}{\mathrm{d}t} \langle {\XTP}^2\rangle =& \frac{2d \tau^*}{\tau} \left\{  p_1[\sigma^2(1-k_{\ee_1}(t))+2\sigma (\moy{X_t}-g_{\ee_1}(t))]  \right.\nonumber\\
&\left.+ p_{-1}[\sigma^2(1-k_{\ee_{-1}}(t))-2\sigma (\moy{X_t}-g_{\ee_{-1}}(t))]  \right\},
\end{align}
which is equivalent to Eq. (\ref{X2mean}):
\begin{align}
\frac{\mathrm{d}}{\mathrm{d}t} \langle {\XTP}^2\rangle =& \frac{2 \sigma}{\tau}\left\{p_1  \left[\langle \XTP \rangle-g_{\ee_1}(t)\right]-p_{-1} \left[\langle \XTP \rangle-g_{\ee_{-1}}(t)\right]\right \}   \nonumber\\
&+\frac{\sigma^2}{\tau}\left\{p_1  \left[1 -k_{\ee_1}(t)\right]+p_{-1} \left[1-k_{\ee_{-1}}(t)\right]\right\}.
%\label{X2mean}
\end{align}

\section{Evolution equations of $\gt_{\rr}$ and $\wt_{\rr}$}
\label{sec:master_eq_fluc_gt_gt}

In this appendix, we start from the master equation (\ref{eqmaitresse}) in order to derive the evolution equations satisfied by $\gt_{\rr}(t)=\moy{(X_t-\moy{X_t}) \eta_{\XX_t+\rr}}$ (Eq. (\ref{eqgtnoapprox})) and $w_{\rr}(u;t) = \moy{\ex{\ii u X_{t}}\eta_{\RTP + \rr}}$ (Eq. (\ref{wgeneral})). The derivations of these two evolution equations are similar, and we will present a general method to derive the evolution equation of the following correlation function:
\begin{equation}
\label{def_ftilde}
\widetilde{f}_{\boldsymbol{r}} (t)\equiv \moy{\mathcal{F}(X_t)\eta_{\RTP + \rr}} = \sum_{\XX,\eta} \mathcal{F}(X)\eta_{\XX + \rr} P(\XX,\eta;t),
\end{equation}
where $\mathcal{F}$ is a generic function of the TP position. The equations satisfied by $\gt_{\rr}$ and $\wt_{\rr}$ will be obtained by taking $\mathcal{F}(X) = X-\moy{X_t}$ and $\mathcal{F}(X) = \ex{\ii u X}$ respectively. Parenthetically, with this method,  one retrieves the equation satisfied by $k_{\rr}(t)$ (Eq. (\ref{eqknoapprox}))  by taking $\mathcal{F}(X) = 1$.\\

 We multiply the master equation (\ref{eqmaitresse}) by $\mathcal{F}(X) \eta_{\XX+\rr}$ and average over all the bath configurations $\eta$ and all the positions of the TP $\XX$. We consider separately each term of the master equation:
\begin{itemize}
\item the left-hand-side term of (\ref{eqmaitresse}) gives the contribution:
  \begin{align}
C_\mathrm{L}  = & \sum_{\XX,\eta} \mathcal{F}(X)\eta_{\XX+\rr}  2d\tau^*\partial_t P(\XX,\eta;t) \\
 = & 2d\tau^*  \sum_{\XX,\eta} \mathcal{F}(X)\eta_{\XX+\rr}  \partial_t P(\XX,\eta;t)\\
=& \sum_{\XX,\eta}  \eta_{\XX+\rr}  \left[  \partial_t \left( \mathcal{F}(X) P(\XX,\eta;t)  \right) -P(\XX,\eta;t) \partial_t\mathcal{F}(X) \right] 
\end{align}
In the two cases we will consider, we note that  $\partial_t \mathcal{F}(X)$ is independent of $\XX$ and $\eta$, so that we obtain
 \begin{align}
C_\mathrm{L}  = &  2d\tau^* \partial_t \left[ \sum_{\XX,\eta}\mathcal{F}(X)  \eta_{\XX+\rr} P(\XX,\eta;t)  \right] -  2d\tau^*  \partial_t \mathcal{F}(X)  \sum_{\XX,\eta}  \eta_{\XX+\rr}P(\XX,\eta;t) \\
     =& 2d\tau^*  \partial_t \widetilde{f}_{\rr} (t)- 2d\tau^*    k_{\rr}(t)\partial_t \mathcal{F}(X) .
\end{align}
where we defined $\widetilde{f}_{\rr} $ in Eq. (\ref{def_ftilde}). We consider separately the two different possible expressions of $\mathcal{F}$:
\begin{itemize}
\item if $\mathcal{F}(X)=X-\moy{X_t}$, then 
\begin{equation}
\label{ }
\partial_t \mathcal{F}(X) = -\frac{\mathrm{d} \moy{X_t}}{\mathrm{d} t},
\end{equation}
and, using Eq. (\ref{eq:def_velocity_general}), we obtain
\begin{equation}
\label{eq:gt_CL}
C_\mathrm{L}  = 2d\tau^*  \partial_t \widetilde{g}_{\rr} (t)+ 2d\tau^*\frac{\sigma}{\tau}\left\{p_1  \left[1-k_{\ee_1}(t)\right]-p_{-1} \left[1-k_{\ee_{-1}}(t)\right] \right\} k_{\rr}(t).
\end{equation}
\item if $\mathcal{F}(X)=\ex{\ii u X}$, then $\partial_t \mathcal{F}(X) = 0$, and we have the following expression of $C_\mathrm{L}$:
\begin{equation}
\label{eq:wt_CL}
C_\mathrm{L} = 2d\tau^*  \partial_t \widetilde{w}_{\rr} (t).
\end{equation}

\end{itemize}

\item the first term of the right-hand-side of the master equation becomes
\begin{equation}
C_1  =  \sum_{\XX} \mathcal{F}(X) \sum_{\mu=1}^d\sum_{{\rr'}\neq\XX-\ee_\mu,\XX} \sum_{\eta} \eta_{\XX+\rr} \left[ P(\XX,\eta^{{\rr'},\mu};t)-P(\XX,\eta;t)\right] .
\end{equation}
With an appropriate change of variable in the sum over $\eta$, we obtain
\begin{equation}
\label{eq:relation_eta_chap}
\sum_{\eta} \eta_{\XX+\rr} P(\XX, \eta^{\rr',\mu};t) = \sum_{\eta} \left( \eta^{\rr',\mu}\right)_{\XX+\rr} P(\XX, \eta ;t),
\end{equation}
so that $C_1$ is written
\begin{equation}
\label{ }
C_1  =  \sum_{\XX} \mathcal{F}(X) \sum_{\mu=1}^d\sum_{{\rr'}\neq\XX-\ee_\mu,\XX} \sum_{\eta} \left[  \left( \eta^{\rr',\mu}\right)_{\XX+\rr} -\eta_{\XX+\rr} \right] P(\XX, \eta ;t).
\end{equation}
We consider the sum over $\rr'$
\begin{eqnarray}
\fl\sum_{{\rr'}\neq\XX-\ee_\mu,\XX}  \left[  \left( \eta^{\rr',\mu}\right)_{\XX+\rr} -\eta_{\XX+\rr} \right] \nonumber \\
\fl  = \sum_{\rr'}  \left[  \left( \eta^{\rr',\mu}\right)_{\XX+\rr} -\eta_{\XX+\rr} \right] - \left[  \left( \eta^{\XX-\ee_\mu,\mu}\right)_{\XX+\rr} -\eta_{\XX+\rr} \right] - \left[  \left( \eta^{\XX,\mu}\right)_{\XX+\rr} -\eta_{\XX+\rr} \right]. \nonumber\\
\label{eq:sum_rprime}
\end{eqnarray}
Recalling that $\eta^{{\rr},\mu}$ a
configuration obtained from $\eta$ by exchanging the occupation
variables of two neighboring 
sites ${\rr}$ and $\rr+\ee_\mu$, we obtain the general relation
\begin{equation}
\label{eq:eta_complique}
(\eta^{\rr',\mu})_{\boldsymbol{x}} = \begin{cases}
 \eta_{\boldsymbol{x}}     & \text{if $\rr'\neq \boldsymbol{x},\boldsymbol{x}-\ee_\mu$ }, \\
 \eta_{\boldsymbol{x}-\ee_\mu}     & \text{if $\rr' = \boldsymbol{x}-\ee_\mu$ }, \\
 \eta_{\boldsymbol{x}+\ee_\mu}     & \text{if $\rr'= \boldsymbol{x}$ }.
 \end{cases}
\end{equation}
We then consider separately the two cases:
\begin{itemize}
  \item if $\rr=\ee_\nu$ ($\nu\in \{\pm1,\dots,\pm d\}$), using Eqs. (\ref{eq:sum_rprime}) and (\ref{eq:eta_complique}), we obtain
  \begin{equation}
\label{ }
\sum_{{\rr'}\neq\XX-\ee_\mu,\XX}  \left[  \left( \eta^{\rr',\mu}\right)_{\XX+\rr} -\eta_{\XX+\rr} \right] = \nabla_{\mu} \eta_{\XX+\rr}+\nabla_{-\mu} \eta_{\XX+\rr},
\end{equation}
where the operator $\nabla_\mu$ was defined in the main text (Eq. (\ref{eq:def_nabla})). We finally obtain
\begin{eqnarray}
C_1 & = & \sum_{\XX,\eta}\mathcal{F}(X) \sum_{\mu=1}^d \left(  \nabla_{\mu} \eta_{\XX+\rr}+\nabla_{-\mu} \eta_{\XX+\rr} \right) P(\XX, \eta ;t) \\
 & = & \sum_{\mu} \moy{ \mathcal{F}(X)  \nabla_{\mu} \eta_{\XX+\rr}}\\
 &=& \sum_{\mu} \nabla_{\mu} \widetilde{f}_{\rr}(t), \label{eq:C1_gt_bulk}
\end{eqnarray}
where the sum over $\mu$ runs over $\{ \pm1 ,\dots, \pm d\}$.
  \item if $\rr=\ee_\nu$, using again Eqs. (\ref{eq:sum_rprime}) and (\ref{eq:eta_complique}), we obtain
  \begin{equation}
\label{ }
\sum_{\mu = 1}^d \sum_{{\rr'}\neq\XX-\ee_\mu,\XX}  \left[  \left( \eta^{\rr',\mu}\right)_{\XX+\ee_\nu} -\eta_{\XX+\ee_\nu} \right] = \sum_{\mu} \nabla_{\mu} \eta_{\XX+\ee_\nu} - \nabla_{-\nu} \eta_{\XX+\ee_\nu}.
\end{equation}
  Then, $C_1$ becomes
  \begin{eqnarray}
\label{ }
C_1 &=& \moy{\mathcal{F}(X)  \left( \sum_{\mu} \nabla_{\mu}  - \nabla_{-\nu}  \right)\eta_{\XX+\ee_\nu}} \\
&=&  \sum_{\mu} \nabla_{\mu}  \widetilde{f}_{\ee_\nu}(t) - \nabla_{-\nu}  \widetilde{f}_{\ee_\nu}(t) \label{eq:C1_gt_boundary}
\end{eqnarray}
  
\end{itemize}
Finally, for any value of $\rr$, Eqs. (\ref{eq:C1_gt_bulk}) and (\ref{eq:C1_gt_boundary}) are recast under the equation
\begin{equation}
\label{eq:gt_C1}
C_1 = \left( \sum_{\mu} \nabla_{\mu} - \delta_{\rr,\ee_\mu} \nabla_{-\mu} \right) \widetilde{f}_{\rr}(t).
\end{equation}
\item we then study the second term of the right-hand-side of the master equation (\ref{eqmaitresse}), which yields the contribution
\begin{align}
\label{ }
C_2 =& \frac{2d\tau^*}{\tau}\sum_{\XX,\eta} \sum_{\mu}p_\mu\mathcal{F}(X)  \eta_{\XX+ \rr} \nonumber\\
&\times \left[\left(1-\eta_{\XX} \right)P(\XX-\ee_{\mu},\eta;t) -\left(1-\eta_{\XX+\ee_{\mu}}\right)P(\XX,\eta;t)\right] \\
=& \frac{2d\tau^*}{\tau} \left \{ \sum_{\XX,\eta} \sum_\mu p_\mu \mathcal{F}(X) \eta_{\XX+ \rr} \left(1-\eta_{\XX} \right)P(\XX-\ee_{\mu},\eta;t) \right. \nonumber\\
&\left.-\sum_{\XX,\eta} \sum_\mu p_\mu \mathcal{F}(X) \eta_{\XX+ \rr} \left(1-\eta_{\XX+\ee_\mu} \right)P(\XX,\eta;t) \right\}
\end{align}
With the change of variable $\XX\leftarrow \XX +\ee_\mu$ in the first sum, and recalling that $X=\XX+\ee_1$, we obtain
\begin{align}
\label{ }
C_2 =& \frac{2d\tau^*}{\tau} \left \{ \sum_{\XX,\eta} \sum_\mu p_\mu\mathcal{F}(X+\sigma \ee_\mu \cdot \ee_1) \eta_{\XX+ \rr+\ee_\mu} \left(1-\eta_{\XX+\ee_\mu} \right)P(\XX,\eta;t) \right. \nonumber\\
&\left.-\sum_{\XX,\eta} \sum_\mu p_\mu \mathcal{F}(X) \eta_{\XX+ \rr} \left(1-\eta_{\XX+\ee_\mu} \right)P(\XX,\eta;t) \right\} .
\end{align}
From this relation, we consider separately the different expressions of $\mathcal{F}$:
\begin{itemize}
\item in the situation where $\mathcal{F}(X) = X-\moy{X_t}$, we obtain
\begin{align}
\label{ }
C_2 =& \frac{2d\tau^*}{\tau} \left \{ \sum_{\XX,\eta} \sum_\mu p_\mu (X+\sigma \ee_\mu \cdot \ee_1 - \moy{X_t}) \eta_{\XX+ \rr+\ee_\mu} \left(1-\eta_{\XX+\ee_\mu} \right)P(\XX,\eta;t) \right. \nonumber\\
&\left.-\sum_{\XX,\eta} \sum_\mu p_\mu (X - \moy{X_t})  \eta_{\XX+ \rr} \left(1-\eta_{\XX+\ee_\mu} \right)P(\XX,\eta;t) \right\} \\
=& \frac{2d\tau^*}{\tau} \left \{ \sum_{\XX,\eta} \sum_\mu p_\mu (X - \moy{X_t}) \nabla_\mu \eta_{\XX+ \rr} \left(1-\eta_{\XX+\ee_\mu} \right)P(\XX,\eta;t) \right. \nonumber\\
&\left.+\sigma \sum_{\XX,\eta} \sum_\mu p_\mu (\ee_\mu \cdot \ee_1)  \eta_{\XX+ \rr+\ee_\mu} \left(1-\eta_{\XX+\ee_\mu} \right)P(\XX,\eta;t) \right\}.
\end{align}
As $\ee_\mu\cdot \ee_1$ is equal to $\pm1$ for $\mu=\pm1$ and 0 otherwise, we finally obtain
\begin{eqnarray}
\fl C_2 =  \frac{2d\tau^*}{\tau}  \sum_{\mu} p_\mu \moy{(X_t-\moy{X_t}) (1-\eta_{\XX_t+\ee_\mu}) \nabla_{\XX_t +\rr}} \nonumber\\
 +  \frac{2d\tau^*}{\tau}  \sigma \left[ p_1 \moy{(1-\eta_{\XX_t+\ee_1})\eta_{\XX_t+\rr+\ee_1}}   -p_{-1} \moy{(1-\eta_{\XX_t+\ee_{-1}})\eta_{\XX_t+\rr+\ee_{-1}}} \right]. \nonumber\\
\label{eq:gt_C2}
\end{eqnarray}

\item in the situation where $\mathcal{F}(X) =\ex{\ii u X}$, we obtain
\begin{align}
\label{ }
C_2 =& \frac{2d\tau^*}{\tau} \left \{ \sum_{\XX,\eta} \sum_\mu p_\mu \ex{\ii u X} \ex{\ii u \sigma (\ee_\mu \cdot \ee_1)} \eta_{\XX+ \rr+\ee_\mu} \left(1-\eta_{\XX+\ee_\mu} \right)P(\XX,\eta;t) \right. \nonumber\\
&\left.-\sum_{\XX,\eta} \sum_\mu p_\mu  \ex{\ii u X}  \eta_{\XX+ \rr} \left(1-\eta_{\XX+\ee_\mu} \right)P(\XX,\eta;t) \right\} \\
=& \frac{2d\tau^*}{\tau}  \sum_{\XX,\eta} \sum_\mu p_\mu  \ex{\ii u X} \left(1-\eta_{\XX+\ee_\mu} \right) \left[  \ex{\ii u \sigma (\ee_\mu \cdot \ee_1)}   \eta_{\XX+\rr+\ee_\mu}- \eta_{\XX+\rr}    \right]P(\XX,\eta;t)  \\
=&  \frac{2d\tau^*}{\tau}  \sum_{\XX,\eta} \sum_\mu p_\mu  \ex{\ii u X} \left(1-\eta_{\XX+\ee_\mu} \right) \nabla_\mu \eta_{\XX+\rr}    P(\XX,\eta;t) \nonumber\\
&  +  \frac{2d\tau^*}{\tau}  \sum_{\XX,\eta} \sum_\mu p_\mu  \ex{\ii u X} \left(1-\eta_{\XX+\ee_\mu} \right)  \left( \ex{\ii u \sigma (\ee_\mu \cdot \ee_1)}-1 \right)  \eta_{\XX+\rr+\ee_\mu}P(\XX,\eta;t)  .
\end{align}
As $\ee_\mu\cdot \ee_1$ is equal to $\pm1$ for $\mu=\pm1$ and 0 otherwise, we finally obtain
\begin{eqnarray}
\fl C_2 =  \frac{2d\tau^*}{\tau}  \sum_{\mu} p_\mu \moy{\ex{\ii u X_t}(1-\eta_{\XX_t+\ee_\mu}) \nabla_{\XX_t +\rr}} \nonumber\\
 +  \frac{2d\tau^*}{\tau}  \sum_{\epsilon=\pm1}  p_\epsilon \left( \ex{\ii u \epsilon \sigma} -1 \right) \moy{\ex{\ii u X_t}(1-\eta_{\XX_t+\ee_\epsilon})\eta_{\XX_t+\rr+\ee_\epsilon}}.  \nonumber\\
\label{eq:wt_C2}
\end{eqnarray}

\end{itemize}

\item the third term yields the contribution $C_3$:
\begin{align}
\label{ }
C_3 =& 2dg  \sum_{\XX,\eta} \mathcal{F}(X) \eta_{\XX_t+\rr}  \sum_{{\rr'}\neq \XX} \left[\left(1-\eta_{\rr'}\right)P(\XX, \hat{\eta}^{{\rr'}};t)-\eta_{\rr'}P(\XX,\eta;t)\right]\\
=& 2dg \sum_{\XX}  \sum_{{\rr'}\neq \XX}   \mathcal{F}(X)  \sum_\eta \eta_{\rr'}\left[ \left( {\hat{\eta}}^{\rr'}\right)_{\XX+\rr}-\eta_{\XX+\rr}   \right]P(\XX,\eta;t),
\end{align}
where we used again Eq. (\ref{eq:relation_eta_chap}). By the definition of ${\hat{\eta}}^{\rr'}$ (configuration obtained from $\eta$ with the change $\eta_{\rr'}\leftarrow 1-\eta_{\rr'}$), we write
\begin{equation}
\label{ }
 \left(  {\hat{\eta}}^{\rr'}\right)_{\boldsymbol{x}}=
\begin{cases}
\eta_{\boldsymbol{x}}   & \text{if $\boldsymbol{x}\neq\rr'$}, \\
1- \eta_{\boldsymbol{x}}       & \text{if $\boldsymbol{x}=\rr'$}.
\end{cases}
\end{equation}
The expression of $C_3$ becomes
\begin{equation}
\label{ }
C_3 = 2dg  \sum_{\XX}  \mathcal{F}(X)  \sum_{\eta} \eta_{\XX+\rr} (1-2\eta_{\XX+\rr}) P(\XX,\eta;t).
\end{equation}
As $\eta_{\XX+\rr} \in \{ 0,1\}$, then $(\eta_{\XX+\rr})^2=\eta_{\XX+\rr} $, and we finally obtain
\begin{equation}
\label{eq:gt_C3}
C_3=-2dg \widetilde{f}_{\rr}(t)
\end{equation}
\item the computation of the contribution $C_4$ of the fourth term of the master equation is similar to that of $C_3$, and we obtain
\begin{equation}
\label{ }
C_4 = 2df \sum_{\XX} \mathcal{F}(X)  \sum_{\eta} (1-\eta_{\XX+\rr}) (1-2\eta_{\XX+\rr}) P(\XX,\eta;t).
\end{equation}
Using again $(\eta_{\XX+\rr})^2=\eta_{\XX+\rr} $, we obtain
\begin{equation}
\label{ }
(1-\eta_{\XX+\rr}) (1-2\eta_{\XX+\rr}) = 1-\eta_{\XX+\rr},
\end{equation}
and, finally,
\begin{equation}
\label{eq:gt_C4}
C_4 = 2df \left\{    \moy{\mathcal{F}(X_t)} - \widetilde{f}_{\rr} (t)\right\}.
\end{equation}
Note that when $\mathcal{F}(X) = X-\moy{X_t}$, then $ \moy{\mathcal{F}(X_t)} =0$.
\end{itemize}

Finally, writing from the master equation (\ref{eqmaitresse}) we obtain  the relation 
\begin{equation}
\label{ }
C_{\mathrm{L}} = C_1+C_2+C_3+C_4,
\end{equation}
and we obtain the evolution equations satisfied by $\gt_{\rr}(t)$ and $\wt_{\rr}(t)$:
\begin{itemize}
\item using  (\ref{eq:gt_CL}), (\ref{eq:gt_C1}), (\ref{eq:gt_C2}), (\ref{eq:gt_C3}) and (\ref{eq:gt_C4}), we get
\begin{align}
2d\tau^*\partial_t\gt_{\rr}(t)=&\sum_\mu\left(\nabla_\mu-\delta_{{\rr},{\boldsymbol e_\mu}}\nabla_{-\mu}\right)\gt_{\rr}(t)-2d(f+g)\gt_{\rr}(t)\nonumber\\
&+\frac{2d\tau^*}{\tau}\sum_{\mu} p_\mu \moy{\delta\XTP(1-\eta_{\RTP+\ee_\mu})\nabla_\mu\eta_{\RTP+\rr}}\nonumber\\
&+\frac{2d\tau^*}{\tau}\sigma\left[p_1 \moy{ (1-\eta_{\RTP+\ee_1})\eta_{\RTP+\rr+\ee_1}} - p_{-1}\moy{ (1-\eta_{\RTP+\ee_{-1}})\eta_{\RTP+\rr+\ee_{-1}}}\right]\nonumber\\
&-\frac{2d\tau^*}{\tau}\sigma\left\{p_1[1-k_{\ee_1}(t)]-p_{-1}[1-k_{\ee_{-1}}(t)]\right\}k_{\rr}(t).
%\label{eqgtnoapprox}
\end{align}
which is equivalent to Eq. (\ref{eqgtnoapprox}), presented in the main text.

\item using  (\ref{eq:wt_CL}), (\ref{eq:gt_C1}), (\ref{eq:wt_C2}), (\ref{eq:gt_C3}) and (\ref{eq:gt_C4}), we get
\begin{eqnarray}
\fl 2d\tau^*\partial_tw_{\rr}(u;t)=\left(\sum_\mu\nabla_\mu-\delta_{{\rr},{\emuu}}\nabla_{-\mu}\right)w_{\rr}(u;t)-2d(f+g)w_{\rr}(u;t)+2df\moy{\ex{\ii u X_{t}}}\nonumber\\
+\frac{2d\tau^*}{\tau}\sum_{\mu}p_\mu \moy{ \ex{\ii u X_{t}}(1-\eta_{\RTP+\emuu})\nabla_\mu\eta_{\RTP+{\rr}}} \nonumber\\
+\frac{2d\tau^*}{\tau}\sum_{\epsilon=\pm1}p_\epsilon\left(\ex{\ii u \epsilon\sigma}-1\right)\moy{\ex{\ii u X_{t}}(1-\eta_{\RTP+\ee_\epsilon})\eta_{\RTP+{\rr}+\ee_\epsilon}}
\end{eqnarray}
which is equivalent to Eq. (\ref{wgeneral}), presented in the main text.

\end{itemize}

\section{Explicit expressions of $\gt_1$ and $\gt_{-1}$ in one dimension}
\label{sec:J1}

In this appendix, we solve the linear system of four equations satisfied by $\alpha$, $\beta$, $\gt_1$ and $\gt_{-1}$, obtained by substituting Eq. (\ref{npositive}) into Eq. (\ref{CL1}),  Eq. (\ref{nnegative}) into Eq. (\ref{CL2}) , and writing 
Eq. (\ref{npositive}) for $n=1$ and Eq. (\ref{nnegative}) for $n=-1$. From Eq. (\ref{npositive}) written with $n=1$, and using the definition of $W$ from Eq. (\ref{def_W}), we obtain the following expression of $\alpha$ in terms of $\gt_1$ and $\gt_{-1}$:
\begin{equation}
\label{solalpha}
\alpha = \frac{a_0}{A_1 r_1-A_{-1}/r_1} + \left( \frac{1}{r_1} + \frac{a_1}{A_1 r_1- A_{-1}/r_1}   \right) \gt_1 + \frac{a_{-1}}{A_1 r_1- A_{-1}/r_1} \gt_{-1},
\end{equation}
where the coefficients $a_0$ and $a_{\pm 1}$ are defined by
\begin{eqnarray}
\fl a_0  =  K_+ \frac{2\tau^*}{\tau}\sigma \left[ p_1 (1-\rho-K_+ r_1)(r_1-1) - p_{-1}\left(1-\rho-\frac{K_-}{r_1} \right)(r_1^{-1}-1)\right] \\
\fl a_1 =  -\frac{2\tau^*}{\tau}p_1 K_+ (r_1-1) \\
\fl a_{-1}  =  -\frac{2\tau^*}{\tau}p_{-1} K_+ (r_1^{-1}-1). \\
\end{eqnarray}
Similarly, from Eq. (\ref{nnegative}) written with $n=-1$, and using the definition of $W'$ from Eq. (\ref{def_Wp}), we obtain the following expression of $\beta$ in terms of $\gt_1$ and $\gt_{-1}$:
\begin{equation}
\label{solbeta}
\beta = -\frac{b_0}{A_1 r_2- A_{-1}/r_2} - \frac{b_{1}}{A_1 r_2- A_{-1}/r_2}  \gt_1 + \left( r_2- \frac{b_{-1}}{A_1 r_2- A_{-1}/r_2}   \right)\gt_{-1},
\end{equation}
where the coefficients $b_0$ and $b_{\pm 1}$ are defined by
\begin{eqnarray}
\fl b_0  =  K_- \frac{2\tau^*}{\tau}\sigma \left[ p_1 (1-\rho-K_+ r_1)(r_2-1) - p_{-1}\left(1-\rho-\frac{K_-}{r_1} \right)(r_2^{-1}-1)\right] \\
\fl b_1 =  -\frac{2\tau^*}{\tau}p_1 K_- (r_2-1) \\
\fl b_{-1}  =  -\frac{2\tau^*}{\tau}p_{-1} K_- (r_2^{-1}-1). \\
\end{eqnarray}
 In the boundary conditions (Eqs. (\ref{CL1}) and (\ref{CL2})), replacing $\gt_{2}$ and $\gt_{-2}$ by
 \begin{eqnarray}
\fl \widetilde{g}_2=\alpha {r_1}^2-\frac{2W}{A_1r_1-A_{-1}r_1^{-1}}{r_1}^2 \\
\fl \widetilde{g}_{-2}=\frac{\beta}{{r_2}^2}+\frac{2W'}{A_1r_2-A_{-1}r_2^{-1}}  \frac{1}{{r_2}^2},
\end{eqnarray}
and using the expression of $\alpha$ and $\beta$ in terms of $\gt_{1}$ and $\gt_{-1}$ (Eqs. (\ref{solalpha}) and (\ref{solbeta})), we obtain the following closed system of linear equations satisfied by $\gt_{1}$ and $\gt_{-1}$:
\begin{equation}
\label{lineargt}
\begin{cases} 
A \gt_1 + B \gt_{-1} = C \\
D \gt_1 + E \gt_{-1} = F 
  \end{cases}
\end{equation}
where we define
\begin{align}
 A\equiv&\frac{2\tau^*}{\tau}\sigma p_1K_+(r_1-1)\frac{A_1r_1^2}{A_1r_1-A_{-1}r_1^{-1}}+A_1r_1 \nonumber\\
&-\left(A_{-1}+2(f+g)+\frac{2\tau^*}{\tau}p_1(\rho+K_+r_1^2)\right),
\end{align}
\begin{equation}
 B\equiv\frac{2\tau^*}{\tau}\sigma p_{-1}K_+(r_1^{-1}-1)\frac{A_1r_1^2}{A_1r_1-A_{-1}r_1^{-1}}+\frac{2\tau^*}{\tau}p_{-1}(\rho+K_+r_1),
\end{equation}
\begin{eqnarray}
\fl C\equiv\frac{2\tau^*}{\tau}\sigma (p_1(1-\rho-K_+r_1)-p_{-1}(1-\rho-K_-r_2^{-1}))(\rho+K_+r_1)\nonumber\\
-\frac{2\tau^*}{\tau}\sigma p_1(1-\rho-K_+r_1)(\rho+K_+r_1^2)\nonumber\\
\fl +\frac{2\tau^*}{\tau}\sigma\left(p_1(1-\rho-K_+r_1)K_+r_1-p_{-1}\left(1-\rho-\frac{K_-}{r_2}\right)\frac{K_+}{r_1}\right)\frac{A_1r_1^2}{A_1r_1-A_{-1}r_1^{-1}},
\end{eqnarray}
\begin{eqnarray}
D\equiv-\frac{2\tau^*}{\tau}\sigma p_{1}K_-(r_2-1)\frac{A_{-1}r_2^{-2}}{A_1r_2-A_{-1}r_2^{-1}}+\frac{2\tau^*}{\tau}p_{1}(\rho+K_-r_2^{-1}),
\end{eqnarray}
\begin{align}
 E\equiv&-\frac{2\tau^*}{\tau}\sigma p_{-1}K_-(r_2^{-1}-1)\frac{A_{-1}r_2^{-2}}{A_1r_2-A_{-1}r_2^{-1}}+A_{-1}r_2^{-1}\nonumber\\
&-\left(A_{1}+2(f+g)+\frac{2\tau^*}{\tau}p_{-1}(\rho+K_-r_2^{-2})\right)
\end{align}
and
\begin{eqnarray}
\fl F\equiv\frac{2\tau^*}{\tau}\sigma (p_1(1-\rho-K_+r_1)-p_{-1}(1-\rho-K_-r_2^{-1}))(\rho+K_-r_2^{-1})\nonumber\\
+\frac{2\tau^*}{\tau}\sigma p_{-1}(1-\rho-K_-r_2^{-1})(\rho+K_-r_2^{-2})\nonumber\\
\fl -\frac{2\tau^*}{\tau}\sigma\left(p_1(1-\rho-K_+r_1)K_-r_2-p_{-1}\left(1-\rho-\frac{K_-}{r_2}\right)\frac{K_-}{r_2}\right)\frac{A_{-1}r_2^{-2}}{A_1r_2-A_{-1}r_2^{-1}}.
\end{eqnarray}

Solving the linear system given in Eq. (\ref{lineargt}), it is finally found that:
\begin{equation}
\label{g1}
\widetilde{g}_{1}=\frac{CE-BF}{AE-BD} \;\;{\rm and }\;\;\widetilde{g}_{-1}=\frac{AF-CD}{AE-BD}.
\end{equation}

$K_+$, $K_-$, $A_1$ and $A_{-1}$ have been determined in Section \ref{velocity_1D}. Using their expressions, we compute $\gt_1$ and $\gt_{-1}$, and finally compute $K$ with the formula:
\begin{equation}
\label{Ktr}
K=\frac{\sigma^2}{2\tau}\left(p_1(1-\rho-K_+r_1)+p_{-1}\left(1-\rho-K_-r_2^{-1}\right)\right)-\frac{\sigma}{\tau}\left(p_1\widetilde{g}_{1}-p_{-1}\widetilde{g}_{-1}\right).
\end{equation}
Note that $\alpha$ and $\beta$ can be deduced straightforwardly from their relations with $\gt_1$ and $\gt_{-1}$ (Eqs. (\ref{solalpha}) and (\ref{solbeta})).

\section{Explicit expressions of $\mt_1$ and $\mt_{-1}$ in one dimension}
\label{app:cumul3}

In the main text we showed that, for $n>0$, $\mt_n$ had the following form

\begin{equation}
	\mt_n=\Gamma r_1^n+(a_+ n^2+b_+ n)r_1^n,
	\label{solmpos}
	\end{equation}
	with
	\begin{eqnarray}
	a_+&=&\frac{1}{2}\frac{C_2^+}{A_1r_1-A_{-1}r_1^{-1}},\\
	b_+&=&\frac{1}{A_1r_1-A_{-1}r_1^{-1}}[C_1^+-a_+(A_1r_1+A_{-1}r_1^{-1})],
	\end{eqnarray}

and where the quantities $C_1^+$ and $C_2^+$ are defined by
\begin{eqnarray}
	\fl C_1^+ = \frac{2\tau^*}{\tau}\left[ p_1 \mt_1K_+(r_1-1)+p_{-1}\mt_{-1}K_+(r_1^{-1}-1)\right] \nonumber \\
	\fl -\frac{2\tau^*}{\tau}p_1\sigma\left\{2[(1-k_1)(\alpha r_1-\alpha' r_1-\alpha)-\gt_1 K_+(r_1-1)]+\sigma(1-k_1)K_+(r_1-1)\right\} \nonumber \\
	\fl-\frac{2\tau^*}{\tau}p_{-1}\sigma\left\{-2[(1-k_{-1})(\alpha r_1^{-1} +\alpha' r_1^{-1}-\alpha)-\gt_{-1} K_+(r_1^{-1}-1)]+\sigma(1-k_{-1})K_+(r_1^{-1}-1)\right\} \nonumber \\
	\fl\equiv  \frac{2\tau^*}{\tau}\left[ p_1 \mt_1K_+(r_1-1)+p_{-1}\mt_{-1}K_+(r_1^{-1}-1)\right]+\psi_+,
\end{eqnarray}
and
\begin{eqnarray}
	C_2^+ &=& -\frac{4\tau^*}{\tau}\sigma\alpha'[p_1(1-k_1)(1-r_1)-p_{-1}(1-k_{-1})(1-r_1^{-1})].
\end{eqnarray}

The expression of the coefficient $\alpha$ was given in  \ref{sec:J1} (Eq. (\ref{solalpha})) and we defined $\alpha'\equiv W/(A_1 r_1-A_{-1}{r_1}^{-1})$.

For $n<0$, a similar resolution leads to the expression

\begin{equation}
	\mt_n=\delta r_2^n+(a_- n^2+b_- n)r_2^n,
	\label{solmneg}
	\end{equation}
	with
	\begin{eqnarray}
	a_-&=&\frac{1}{2}\frac{C_2^-}{A_1r_2-A_{-1}r_2^{-1}},\\
	b_-&=&\frac{1}{A_1r_2-A_{-1}r_2^{-1}}[C_1^- -a_-(A_1r_2+A_{-1}r_2^{-1})],
	\end{eqnarray}
and where the quantities $C_1^-$ and $C_2^-$ are defined by
\begin{eqnarray}
	\fl C_1^- = \frac{2\tau^*}{\tau}\left[ p_1 \mt_1K_-(r_2-1)+p_{-1}\mt_{-1}K_-(r_2^{-1}-1)\right] \nonumber \\
	\fl-\frac{2\tau^*}{\tau}p_1\sigma\left\{2[(1-k_1)(\beta r_2-\beta' r_2-\beta)-\gt_1 K_-(r_2-1)]+\sigma(1-k_1)K_-(r_2-1)\right\} \nonumber \\
	\fl-\frac{2\tau^*}{\tau}p_{-1}\sigma\left\{-2[(1-k_{-1})(\beta r_2^{-1} +\beta' r_2^{-1}-\beta)-\gt_{-1} K_-(r_2^{-1}-1)]+\sigma(1-k_{-1})K_-(r_2^{-1}-1)\right\} \nonumber\\
	\fl \equiv \frac{2\tau^*}{\tau}\left[ p_1 \mt_1K_-(r_2-1)+p_{-1}\mt_{-1}K_-(r_2^{-1}-1)\right]+\psi_-,\\
\end{eqnarray}
and
\begin{eqnarray}
	C_2^- = -\frac{4\tau^*}{\tau}\sigma\beta'[p_1(1-k_1)(1-r_2)-p_{-1}(1-k_{-1})(1-r_2^{-1})].
	\end{eqnarray}

The expression of the coefficient $\beta$ was given in  \ref{sec:J1} (Eq. (\ref{solbeta})) and we defined $\beta'=W'/(A_1 r_2-A_{-1}{r_2}^{-1})$.

As explained in the main text, we finally a linear system of four equations with unknowns $\widetilde{m}_1$, $\widetilde{m}_{-1}$, $\Gamma$ and $\delta$ (Eq. (\ref{systemem})), involving the following quantities:
\begin{equation}
M_{13}=\frac{2\tau^*}{\tau}\frac{r_1}{A_1r_1-A_{-1}r_1^{-1}}[p_1K_+(r_1-1)-1]
\end{equation}
\begin{equation}
M_{14}=\frac{2\tau^*}{\tau}\frac{r_1}{A_1r_1-A_{-1}r_1^{-1}}p_1K_+(r_1^{-1}-1)
\end{equation}
\begin{equation}
M_{23}=-\frac{2\tau^*}{\tau}\frac{1}{r_2}\frac{1}{A_1r_2-A_{-1}r_2^{-1}}p_1K_-(r_2-1)
\end{equation}
\begin{equation}
M_{24}=-\frac{2\tau^*}{\tau}\frac{1}{r_2}\frac{1}{A_1r_2-A_{-1}r_2^{-1}}[p_{-1}K_-(r_2^{-1}-1)+1]
\end{equation}
\begin{equation}
M_{33}=-\left(A_{-1}+2(f+g)+\frac{2\tau^*}{\tau}p_1k_2\right) +\frac{2\tau^*}{\tau}\frac{2A_1}{A_1r_1-A_{-1}r_1^{-1}}p_1K_+(r_1-1)
\end{equation}
\begin{equation}
M_{34}=\frac{2\tau^*}{\tau}p_{-1}k_1 +\frac{2\tau^*}{\tau}\frac{2A_1}{A_1r_1-A_{-1}r_1^{-1}}p_{-1}K_+(r_1^{-1}-1)
\end{equation}
\begin{equation}
M_{43}=\frac{2\tau^*}{\tau}p_{-1}k_1 -\frac{2\tau^*}{\tau}\frac{2A_{-1}}{A_1r_2-A_{-1}r_2^{-1}}p_{1}K_-(r_2-1)
\end{equation}
\begin{equation}
M_{44}=-\left(A_{1}+2(f+g)+\frac{2\tau^*}{\tau}p_{-1}k_{-2}\right) -\frac{2\tau^*}{\tau}\frac{2A_{-1}}{A_1r_2-A_{-1}r_2^{-1}}p_{-1}K_-(r_2^{-1}-1)
\end{equation}
\begin{equation}
Y_1=\frac{2a_+A_{-1}-\psi_+r_1}{A_1r_1-A_{-1}r_1^{-1}}
\end{equation}
\begin{equation}
Y_2=-\frac{2a_-A_{1}-\psi_- r_2^{-1}}{A_1r_2-A_{-1}r_2^{-1}}
\end{equation}
\begin{equation}
Y_3=-\varphi_1-2A_1\frac{a_+(A_1r_1-3A_{-1}r_1^{-1})+\psi_-}{A_1r_1-A_{-1}r_1^{-1}}
\end{equation}
\begin{equation}
Y_4=-\varphi_{-1}-2A_{-1}\frac{a_-(3A_1r_2-A_{-1}r_2^{-1})+\psi_-}{A_1r_2-A_{-1}r_2^{-1}}
\end{equation}

\section{Equations satisfied by $\wt_{\pm1}(u)$: additional definitions}
\label{detailedimplicit}

In this appendix, we give the explicit expressions of the coefficients $B_j$, $C_j$, $D_j$, $E_j$ involved in Eq. (\ref{eq:system_w_alpha}).

\begin{equation}
B_1=A_1+\frac{2\tau^*}{\tau}p_1(\ex{\ii u\sigma}-1)(1-k_1)
\end{equation}
\begin{equation}
B_2=A_1+A_{-1}+2(f+g)+\frac{2\tau^*}{\tau}p_1(\ex{\ii u\sigma}-1)[1-\wt_1(u)]+\frac{2\tau^*}{\tau}p_{-1}(\ex{-\ii u\sigma}-1)[1-\wt_{-1}(u)]
\end{equation}
\begin{equation}
B_3=A_{-1}+\frac{2\tau^*}{\tau}p_{-1}(\ex{-\ii u\sigma}-1)(1-k_{-1})
\end{equation}
\begin{equation}
B_4=\frac{2\tau^*}{\tau}\left\{p_1[k_1-\wt_1(u)](r_1\ex{\ii u\sigma}-1)+p_{-1}[k_{-1}-\wt_{-1}(u)](r_1^{-1}\ex{-\ii u\sigma}-1)\right\}
\end{equation}
\begin{equation}
B_5=2f+\frac{2\tau^*}{\tau}\rho\left\{p_1[k_1-\wt_1(u)](\ex{\ii u\sigma}-1)+p_{-1}[k_{-1}-\wt_{-1}(u)](\ex{-\ii u\sigma}-1)\right\}\end{equation}
\begin{equation}
C_4=\frac{2\tau^*}{\tau}\left\{p_1[k_1-\wt_1(u)](r_2\ex{\ii u\sigma}-1)+p_{-1}[k_{-1}-\wt_{-1}(u)](r_2^{-1}\ex{-\ii u\sigma}-1)\right\}
\end{equation}
\begin{equation}
D_1 = A_1+\frac{2\tau^*}{\tau}p_1(\ex{\ii u\sigma}-1)(1-k_1)
\end{equation}
\begin{equation}
D_2=A_{-1}+2(f+g)+\frac{2\tau^*}{\tau}p_1\ex{\ii u\sigma}k_2+\frac{2\tau^*}{\tau}\left[p_1(\ex{\ii u\sigma}-1)+p_{-1}(\ex{-\ii u\sigma}-1)\right]
\end{equation}
\begin{equation}
D_3=\frac{2\tau^*}{\tau}p_{-1}k_1
\end{equation}
\begin{equation}
D_4=2f+\frac{2\tau^*}{\tau}\left[p_1\ex{\ii u\sigma}k_1 k_2-p_{-1}k_1 k_{-1}\right]
\end{equation}
\begin{equation}
D_5=\frac{2\tau^*}{\tau}p_1(\ex{\ii u\sigma}-1)
\end{equation}
\begin{equation}
D_6=\frac{2\tau^*}{\tau}p_{-1}(\ex{-\ii u\sigma}-1)
\end{equation}
\begin{equation}
E_1=\frac{2\tau^*}{\tau}p_1k_{-1}
\end{equation}
\begin{equation}
E_2=A_1+2(f+g)+\frac{2\tau^*}{\tau}p_{-1}\ex{-\ii u\sigma}k_{-2}+\frac{2\tau^*}{\tau}\left[p_1(\ex{\ii u\sigma}-1)+p_{-1}(\ex{-\ii u\sigma}-1)\right]
\end{equation}
\begin{equation}
E_3 = A_{-1}+\frac{2\tau^*}{\tau}p_{-1}(\ex{-\ii u\sigma}-1)(1-k_{-1})
\end{equation}
\begin{equation}
E_4=2f+\frac{2\tau^*}{\tau}\left[p_{-1}\ex{-\ii u\sigma}k_{-1} k_{-2}-p_{1}k_1 k_{-1}\right]
\end{equation}

\section{Algorithm and numerical methods}
\label{app:numericalmeth}

To sample exactly the master equation (\ref{eqmaitresse}), we generate a sequence of random numbers $(\tau, x, \mu)$ with the joint probability density function $p(\tau, x, \mu)$ where $p(\tau, x, \mu)\dd \tau$ is the probability at time $t$ that the next event occurs in the infinitesimal time interval $[t+\tau, t+\tau+\dd\tau]$, at site $x$, and is of type $\mu$ (i.e. a diffusion event, an absorption event or a desorption event). We write
\begin{equation}
\label{ }
p(\tau, x, \mu)=p_1(\tau)p_2(x|\tau)p_3(\mu|,x,\tau)
\end{equation}
where
\begin{itemize}
	\item $p_1(\tau)\dd\tau$ is the probability at time $t$ that the next event occurs in the time interval $[t+\tau,t+\tau+\dd\tau]$.
	\item $p_2(x|\tau)\dd\tau$ is the probability that the next event occurs at site $x$, knowing that it occurs during the  time interval $[t+\tau,t+\tau+\dd\tau]$.
	\item $p_3(\mu|x,\tau)\dd\tau$ is the probability that the next event is of type $\mu$ knowing that it occurs during the  time interval $[t+\tau,t+\tau+\dd\tau]$ and at site $x$.
\end{itemize}
Writing $c_{x,\mu}$ the transition rate of event $\mu$ at site $x$, we define
\begin{eqnarray}
r_x & \equiv & \sum_\mu c_{x,\mu}, \\
R & \equiv & \sum_x r_x.
\end{eqnarray}
$r_x$ is then the total rate of the events at site $x$, and $R$ is the total rate of all the events on all lattice sites. The quantities $\tau$, $x$ and $\mu$ are then respectively drawn from the following distributions:
\begin{eqnarray}
p_1(\tau) & = & R\ex{-R\tau}, \\
p_2(x|\tau) & = & \frac{r_x}{R},\\
p_3(\mu|x, \tau) &=& \frac{c_{x,\mu}}{r_x}.
\end{eqnarray}
The algorithm is as follows. We build a lattice of length $2L+1$ (the spacing of the lattice $\sigma$ is taken equal to 1). The boundary conditions are periodic, and $L$ is chosen to be large enough so that we can consider the lattice as infinite (in the results presented below, $L\geq250$). The initial condition is the following: the TP is initially at the origin and at each site different from the origin, a particle is set with probability $\rho$. We chose a final simulation time $t_\mathrm{max}$. At each step of the simulation, and as long as $t<t_\mathrm{max}$, the algorithm follows these steps:
\begin{enumerate}
	\item Set $R=0$.
	\item For each $x\in[0,2L+1]$, compute $r_x$:
	\begin{itemize}
		\item if the site $x$ is occupied by a bath particle, three events are possible : a jump to the left, a jump to the right, or a desorption event (respectively labeled by 1, 2 and 3). The associate rates $c_{x,\mu}$ are
		\begin{eqnarray}
		c_{x,1} & = & \frac{1}{2}(1-g)(1-\eta_{x-1}) \\
		c_{x,2} & = & \frac{1}{2}(1-g)(1-\eta_{x+1}) \\
		c_{x,3} & = & g
		\end{eqnarray}
		so that $r_x= \frac{1}{2}(1-g)(1-\eta_{x-1})+\frac{1}{2}(1-g)(1-\eta_{x+1})+g$.
		\item if the site $x$ is occupied by the TP, two events are possible : a jump to the left, or a jump to the right. The local rate is then $r_x=(1-p_1)(1-\eta_{x-1})+p_1(1-\eta_{x+1})$.
		\item if the site is empty, the only possible event is an absorption event, and $r_x=f$.
	\end{itemize}
	\item Compute the total rate $R=\sum_{x}r_x$.
	\item Draw $\tau$ from the distribution $p_1(\tau)=R\ex{-R\tau}$.
	\item Draw $x$ from the distribution $p_2(x|\tau)=r_x/R$.
	\item Draw the event $\mu$ from the distribution $p_3(\mu|x,\tau)=c_{x,\mu}/r_x$.
	\item Update the lattice occupation after the realization of the event (note that such an algorithm is rejection-free).
	\item Increase the time $t \leftarrow t+\tau$.
\end{enumerate}
We finally keep track of the TP position $\XTP$ with time. With a large number of realizations, we  sample the p.d.f. of $\XTP$.

\section{Range of parameters for which $\KTP>1/2$}
\label{sec:crit_2}

In section \ref{sec:crit_1}, we determined the critical value of the desorption rate $g_\mathrm{c}$ allowing the emergence of a nontrivial maximum value for the function $K(\rho)$ for a given value of the bias $\delta$. As a complementary approach, for a fixed value of the bias, we study the domain of the plane $(g,\rho)$ in which the diffusion coefficient is greater than $1/2$ , which is its value when there is no bath particle.

\begin{figure}
	\begin{center}
		\includegraphics[width=11cm]{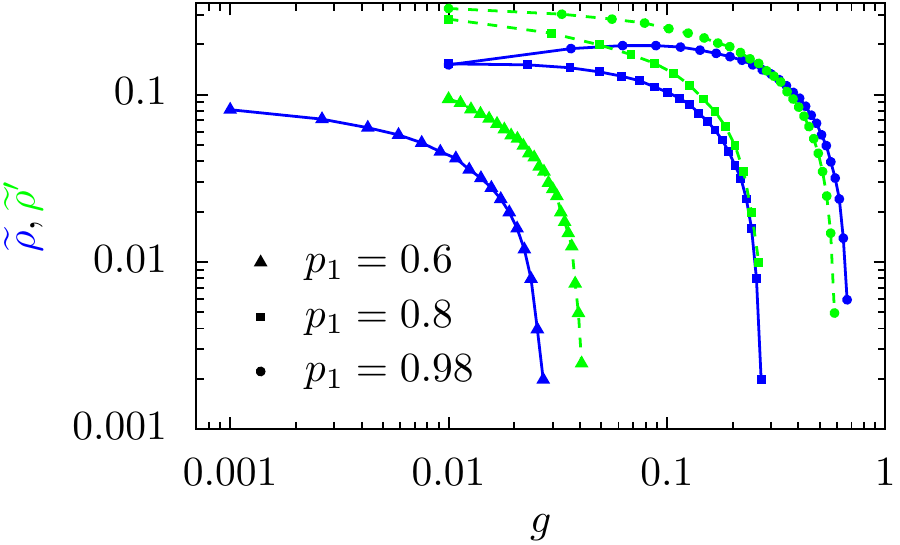}
		\caption{Critical value of the density $\widetilde{\rho}$ below which the diffusion coefficient is greater than $1/2$, as a function of $g$ and for different values of the bias $p_1$ (curves in blue).  Critical value of the density $\widetilde{\rho}'$ below which $\gt_{-1}-\gt_{-2} \geq 0$, as a function of $g$ and for different values of the bias $p_1$ (curves in green).}
		\label{fig:g_rhotilde}
	\end{center}
\end{figure}

In the case where the curve $\KTP(\rho)$ displays an extremum value, the value of the diffusion coefficient is greater than $1/2$ in the range $[0,\widetilde{\rho}]$, where $\widetilde{\rho}$ is a function of $p_1$ and $g$ only. From the analytical curves $\KTP(\rho)$, we can then deduce the values of $\widetilde{\rho}$ as a function of $g$ for different values of $p_1$, which are plotted on Fig. \ref{fig:g_rhotilde}.

As it was shown on Fig. \ref{fig:gtilde_vary_g}, for a given value of $p_1$ and $g$, there exists a value of $\rho$ below which the difference $\gt_{-1}-\gt_{-2}$ is negative, i.e. below which there is a minimum for $\gt_n$ in the range $n<0$. This critical value of $\rho$ will be denoted by $\widetilde{\rho}'$. In order to compare the domain of parameters giving respectively $\KTP>1/2$ and $\gt_{-1}-\gt_{-2} \geq 0$, we also represent on Fig. \ref{fig:g_rhotilde} the curves of $\widetilde{\rho}'$ as a function of $g$ for different values of $p_1$.

The domains in the plane $(\rho,g)$ for which $K>1/2$ and for which  $\gt_{-1}-\gt_{-2}<0$ then seem to be correlated.

\section*{References}

\bibliographystyle{unsrt}
\bibliography{../../library}

\end{document}